\documentclass[journal]{IEEEtran} %
\usepackage{setspace}
\usepackage{graphicx}
\usepackage{balance}
\usepackage{cite}
\usepackage{amsthm}
\usepackage{amssymb}
\usepackage{amsmath}
\usepackage{amsfonts}
\usepackage{amsmath}
\usepackage{graphicx}
\usepackage{mathtools}
\usepackage{multirow}
\usepackage{epstopdf}
\usepackage{subfigure}
\usepackage{array}
\usepackage[table]{xcolor}
\usepackage{caption}
\graphicspath{{Figures/}}
\usepackage{algorithm}
\usepackage{algorithmic}
\usepackage{makecell} 
\usepackage{arydshln}
\usepackage{adjustbox}
\usepackage{enumitem}
\usepackage{rotating}
\usepackage{tabularx}
\usepackage{csquotes}

\usepackage{threeparttable}
\usepackage{pifont}
\usepackage{array}
 \usepackage{bbding}
\usepackage{wasysym}
\usepackage{color, colortbl}
\usepackage{booktabs}

\DeclarePairedDelimiter\floor{\lfloor}{\rfloor}


\newenvironment{abbreviations}{\begin{list}{}{%
			\setlength{\labelwidth}{3cm}\setlength{\leftmargin}{\labelwidth+\labelsep}%
			\setlength{\itemsep}{0pt}}}{\end{list}}
		
\theoremstyle{boldremark}
\newtheorem{remark}{Remark}

\IEEEpeerreviewmaketitle

\begin{document}  

	\title{Multidimensional Index Modulation for 5G and Beyond Wireless Networks}
	

	\author{Seda~Do\u{g}an, Armed~Tusha, Ertu{g}rul Ba{s}ar\IEEEmembership{, Senior {Member}, IEEE} and H\"{u}seyin Arslan\IEEEmembership{, Fellow, IEEE}
		\thanks{Seda~Do\u{g}an and Armed~Tusha are with the Communications, Signal Processing, and Networking Center (CoSiNC),
			Department of Electrical and Electronics Engineering, Istanbul Medipol University, 34810, Istanbul, Turkey (e-mail: {sdogan, atusha}@st.medipol.edu.tr).}
		\thanks{Ertu{g}rul Basar is with the Communications Research and Innovation Laboratory (CoreLab), Department of Electrical and Electronics Engineering, Ko\c{c}¸ University, Sariyer 34450, Istanbul, Turkey (e-mail: ebasar@ku.edu.tr).}
		\thanks{H\"{u}seyin Arslan is with the Communications, Signal Processing, and Networking Center (CoSiNC),
			Department of Electrical and Electronics Engineering, Istanbul Medipol University, Istanbul, 34810, Turkey and also with the Department of Electrical Engineering, University of South Florida, Tampa, FL, 33620, USA (e-mail: huseyinarslan@medipol.edu.tr).}}
		
\markboth{\MakeLowercase{\MakeUppercase{D}o\u{g}an} \MakeLowercase{\textit{et al.}:  \MakeUppercase{I}ndex \MakeUppercase{M}odulation for \MakeUppercase{5G} and \MakeUppercase{B}eyond}}{}
	
%
	
	\maketitle
	
	\begin{abstract}

Index modulation (IM) provides a novel way for the transmission of additional data bits via the indices of the available transmit entities compared to classical communication schemes. This study examines the flexible utilization of existing IM techniques in a comprehensive manner to satisfy the challenging and diverse requirements of 5G and beyond services. After spatial modulation (SM), which transmits information bits through antenna indices, application of IM to orthogonal frequency division multiplexing (OFDM) subcarriers has opened the door for the extension of IM into different dimensions, such as radio frequency (RF) mirrors, time slots, codes, and dispersion matrices. Recent studies have introduced the concept of multidimensional IM by various combinations of one-dimensional IM techniques to provide higher spectral efficiency (SE) and better bit error rate (BER) performance at the expense of higher transmitter (Tx) and receiver (Rx) complexity.
Despite the ongoing research on the design of new IM techniques and their implementation challenges, proper use of the available IM techniques to address different requirements of 5G and beyond networks is an open research area in the literature. For this reason, we first \textcolor{black}{provide the dimensional-based categorization of available IM domains and review the existing IM types regarding this categorization.} Then, we develop a framework that investigates the efficient utilization of these techniques and establishes a link between the IM schemes and 5G services, namely enhanced mobile broadband (eMBB), massive machine-type communications (mMTC), and ultra-reliable low-latency communication (URLLC). Additionally, this work defines key performance indicators (KPIs) to quantify the advantages and disadvantages of IM techniques in time, frequency, space, and {code} dimensions. Finally, future recommendations are given regarding the design of flexible IM-based communication systems for 5G and beyond wireless networks.
	
	\end{abstract}


\section*{Nomenclature}
\begin{abbreviations}
	\item[{3GPP}]{3rd Generation Partnership Project}
	\item[{4G}]{4th Generation}
	\item[{5G}]{5th Generation}
    \item[{6G}]{6th Generation}
	\item[{BER}]{Bit Error Rate}
	\item[{BLER}]{Block Error Rate}
	\item[{BPSK}]{Binary Phase Shift Keying}
	\item[{BS}]{Base Station}
	\item[{CFO}]{Carrier Frequency Offset}
	\item[{CIM-SM}]{Code Index Modulation with SM}
    \item[{CIM-SS}]{Code Index Modulation Spread Spectrum}
    \item[{CI-OFDM-IM}]{Coordinate Interleaved OFDM-IM}
    \item[{CFIM}]{Code-Frequency Index Modulation}
	\item[{CP}]{Cyclic Prefix}
	\item[{CR}]{Cognitive Radio}
	\item[{CS}]{Compressed Sensing}
	\item[{CSI}]{Channel State Information}
	\item[{DL}]{Downlink}
	\item[{DM}]{Dispersion Matrix}
	\item[{DMBM}]{Differential Media-based Modulation}
	\item[{DM-OFDM}]{Dual-Mode OFDM}
	\item[{DM-SCIM}]{Dual-Mode Single Carrier with IM}
	\textcolor{black}{\item[{DP-SM}]{	Dual Polarized SM }}
	\item[{DS-SS}]{Direct Sequence Spread Spectrum}
	\item[{DSM}]{Differential Spatial Modulation}
	\item[{DSTSK}]{Differential Space-Time Shift Keying}
	\item[{EE}]{Energy Efficiency}
	\item[{eMBB}]{Enhanced Mobile Broadband}
	\item[{ESIM-OFDM}]{Enhanced Subcarrier Index Modulation OFDM}
	\item[{ESM}]{Enhanced Spatial Modulation}
	\item[{FD}]{Full-Duplex}
    \item[{FFT}]{Fast Fourier Transform}
	\item[{FSK}]{Frequency Shift Keying}
	\item[{FTN-IM}]{Faster-than-Nyquist Signaling with IM }
	\item[{GB}]{Grant-Based}
	\item[{GCIM-SS}]{Generalized CIM-SS}
	\item[{GF}]{Grant-Free}
	\item[{GFDM}]{Generalized Frequency Division
		Multiplexing}
	\item[{GFDM-IM}]{GFDM with IM}
	\item[{GFDM-SFIM}]{GFDM with Space-Frequency IM}
	\textcolor{black}{\item[{GPQSM}]{Generalized Precoding-aided QSM}}
	\textcolor{black}{\item[{GPSM}]{Generalized Precoding-aided SM}}
    \item[{GSFIM}]{Generalized Space-Frequency IM}
	\item[{GSM}]{Generalized Spatial Modulation}
	\item[{GSSK}]{Generalized Space Shift Keying}
	\item[\textcolor{black}{GSTFIM}]{\textcolor{black}{Generalized Space-Time-Frequency IM}}
	\item[{GSTSK}]{Generalized Space Time Shift Keying}
	\item[IAI]{Inter-Antenna Interference}
	\item[IAS]{Inter-Antenna Synchronization}
    \item[ICI]{Inter-Carrier Interference}
	\item[{IFFT}]{Inverse Fast Fourier Transform}
	\item[{IM}]{Index Modulation}
	\item[{IMMA}]{Index Modulation-based Multiple Access}
	\textcolor{black}{\item [IM-OFDM-SS] {Index Modulated OFDM Spread Spectrum}}
	\item[{IoT}]{Internet-of-Things}
	\item[{I/Q}]{In-phase and Quadrature}
	\item[{ISI}]{Inter-Symbol Interference}
	\item[{ISM-OFDM}]{SM-OFDM with Subcarrier IM}
	\item[{IUI}]{Inter-User Interference}
	\textcolor{black}{\item[{JA-MS-STSK}]{Joint Alphabet MS-STSK}}
    \textcolor{black}{\item[{JA-STSK}]{Joint Alphabet STSK}}
	\item[{KPI}]{Key Performance Indicator}
	\item[{LLR}]{Log-Likelihood Ratio}
	\textcolor{black}{\item[{LMG-SSTSK}]{ Layered Multi-Group Steered STSK}}
	\textcolor{black}{\item[{LMS-GSTSK}]{ Layered Multi-Set GSTSK}}
	\item[{LTE}]{Long Term Evolution}
	\textcolor{black}{\item[{L-OFDM-IM}]{ Layered OFDM-IM}}
	\item[MAC]{Medium Access Control}
	\item[{MA-SM}]{Multiple Active Spatial Modulation}
	\item[{MBM}]{Media-based Modulation}
	\item[{MIMO}]{Multiple-Input Multiple-Output}
	\item[{ML}]{Maximum Likelihood}
	\item[{MM-OFDM}]{Multiple-Mode OFDM}
	\item[{mMTC}]{Massive Machine-Type Communications}
	\item[{mmWave}]{Millimeter Wave}
	\item[{MRC}]{Maximum Ratio Combining}
	\item[{MSF-STSK}]{Multi-Space-Frequency STSK}
	\item[{MS-STSK}]{Multi-Set STSK}
	\item[{NB-IoT}]{Narrowband Internet-of-Things}
	\item[{NOMA}]{Non-Orthogonal Multiple Access}
	\item[{NR}]{New Radio}
	\item[{OFDM}]{Orthogonal Frequency Division
		Multiplexing}
	\item[{OFDMA}]{Orthogonal Frequency Division
		Multiple Access}
	\item[{OFDM-GIM}]{OFDM with Generalized IM}
	\item[{OFDM-IM}]{OFDM with Index Modulation}
	\item[{OFDM-I/Q-IM}]{OFDM with I/Q Index Modulation}
	\item[{OFDM-ISIM}]{OFDM with Interleaved Subcarrier IM}
    \item[{OFDM-STSK}]{OFDM with STSK}
    \item[{OFDM-STSK-IM}]{OFDM-STSK with Frequency IM}
	\item[{PAPR}]{Peak-to-Average Power Ratio}
	\item[{PHY}]{Physical Layer}
	\item[{PLS}]{Physical Layer Security}
	\item[\textcolor{black}{PM}] {\textcolor{black}{Polarization Modulation}} 
	\item[\textcolor{black}{PolarSK}] {\textcolor{black}{Polarization Shift Keying}} 
	\item[{PSK}]{Phase Shift Keying}
	\item[\textcolor{black}{PSM}] {\textcolor{black}{Precoded Spatial Modulation}} 
	\item[{PSK}]{Phase Shift Keying}
	\item[\textcolor{black}{PU}] {\textcolor{black}{Primary User}}
	\item[{QAM}]{Quadrature Amplitude Modulation}
	\item[\textcolor{black}{QCM}] {\textcolor{black}{Quadrature Channel Modulation}} 
	\item[{QSM}]{Quadrature Spatial Modulation}
	\item[{RF}]{Radio Frequency}
	\item[{Rx}]{Receiver}
	\item[{SC}]{Single Carrier}
	\item[{SC-FDMA}]{Single Carrier Frequency Division
		Multiple Access}
    \item[{SC-IM}]{Single Carrier with Index Modulation}
    \item[{SCS}]{Subcarrier Spacing}
	\item[{SD}]{Spatial Diversity}
	\item[{SE}]{Spectral Efficiency}
    \item[{SFSK}]{Space-Frequency Shift Keying}
	\item[{SIM-OFDM}]{Subcarrier Index Modulation OFDM}
	\item[{SM}]{Spatial Modulation}
	\item[{SM-MBM}]{SM with MBM}
	\item[{SMX}]{Spatial Multiplexing}
	\item[{SPSK}]{Space-Polarization Shift Keying }
	\item[{SSK}]{Space Shift Keying}
	\item[{STBC}]{Space-Time Block Coding}
	\item[\textcolor{black}{STBC-QSM}] {\textcolor{black}{Space-Time Block Coded QSM}} 
	\item[{STBC-SM}]{Space-Time Block Coded SM}
	\item[\textcolor{black}{STCM}] {\textcolor{black}{Space-Time Channel Modulation}} 
	\item[{STFSK}]{Space-Time-Frequency Shift Keying}
	\item[{STSK}]{Space-Time Shift Keying}
	\item[\textcolor{black}{ST-MBM}] {\textcolor{black}{Space-time MBM}}
	\item[\textcolor{black}{ST-QSM}] {\textcolor{black}{Space-time QSM}}
    \item[\textcolor{black}{SU}] {\textcolor{black}{Secondary User}}
	\item[\textcolor{black}{SURLLC}] {\textcolor{black}{Secure URLLC}}
	\item[\textcolor{black}{TCM}] {\textcolor{black}{Trellis Coded Modulation}} 
	\item[\textcolor{black}{TCSM}] {\textcolor{black}{Trellis Coded SM}} 
	\item[\textcolor{black}{TC-QSM}] {\textcolor{black}{Trellis Coded QSM}} 
	\item[{TI-MBM}]{Time-Indexed MBM}
	\item[{TI-SM}]{Time-Indexed SM}
	\item[{TI-SM-MBM}]{Time-Indexed SM-MBM}
	\item[{TTI}]{Transmission Time Interval}
	\item[{Tx}]{Transmitter}
	\item[{UE}]{User Equipment}
    \item[{UL}]{Uplink}
	\item[{URLLC}]{Ultra-Reliable Low-Latency Communication}
	\item[{V2X}]{Vehicle-to-Everything}
	\item[\textcolor{black}{VLC}] {\textcolor{black}{Visible Light Communication}} 
	\item [{V-BLAST}] {Vertical Bell Laboratories Layered Space-Time}
	\item[\textcolor{black}{ZTM-OFDM-IM}] {\textcolor{black}{Zero-Padded Tri-Mode IM-aided OFDM}}

\end{abbreviations}

	\section{Introduction} \label{introduction}
	
\textcolor{black}{The rapid growth of smart devices and services, such as sensors, smartphones, ultra-high-definition video streaming, wearable electronics, autonomous driving, drones,  Internet-based smart homes, and a broad range of augmented reality \& virtual reality applications, \textcolor{black}{leads to} enormous data traffic that cannot be handled by 4th generation (4G) Long Term Evolution (LTE)-based communication systems \cite{8985528}. Nearly ten-fold increase in {the} global mobile data traffic is envisioned from 2020 (57 exabytes/month) to 2030 (5016 exabytes/month) \cite{itu2030, Cisco2, Cisco3}. In an effort to support this overwhelming data volume and variety in 5th generation (5G) New Radio (NR) systems, International Telecommunication Union classifies numerous applications and use-cases into three main services, named enhanced mobile broadband (eMBB), massive machine-type communication (mMTC), and ultra-reliable low-latency communication (URLLC) \cite{Ref4,ITU2}. eMBB use-case is a continuation of 4G LTE systems with moderate reliability and high data rate requirements. In mMTC, providing service to a massive number of {user equipments (UEs)} is the main priority, while URLLC is the most challenging service for 5G New Radio (NR) systems due to the strict requirements for ultra-reliability with low-latency \cite{Ref2, Ref4, ITU2, 8412469}}. \textcolor{black}{In line with this trend, securing communication is essential for wireless networks, but it is disregarded during 5G standardizations. Thus, security is one of the pivotal requirements that need to be satisfied in the 6th generation (6G) and beyond networks, especially for {scenarios with} URLLC \cite{6gbe}.  In short, a surprisingly diverse range of requirements poses two main challenges for researchers and engineers worldwide: \textit{1) providing service in the presence of intensive data traffic over the current communication systems,} and \textit{2) supporting a wide range of applications and use-cases.}} 

\subsection{IM Can Revive Wireless Networks}

Many researchers are putting tremendous effort on finding solutions to the aforementioned problems. In order to accomplish the former \textit{(1)}, spectrum-efficient approaches have been proposed by academia and industry, such as massive multiple-input multiple-output (MIMO) signaling, millimeter wave (mmWave) communications, and non-orthogonal multiple access (NOMA) schemes \cite{6515173,8114722}. Besides high spectral efficiency (SE), 5G NR and beyond communication systems require a much more flexible structure for the latter \textit{(2)}. In this spirit, plenty of work has been done to achieve flexibility in the medium access control (MAC) layer and physical layer (PHY) for the future generation systems \cite{Ref5,3ggp1,3gpp2}. In order to attain {a high} degree of freedom in {the} MAC layer, various radio resource management and multi-user scheduling techniques have been studied in the literature \cite{8004168, 7565189,8263598}. \textcolor{black}{From the perspective of {the} PHY design, {multi-}numerology concept has been adopted for conventional orthogonal frequency division multiplexing (OFDM) systems \cite{3gpp2,3gpp3}. Variable subcarrier spacings (SCSs) up to 120 kHz and mini-slot design that can consist of 2, 4 or 7 OFDM symbols have been introduced to meet different latency constraints.}

In addition to the waveform-based approach, the use of different modulation options in the PHY has been also considered as the source of flexibility to support various {UE demands}. Three traditional modulation schemes, quadrature amplitude modulation (QAM), frequency shift keying (FSK), and phase shift keying (PSK) offer different performance under a variety of radio channel conditions \cite{1094370, 789668}. Specifically, transmission with lower order modulations provides robustness against channel impairments at the cost of decrease in SE, while the use of higher modulation orders maximizes achievable data rate under {satisfactory} channel conditions. Therefore, adaptive modulation selection {with respect to} the channel conditions has been adopted in {modern} communication systems \cite{1094370}.
However, flexibility stemming from the adaptive selection of modulation schemes is limited by {the} modulation order {in these traditional schemes}. On the other hand, recently reputed index modulation (IM) techniques have drawn substantial attention from the researchers because of their {inherently} flexible structure and promising advantages in terms of SE, energy efficiency (EE), complexity and reliability \cite{SM, 7509396, 8004416}.

\begin{figure}
	\centering
	{\includegraphics[scale = 0.59]{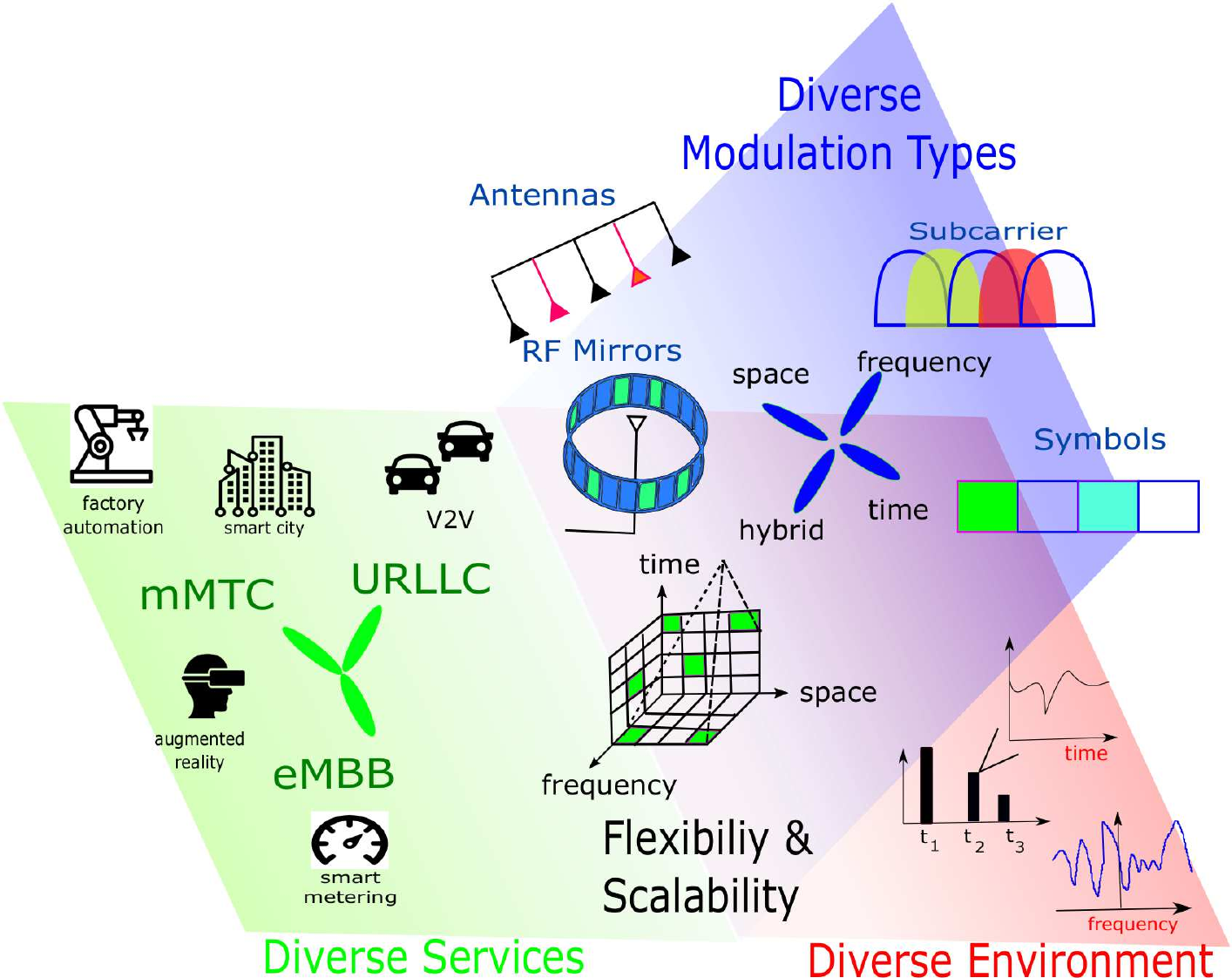}}
	\caption{Diverse IM variants for various services and channel conditions.}
	\label{Fig:1}
\end{figure}

The main idea of IM is the utilization of the available transmit entities, such as antenna indices in space, subcarrier indices in frequency, and slot indices in time, to convey additional information bits {along with} the conventional $M$-ary symbols \cite{SM, 7509396, 8004416}. Application of IM in various domains {enables an attractive} trade-off among SE, EE, transceiver complexity, interference immunity, and transmission reliability \cite{7544555, MEMISOGLU}. {Therefore}, the concept of IM has introduced new research {opportunities} for 5G {and beyond} wireless systems. Inspired by the performance of {one-dimensional} IM types, such as spatial modulation (SM) and OFDM with IM (OFDM-IM), the multidimensional IM concept, which is composed of {various combinations of one-dimensional} IM options, has been introduced in recent studies. Despite {the} ongoing active research on IM techniques, {the following important questions remain unanswered within the context of emerging IM solutions:} how can the vast flexibility of IM be utilized for 5G and beyond systems, and how can IM solutions fulfill the broad range of user and application demands, as delineated in Fig.~\ref{Fig:1}.

\newcolumntype{C}[1]{>{\centering\let\newline\\\arraybackslash\hspace{0pt}}m{#1}}
\begin{table*}[h!]
	\caption{\textcolor{black}{Summary of the existing magazine and survey articles on IM techniques}}
	\centering
	\small
	\renewcommand{\arraystretch}{1.15}
	\begin{tabular}{|C{1.4cm} |C{1.2cm}|C{1cm}|C{4,2 cm}|p{7.9cm}|}   	
		\hline 
		{\textbf{{Type}}}	&{\textbf{{Ref.}}} & {\textbf{{Year}}} & {\textbf{{IM Domain(s)}}} &\makecell{\textbf{{ Main Contributions}}} \\  \hline \hline 
		\multirow{12}{*}{{\makecell{\textcolor{black}{{\normalsize{Magazine}}} \\ \textcolor{black}{\normalsize{Articles}}}}} & \multirow{3}{*}{\textcolor{black}{\cite{6094024}}} & \multirow{3}{*}{\textcolor{black}{2011}} &\multirow{3}{*}{\textcolor{black}{Space}}  &  \textcolor{black}{The principles of SM have been introduced and its advantages \& disadvantages have been compared with the conventional MIMO.} 	     \\  \cline{2-5} \cline{2-5}	
		&\multirow{3}{*}{\cite{7509396} }  & \multirow{3}{*}{2016} & \multirow{3}{*}{\makecell{Space, and Frequency}} & The potentials and implementation of IM techniques including SM and  OFDM-IM for multi-user MIMO and multi carrier communication systems have been investigated.  \\ \cline{2-5} \cline{2-5}  
		&\multirow{2}{*}{\cite{8353362}}& \multirow{2}{*}{2018}&  \multirow{2}{*}{Space, Time, and Frequency} & Space, time, frequency, space-time and space-frequency IM options have been compared in terms of SE and EE. \\ \cline{2-5} \cline{2-5}    
		&\multirow{3}{*}{\cite{8410878} } & \multirow{3}{*}{2018} &\multirow{3}{*}{Space, and Frequency}  &  Space and frequency domain IM techniques have been subsumed by considering vehicular and railway communications. \\  \cline{2-5} \cline{2-5}
		&\multirow{4}{*}{\textcolor{black}{\cite{9076117}}} & \multirow{4}{*}{\textcolor{black}{2020}} &\multirow{4}{*}{\textcolor{black}{Frequency}}  &  \textcolor{black}{Frequency domain IM types including OFDM-IM, DM-OFDM, and ZTM-OFDM-IM have been evaluated for future wireless networks including cognitive radios, relay networks and multi-user communications.} \\ \hline \hline 
		\multirow{20}{*}{{\makecell{\textcolor{black}{\normalsize{Survey}} \\ \textcolor{black}{\normalsize{Articles}}}}}
		&\multirow{2}{*}{\textcolor{black}{\cite{6123167}}}& \multirow{2}{*}{\textcolor{black}{2012}} & \multirow{2}{*}{\textcolor{black}{Space, and Time}} & \textcolor{black}{The potentials of MIMO signaling through STSK have been discussed, and unified STSK framework has been introduced.} \\ \cline{2-5} \cline{2-5}
		&\multirow{3}{*}{\textcolor{black}{\cite{   6678765}}} & \multirow{3}{*}{\textcolor{black}{2014}} &\multirow{3}{*}{\textcolor{black}{Space}}  &  \textcolor{black}{General aspects of SM-MIMO, its experimental evaluation and its integration with the promising communication systems have been presented.} \\  \cline{2-5} \cline{2-5}	  
		&\multirow{2}{*}{\textcolor{black}{\cite{6823072}}} & \multirow{2}{*}{\textcolor{black}{2015}} &\multirow{2}{*}{\textcolor{black}{Space}}  &  \textcolor{black}{Transceiver design, spatial constellation optimization and link adaptation techniques for SM have been assessed.} 	     \\  \cline{2-5} \cline{2-5}	 
		&\multirow{3}{*}{\cite{8004416 }} & \multirow{3}{*}{2016} & \multirow{3}{*}{ \makecell{Space, Time, Frequency,\\ and Channel} } &  Practical application of SM, OFDM-IM, and CM have been elaborated, and practical issues such as ICI, PAPR have been discussed.\\ \cline{2-5} \cline{2-5}
		&\multirow{2}{*}{\cite{8070115} } & \multirow{2}{*}{2017} &  \multirow{2}{*}{Space, Time, and Frequency} & Benefits and fundamental limitations have been discussed for SM, OFDM-IM and SC-IM. \\  \cline{2-5} \cline{2-5}
		&\multirow{3}{*}{\cite{8417419}} & \multirow{3}{*}{2018} & {\multirow{3}{*}{ \makecell{Space, Time, Frequency,\\ and Channel} } } & Potential challenges and open issues for channel domain IM types have been provided in addition to space, time and frequency domain IM techniques. \\ \cline{2-5} \cline{2-5} 
		&\multirow{2}{*}{\textcolor{black}{\cite{8315127}}} & \multirow{2}{*}{\textcolor{black}{2018}} &\multirow{2}{*}{\textcolor{black}{\makecell{Space, Time, Frequency, \\ Code, and Channel}}}  &  \textcolor{black}{50 years history of SM and IM concepts, and the road from permutation modulation to OFDM-IM have been revealed.} \\ \cline{2-5} \cline{2-5}
		&\multirow{4}{*}{\textcolor{black}{\cite{8765384}}} & \multirow{4}{*}{\textcolor{black}{2019}} &\multirow{4}{*}{\textcolor{black}{Space}}  &  \textcolor{black}{ The integration of recent SM variants with promising technologies, such as CS theory, and their application in the future emerging systems, such as optical wireless communications, have been discussed. }\\ \hline 		
	\end{tabular}
	\label{Tab:Summary}
\end{table*} 

\begin{table*}
	\caption{\textcolor{black}{ The presented IM techniques in the existing magazine and survey articles, and the comparison with the proposed survey}}
	\centering
	\setlength{\arrayrulewidth}{0.76pt}
	\renewcommand{\arraystretch}{1.1} 
	\fontsize{7pt}{7.5pt}
	\selectfont
	\begin{tabular}
		{|>{\color{black}}C{3cm}||
			>{\columncolor{blue!8}}>{\color{black}}c|
			>{\columncolor{white}}>{\color{black}}c|
			>{\columncolor{blue!8}}>{\color{black}}c|
			>{\columncolor{white}}>{\color{black}}c|
			>{\columncolor{blue!8}}>{\color{black}}c|
		    >{\columncolor{white}}>{\color{black}}c|
			>{\columncolor{blue!8}}>{\color{black}}c|
			>{\columncolor{white}}>{\color{black}}c|
			>{\columncolor{blue!8}}>{\color{black}}c|
			>{\columncolor{white}}>{\color{black}}c|
			>{\columncolor{blue!8}}>{\color{black}}c|
		    >{\columncolor{white}}>{\color{black}}c|
			>{\columncolor{blue!8}}>{\color{black}}c|
			>{\columncolor{white}}>{\color{black}}c|}
		\hline
		\multirow{2}{*}{\textbf{{IM Techniques}}}
		& \multicolumn{5}{c|}{{\textcolor{black}{Magazine Articles}}}
		& \multicolumn{9}{c|}{{\textcolor{black}{Survey Articles}}}\\
		\cline{2-15}
		& {\textcolor{black}{\cite{6094024}}}
		& \textcolor{black}{\cite{7509396}} 
		& \textcolor{black}{\cite{8353362}}
		& \textcolor{black}{\cite{8410878} }
		&  {\textcolor{black}{\cite{9076117}}}
		&  {\textcolor{black}{\cite{6123167}}} 
		& {\textcolor{black}{\cite{6678765}}} 
		& {\textcolor{black}{\cite{6823072}}} 
		&  \textcolor{black}{\cite{8004416}} 
		& \textcolor{black}{\cite{8070115}} 
		& \textcolor{black}{\cite{8417419}} 
		& {\textcolor{black}{\cite{8315127}}} 
		& {\textcolor{black}{\cite{8765384}}} 
		& This\\
	    \hline  \hline
		{SSK}\cite{956483} & \ding{51}	&  & &  \ding{51}& & \ding{51}& \ding{51}&  \ding{51}  & \ding{51}& \ding{51} & \ding{51} &\ding{51}& \ding{51} & \ding{51}\\
		\hline
		{GSSK}\cite{4699782} & \ding{51}& & & &  &\ding{51} & \ding{51}&  \ding{51} &  \ding{51}&  \ding{51}&\ding{51} &\ding{51} & \ding{51}& \ding{51} \\
		\hline
		{SM}\cite{4382913} & \ding{51} & \ding{51} & \ding{51}& \ding{51}& &\ding{51} & \ding{51}&  \ding{51} & \ding{51} & \ding{51}&  \ding{51} &  \ding{51}   & \ding{51} &  \ding{51}\\
		\hline
		{GSM} \cite{5757786} & &\ding{51} & \ding{51}& \ding{51}&  &\ding{51} & \ding{51} &  \ding{51}  &  \ding{51}&  \ding{51}&  \ding{51} & \ding{51} & \ding{51}& \ding{51}\\
		\hline
		{MA-SM}\cite{6166339} & & \ding{51}& & & & & & \ding{51}& \ding{51} &  \ding{51} & \ding{51} & \ding{51}  & \ding{51}&  \ding{51}\\
		\hline
		{QSM} \cite{6868290} & & \ding{51}& \ding{51}& \ding{51}& & & &  & \ding{51} &  \ding{51} &  \ding{51}&  \ding{51} & \ding{51}&  \ding{51}\\
		\hline
		\textcolor{black}{PSM \cite{5956573}}  &  & & \ding{51}& & & & & & & & \ding{51}  &  & &  \ding{51}\\ 
		\hline
		\textcolor{black}{GPSM \cite{6644231}}  & & & & & & & & &  & & \ding{51}&  & &  \ding{51}\\ 
		\hline
		\textcolor{black}{GPQSM \cite{7467510}}  & & & & & & & & & & &  \ding{51} &  & &  \ding{51}\\ 
		\hline
		\textcolor{black}{TCSM \cite{5508985}}  & & & & &  &  \ding{51}& &  \ding{51}&  \ding{51}& &  &  & \ding{51}&  \ding{51}\\ 
		 \hline
		\textcolor{black}{TC-QSM \cite{TCQSM1}}  & & & & &  & & & & & & &  & &  \ding{51}\\ 
		 \hline
		{ESM}\cite{7084604} & & & & &  & & &  & \ding{51} & & \ding{51} & & \ding{51}& \ding{51}\\
		\hline 
		STBC-SM\cite{5672371} & & & \ding{51}& &  &  & \ding{51}  & \ding{51} & \ding{51} &  \ding{51} &  & \ding{51}& \ding{51}& \ding{51}\\
		\hline
		\textcolor{black}{STBC-QSM\cite{8277677}}  & & &  & & & & & & & & & & &  \ding{51}\\
		\hline	
		{DSM}\cite{6879496} & & & \ding{51}& \ding{51}& & & & & \ding{51}&  \ding{51}&  &  & \ding{51} & \ding{51}\\ 
		\hline
		{RIS-IM} \cite{8981888} & & & & & & & & & & & & & &  \ding{51}\\ 
		\hline
		{{{ISM-OFDM} \cite{7440707}}} & & & & & & & & & \ding{51} & &  & & & \ding{51}\\
	\hline
	{SIM-OFDM} \cite{5449882} & & & & & & & &  & \ding{51} &  \ding{51} & \ding{51} &  \ding{51} & \ding{51}  & \ding{51}\\
	\hline
	{ESIM-OFDM} \cite{6162549} & & & & & & & & &  \ding{51} &  \ding{51} & \ding{51}& \ding{51} &  & \ding{51}\\
	\hline
	{OFDM-IM} \cite{Basar1} & &\ding{51} &\ding{51} & \ding{51}&  \ding{51} & &  & & \ding{51}&  \ding{51}  &  &  \ding{51} & \ding{51} &  \ding{51}\\
	\hline
	{OFDM-ISIM}\cite{6841601}  & & & \ding{51}& \ding{51}& & & & & \ding{51}&  \ding{51}  &  & \ding{51}& &   \ding{51}\\ 
	\hline
    {CI-OFDM-IM} \cite{7086323} & & & & & & & & &  \ding{51}& &  &  \ding{51} & \ding{51} &   \ding{51}\\ 
	\hline
	{OFDM-I/Q-IM}  \cite{7112187, 7230238} & & & \ding{51}& & & & &  & \ding{51} & & & \ding{51} & \ding{51}&   \ding{51}\\ 
	\hline
	{OFDM-GIM} \cite{7112187}&  & \ding{51}& & & & & & &  \ding{51} &  \ding{51} &  \ding{51} &  \ding{51} & &   \ding{51}\\
	\hline
	\textcolor{black}{{OFDM-GIM-I/Q} \cite{7470953}} &  & & & & & & & & &  &  \ding{51} &  & &   \ding{51}\\
	\hline
	{DM-OFDM}\cite{7547943} & & & & & \ding{51} & &  & & \ding{51}  &  \ding{51} &  \ding{51} &  \ding{51} & \ding{51} & \ding{51} \\
	\hline
    {MM-OFDM} \cite{7936676} & & & & & \ding{51} & &  & & \ding{51} & &  \ding{51} &  \ding{51} & \ding{51}& \ding{51}\\
    	\hline
    \textcolor{black}{{ZTM-OFDM-IM}  \cite{8254931}} & & & & & \ding{51} & & & &  & & \ding{51} &  & & \ding{51}\\
	\hline
	\textcolor{black}{{L-OFDM-IM}  \cite{8734769}} & & & & & & & & &  & &  &  & & \ding{51}\\
	 \hline
	{GFDM-IM}  \cite{7848916} & & & & &  & & & & \ding{51}& &  &  & \ding{51} &  \ding{51}\\
	\hline
	{{GSFIM} \cite{7277106}}  & &\ding{51}& \ding{51} & & & & & \ding{51} & \ding{51}& &  & & \ding{51} &  \ding{51} \\
	\hline
    \textcolor{black}{{GSTFIM} \cite{8487049}} & & & & & & & & & & & & &  \ding{51} &  \ding{51} \\
	\hline
	{{GFDM-SFIM} \cite{8277679}} & & & & & & & & &  \ding{51} & &  & & &  \ding{51}\\
	\hline
	{{MBM}\cite{6620786}} & & & & \ding{51}& & & & &  \ding{51} & &  &  \ding{51} & \ding{51} &  \ding{51}\\
	\hline
	{DMBM}\cite{7887715} & & & & & & & & & \ding{51} & & \ding{51} & & &  \ding{51}\\
	\hline	
	\textcolor{black}{QCM \cite{8014408}} & & & & & &  & & &  & &  & & &  \ding{51}\\
	\hline	
	{{TI-MBM} \cite{8110611}} & & & & & & & & &  \ding{51} & &  & & &  \ding{51}\\ 
	\hline
	{{TI-SM-MBM} \cite{8110611}} & & & & & & & & & \ding{51} & &  & & &  \ding{51}\\ 
	\hline
	{SM-MBM} \cite{8110611} & & & & & & & & &  \ding{51} & &  &  & &  \ding{51}\\
	\hline
	{TI-SM} \cite{7925922} & & & & & & & & & \ding{51} & &  & & &  \ding{51} \\
	\hline
	{SC-IM}  \cite{7738501} & & & & & & &  &  &  \ding{51} & \ding{51} & \ding{51} &  \ding{51} & \ding{51} & \ding{51}\\
	\hline
	{FTN-IM} \cite{7973048} & & & & & & &  & & & \ding{51}&  \ding{51}&  \ding{51} & &  \ding{51}\\
	\hline
	{DM-SCIM}  \cite{8287922}& & & & & & &  & & & \ding{51}& \ding{51} &  \ding{51} & & \ding{51}\\
	\hline
	{{CIM-SS}} \cite{6994807}& & & & & & &  & &  \ding{51} & &  & \ding{51} &  \ding{51} &  \ding{51}\\ 
	\hline
    {{GCIM-SS}} \cite{7317808}& & & & & & &  & &  \ding{51} & &  & \ding{51} &  \ding{51} &  \ding{51}\\ 
	\hline
	CIM-SM \cite{8404646TSP}& & & & & & &  & & &  & & & &  \ding{51}\\
	\hline
	 \textcolor{black}{{CFIM} \cite{8792959}}& & & & & & & & & & & & & &  \ding{51}\\ 
	 \hline
	\textcolor{black}{{IM-OFDM-SS}\cite{8264822} }& & & & & & & & & & & & & &  \ding{51}\\
	\hline
	\textcolor{black}{STCM\cite{7864471}}&  &  & & & & & & & & & & & &  \ding{51}\\
	\hline
    \textcolor{black}{ST-MBM\cite{8668573}}  &  & & & & & & & & & & & & &  \ding{51}\\
	\hline
	{STSK} \cite{5599264} & & &  & & & \ding{51} & \ding{51}& \ding{51}  &\ding{51} &  \ding{51} & \ding{51} & \ding{51} & \ding{51} & \ding{51}\\
	\hline
    {DSTSK} \cite{5599264} & & &  & & & \ding{51} & \ding{51}& \ding{51} & \ding{51} &  \ding{51} &  &  \ding{51} & \ding{51} & \ding{51}\\
	\hline
	{GSTSK} \cite{5703198} & & & & & &  & \ding{51} & \ding{51} & \ding{51}&  \ding{51} &  \ding{51} &  & \ding{51} & \ding{51}\\
	\hline
	\textcolor{black}{LMG-SSTSK\cite{8281511}} &  & & & & & & & & & &  \ding{51}  &  & &  \ding{51}\\ 
	\hline
	\textcolor{black}{LMS-GSTSK\cite{8281511}} &  & & & & & & & & & & & &  &  \ding{51}\\ 
	\hline
	{SFSK}\cite{5688440} & & & & & & & \ding{51} & \ding{51}&  &  \ding{51} & \ding{51} &  & \ding{51} & \ding{51}\\ 
	\hline
	{STFSK}\cite{5688440} & & & & & &  & \ding{51}& \ding{51} & &  \ding{51} &  \ding{51} &  \ding{51} & \ding{51} & \ding{51}\\ 
	\hline
	{MS-STSK} \cite{7494949} & & & & & & & & &  \ding{51} & &  \ding{51} & \ding{51} & &  \ding{51} \\
	\hline
	{MSF-STSK} \cite{7792569} & & & & & & & & & \ding{51} & & \ding{51} &  \ding{51}  & &  \ding{51}\\
	\hline
   {OFDM-STSK} \cite{6399190} & & & & &  &  & \ding{51}& \ding{51}&  & &  &  \ding{51} & & \ding{51}\\ 
	\hline
	{OFDM-STSK-IM}\cite{8322306} & & & & & & & & &  & & & & &  \ding{51}\\
	\hline
	{OFDM-SFSK} \cite{8417812}& & &  & & & & & & &  & & & &  \ding{51}\\
	\hline
    \textcolor{black}{JA-STSK \cite{8003424}}&  & & & & & & & &  & & & & &  \ding{51}\\
	\hline
	\textcolor{black}{JA-MS-STSK \cite{8003424}}&  & & & & & & & &  & & & & &  \ding{51}\\
	\hline
    \textcolor{black}{DP-SM \cite{7795239}} & &  & & & & & & & & &  & & &  \ding{51} \\
	\hline
    \textcolor{black}{PolarSK\cite{8048033}} & & & &  & & & & & &  & & & &  \ding{51} \\
    \hline
     \textcolor{black}{PM\cite{9028167}} & & & & & &  & & & & & & & &  \ding{51} \\
 \hline
	\end{tabular}
\label{Tab:Taxonomy}
\end{table*}

\subsection{Related Works}

\textcolor{black}{Until today, several survey and magazine articles have appeared in the literature to shed light on the prominent members of the IM family, as listed in Table \ref{Tab:Summary}.}
\textcolor{black}{SM represents an early stage of the IM concept and thus \cite{6094024} has introduced the working principle of SM associated with its superiority over the mature MIMO technology in terms of hardware and cost-efficiency. Moreover, beneficial insights have been provided on the exploitation of a wireless channel as a possible modulation unit. Besides SM-based MIMO investigation, in \cite{6123167}, the potential of space-time shift keying (STSK) with MIMO has been elaborated in a comprehensive manner. Specifically, a flexible framework allowing accommodation of multiple sub-mechanisms, i.e., space shift keying (SSK), SM, orthogonal space-time block coding (STBC), Vertical Bell Laboratories Layered Space-Time (V-BLAST), and linear dispersion codes (LDC), has been introduced as a unified STSK scheme. Di Renzo \textit{et al.} not only have presented different aspects of SM-MIMO including its principles, transceiver design, and hardware implementation, but also have paid attention to its integration with the emerging communication systems, such as relay-aided designs, small-cells, cooperative networks, mmWave systems, and visible light communications (VLC) \cite{6678765}. Design guidelines for SM-MIMO have been discussed with the emphasis on receiver design, spatial constellation optimization, and link adaptation techniques in \cite{6823072}. Different from the aforementioned studies, in \cite{7509396}, Basar has evaluated not only the future potentials and implementation feasibility of SM-MIMO architectures, but also frequency domain IM-based multicarrier systems, i.e., OFDM-IM and MIMO-aided OFDM-IM.} Also, the author has reviewed advanced SM technologies, such as generalized SM (GSM), enhanced SM (ESM), and quadrature SM (QSM). In \cite{8004416}, Basar \textit{et al.} have provided an overview of the IM variants present in the literature and elaborated on the advantages of SM, OFDM-IM, and channel modulation (CM). They have assessed the application of these modulation techniques to different networks and systems and reviewed some practical concerns for OFDM-IM, such as peak-to-average power ratio (PAPR), inter-carrier interference (ICI), and achievable rate. Sugiura \textit{et al.} {have discussed} the limitations of IM in space, time, and frequency \cite{8070115}. They have compared single-carrier (SC) transmission with OFDM and examined the importance of time-limited pulses for SM. The challenges that occurred by the acquisition of channel state information (CSI) have been revealed for {the} SM technology. In \cite{8417419}, CM, which is media-based modulation (MBM), has been discussed in addition to space, time, and frequency domain IM variants. In \cite{8410878}, authors have classified space and frequency domain IM techniques for vehicular and railway communications. \textcolor{black}{Cheng \textit{et al.} have compared space, frequency, space-time, and space-frequency domain IM variants in terms of SE and EE \cite{8353362}.} Future directions to further increase the SE of IM techniques have been also suggested.  \textcolor{black}{In \cite{8315127}, Ishikawa \textit{et al.} have shed light on the historical background of permutation modulation, SM, and IM concepts, and have disclosed the road from permutation modulation to OFDM-IM. Research progresses on SM variants, performance enhancement schemes for SM, its integration with promising technologies, such as compressed-sensing (CS) theory and NOMA-aided systems, and its application in emerging systems, such as mmWave and optical wireless communications, have been presented in \cite{8765384}. \textcolor{black}{Recently, Li \textit{et al.} have evaluated frequency domain IM types including OFDM-IM, dual-mode OFDM (DM-OFDM), and zero-padded tri-mode IM-aided OFDM {(ZTM-OFDM-IM)} for future wireless networks including cognitive radio (CR) networks, relay-aided networks and multi-user communications  \cite{9076117}.} The presented IM techniques in the existing magazine and survey articles are given in Table~{\ref{Tab:Taxonomy}}, and compared with the proposed survey.}

\subsection{Contributions}

 Against this background, the main contributions of this paper are listed as follows:
 
 \textcolor{black}{
 	\begin{itemize}
 		\item A comprehensive review of the existing IM approaches in the literature is presented and dimensional-based categorization is performed.
 		\item  To the best of authors' knowledge, this study is the first for both providing a survey and comparison of one-dimensional and multidimensional IM options.  
 		\item To further extend the understanding of these IM schemes, their differences and the trade-off among them are identified with respect to the achievable data rate, power consumption, transmission reliability, and practical implementation. 
 		\item  The reviewed IM techniques are categorized considering the requirements of eMBB, mMTC, and URLLC services to shed light on the application of IM techniques for future use-cases and applications.
 		\item The main benefits and shortcomings of available IM domains are quantified.
 		\item Finally, potential challenges and future directions on the integration of IM concept into the prominent wireless technologies, such as CR networks, cooperative systems, and non-orthogonal communications, are elaborated.      
 	\end{itemize}
 }

\subsection{Paper Organization}

\begin{figure}
	\centering
	{\includegraphics[scale = 0.74]{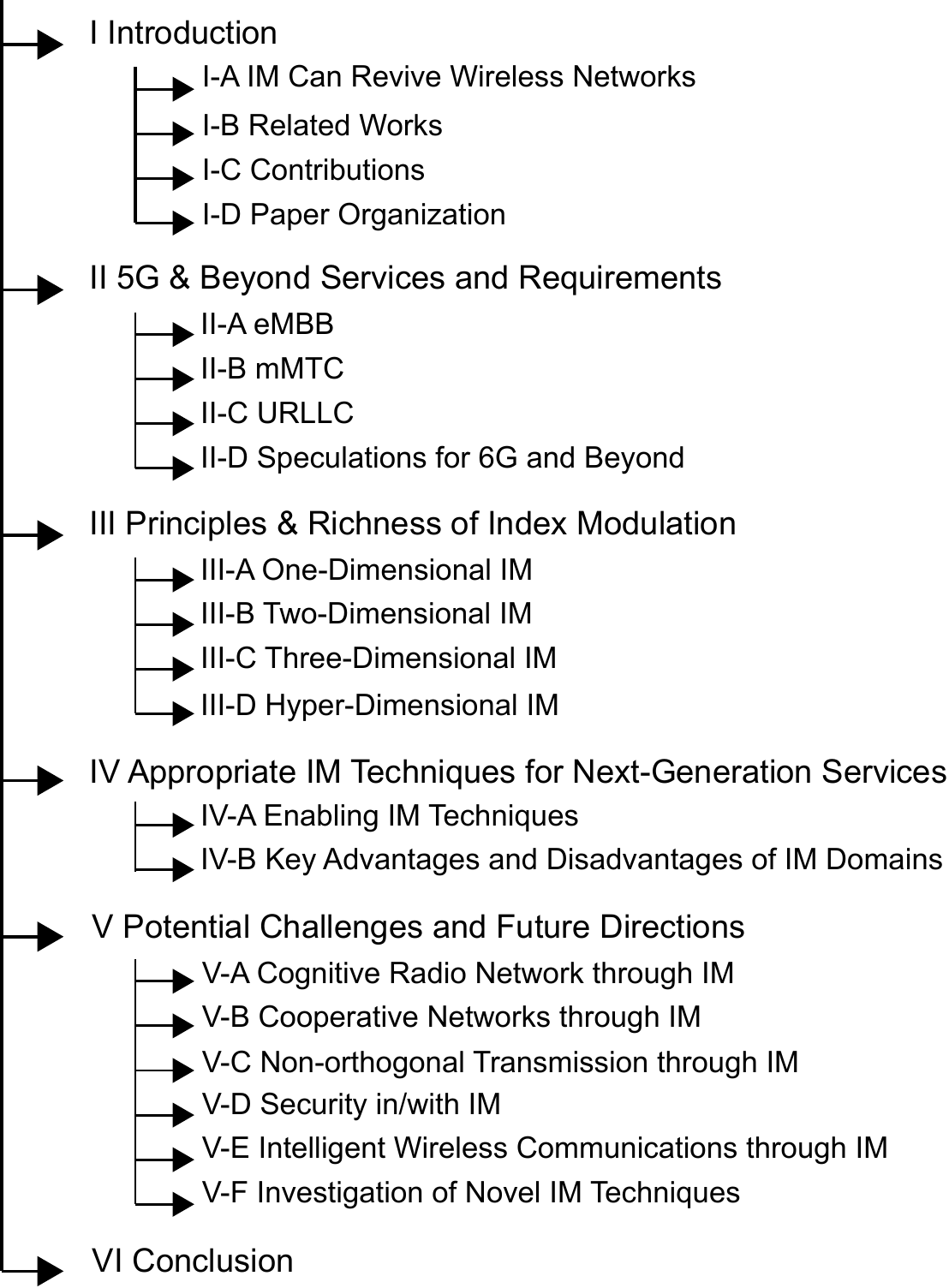}}
	\caption{Organization of the survey.}
	\label{Fig:1a}
\end{figure}

 \textcolor{black}{The organization of the survey is given in Fig.~\ref{Fig:1a} via the chart flow. Section II revises the requirements of 5G and beyond services in wireless networks. Section III presents a comprehensive taxonomy of the existing IM schemes in the literature and then provides useful insights on future multidimensional IM variants. In Section IV, enabling IM techniques for 5G and beyond services are provided, and key advantages \& disadvantages of the available IM domains are revealed. Section V  provides the potential challenges and future directions for IM-aided communication networks. Finally, Section VI concludes the work. \footnote{ \textit{Notation:} Bold, capital and lowercase letters are used for matrices and column vectors, respectively. $(.)^T$ and $(.)^H$ denote transposition and Hermitian transposition, respectively. $\mathrm{E}[.]$ stands for expectation and $\mathbb{C}$ is the ring of complex numbers. $\binom{.}{.}$ denotes the binomial coefficient and $\floor{.}$ is the floor function.}}

  \section{5G and Beyond Services and Requirements} \label{SectionNO2} 
  
In this section, three main services of 5G networks and their use scenarios are briefly explained, along with their widely accepted KPIs and benchmarks in the 3rd Generation Partnership Project (3GPP) standards. Additionally, forecasts for beyond 5G systems are mentioned. The KPIs and their values in the standards are given in Table \ref{tab:LatTable2}.

\begin{table*}
	
	\caption{KPIs for next-generation services}
	\centering
	\small
	\renewcommand{\arraystretch}{1.25}
	\begin{tabular}{|C{2cm}||C{4cm}|p{10cm}|} 
		\hline
		\makecell{\textbf{{Service Type}}}  & 	\makecell{\textbf{{KPIs}}} & \makecell{\textbf{{Definitions}}}\\
		\hline \hline
		\multirow{4}{*}{ eMBB }&\multirow{2}{*}{Data rate}& Supporting peak data rates of 10 Gbit/s and 20 Gbit/s for UL and DL transmission, respectively. \\  \cline{2-3}
		& \multirow{2}{*}{Mobility}& Achieving desired data rate for a given mobility class ($10$~{km}/{h} $\leq V \leq$ $500$~{km}/{h}  ). \\ \cline{2-3} 
		\hline 	\hline 
		\multirow{6}{*}{ mMTC}& Connection capability & Number of mMTC UEs per a cell ($1.000.000$ UEs per km$^2$).\\ \cline{2-3}
		& 	\multirow{2}{*}{Power Consumption} & At least 10 years of lifetime for a device by sending 20 bytes and 200 bytes for UL and DL transmission, respectively.\\ \cline{2-3}
		& 	\multirow{3}{*}{Coverage} & Maximum coupling loss that corresponds to total loss including antenna gain, path loss and shadowing for baseline data rate of 160 bit/s ($max_{CL} = 164$ dB). \\
		\hline
		\hline 
		\multirow{4}{*}{ URLLC}& \multirow{2}{*}{Latency}& The elapsed time for successful transmission and reception of a packet ($0.25$ ms $\leq$ \text{latency} $\leq$ $1$ ms).  \\ \cline{2-3}
		& \multirow{2}{*}{Reliability} & Successful reception of a packet with the reliability range of ($10^{-5}$ $\leq$ BLER $\leq$ $10^{-9}$). \\ \cline{2-3}
		\hline
	\end{tabular}
	\label{tab:LatTable2} 
\end{table*}

\subsection{eMBB}\label{sec:eMMB}

High data rate use-cases and applications, such as virtual reality, broadband internet access, and high definition video streaming, are grouped under the eMBB service, which can be considered as the continuation of the current 4G technology \cite{Nok1}. For these applications, peak data rates up to 10 Gbit/s are required for the uplink (UL) and downlink (DL) transmission of a UE \cite{7833459}. Hence, to {support the increased data rate requirements, bandwidths of at least 100 MHz and 1 GHz are proposed for sub-6 GHz and above 6 GHz bands, respectively. Furthermore, supporting a high data rate transmission for three different mobility classes must be considered: pedestrian speeds up to 10 km/h, vehicular speeds from 10 km/h to 120 km/h, and high-speed vehicular speeds from 120 km/h to 500 km/h.} Therefore, {increasing the SE} via the development of new flexible communication schemes has become a {critical} demand. eMBB applications are expected to perform scheduled transmissions due to their characteristics, namely delay-tolerant and continuous. Hence, {the eMBB service requires a spectrum-efficient waveform and modulation design at the cost of a moderate level of transmission latency and system complexity.}

\subsection{mMTC}\label{sec:mMTC}

In the context of Internet-of-Things (IoT), {a} connection of a massive number of UEs to a network is expected in the next-generation systems \cite{3gppMTC1, Rohde}.
The coexistence of numerous machine-type and mobile UEs in the network {puts} pressure on service providers to {satisfy the} diverse demands \cite{7565189}. Various applications of the IoT, such as smart cities, connected vehicles, smart agriculture, public safety, and asset tracking, require different levels of coverage area, battery life, and connection capability \cite{8360103}. 
Unlike {the} human-oriented {higher} data rate communication in {the} LTE systems, providing {service} for massive machine-type UEs with {lower} data rate is the primary focus {of} mMTC. Although mMTC is latency-tolerant, long waiting time occurs due to the scheduling of a large number of UEs. Therefore, random access mechanisms are proposed as promising solutions for mMTC \cite{6678832}. However, the current OFDM technology requires strict synchronization {between the UEs} to avoid inter-user-interference (IUI) \cite{ShpirtveBen}.  
mMTC use-cases with ultra low power consumption and wide coverage area are grouped into narrowband IoT (NB-IoT) by the 3GPP \cite{7876968, 8120239, Shpirt}. The standards adopt orthogonal frequency division multiple access (OFDMA) and single carrier frequency division multiple access (SC-FDMA) for DL and UL transmission and introduce SCS of 3.75 kHz for UL transmission over random access channels. Narrowband transmission, which leads to a low data rate, is performed to reduce power consumption and to guarantee a lifetime exceeding 10 years. In order to reduce the cost, mMTC UE is equipped with a single antenna, and half-duplex transmission is adopted. Retransmission of a packet is allowed to improve the coverage area at the expense of at most 10ms transmission latency.

\subsection{URLLC} \label{sec:URLLC1}

In 5G and beyond wireless systems, achieving ultra-reliability and low-latency is a crucial as well as a very challenging task. URLLC use-cases and applications need to guarantee {block error rate (BLER) values up to $10^{-9}$ within the latency bounds of 0.25 ms} \cite{3gpp2, 3gpp3, Schulz2017}. The latency refers to the round trip time required for the successful transmission and reception of the transmitted data packet. {In the current systems, the long handshaking process between a UE and base station (BS), data processing time, and transmission time interval (TTI) are the major barriers in achieving low-latency communications} \cite{Ref6}. {Moreover, the smallest resource unit is a subframe that consists of 14 OFDM symbols corresponding to a TTI of 1 ms for 15kHz SCS. This rules out the possibility of any transmission faster than 1ms.} Thus, mini-slot concept is adopted in 5G NR to meet the different latency requirements. For DL transmission, no latency occurs due to the handshaking since the BS manages the communication. 
However, the handshaking process between UE and BS is mandatory in UL transmission to establish reliable communication at the cost of an extra delay that corresponds to 3 TTIs. {This is in addition to the reduction in SE and EE due to the signaling overhead and the processing complexity, respectively.} Hence, the UL latency for 4G LTE systems is almost doubled compared to DL. Moreover, UL with grant-based (GB) transmission further leads to the waste of resources due to its sporadic nature. {In GB access, BS assigns available resources to a UE continuously.} However, UE with URLLC utilizes the resources intermittently. In the literature, grant-free (GF) access is extensively investigated to avoid the latency caused by the handshaking process \cite{Ref2}. However, UEs with GF transmission are exposed to collisions that reduce system reliability. Hence, interference immune multiple accessing schemes are required for achieving URLLC.

\textcolor{black}{\subsection{Speculations for 6G and Beyond} } \label{sec:6gbeyond}

\textcolor{black}{As in 5G systems, underlying applications, and used cases will be the driving factors in 6G and beyond wireless networks \cite{8869705}. For instance, 6G is expected to open the door for a wide range of unprecedented services such as self-driving cars, virtual reality, flying vehicles, human-body, and holographic communications\cite{yastrebova2018future}. Hence, the future of wireless system operators must simultaneously deliver much higher data rates, higher security, and communication reliability within a shorter latency compared with the aforementioned scenarios of 5G. For example, a five times increase of average data consumption per UE and down to 0.1 ms latency are expected by 2024 \cite{mourad2020towards}. Moreover, a service of joint eMBB and URLLC with security constraints, and other combinations of eMBB, mMTC and URLLC are envisioned to represent these new applications and use-cases. In this context, artificial intelligence, machine learning, reconfigurable intelligent surfaces (RIS), unmanned aerial vehicles (UAVs), and terahertz (THz) communications are mainly speculated among potential technologies in beyond 5G \cite{dfsdfafgadgfa}. Extensive research is afforded by both academia and industry for beyond 5G wireless networks in the industrialized countries. In China, several research groups are established to enhance intelligent manufacturing. Horizon 2020 ICT-09-2017 project considers mmWave and THz spectrum as a possible solution for scenarios with joint eMBB and latency limitations. Moreover, in \cite{dang2020should}, IM is considered as complementary technology to conventional OFDM-based multiplexing in order to achieve flexibility in 6G systems.}          

\vspace{3mm}

\section{Principles and Richness of Index Modulation } \label{Sec:Types}

\begin{figure}
	\centering
	\includegraphics[scale = 0.34]{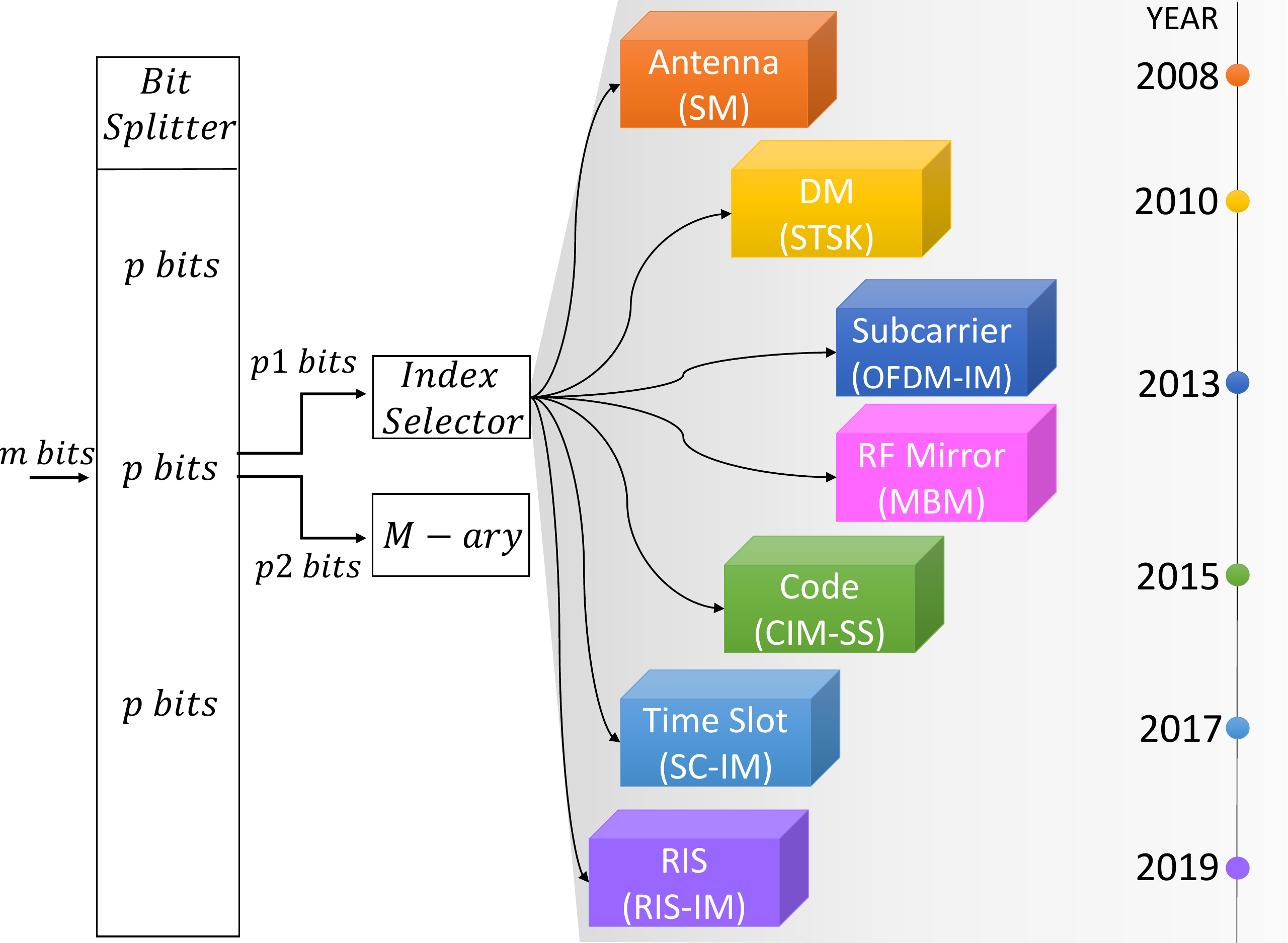}
	\caption{Basic implementation of IM, and the timeline of substantial IM techniques.}
	\label{Fig:MultIM}
\end{figure}

\textcolor{black}{IM deals with the mapping of data bits to information-bearing transmit entities, such as antennas, subcarriers, radio frequency (RF) mirrors, dispersion matrices (DMs), codes, time slots, and different combinations thereof. {In order to convey additional information bits along with conventional $M$-ary symbols, partial activation of the entities in a given domain is performed through IM.} Although the {initial proposal of IM concept dates back to almost the beginning of the century, it has drawn substantial attention from the research community over the last decade \cite{956483}.}  {Fig.~\ref{Fig:MultIM} illustrates the timeline of the substantial IM variants in the literature}. }

\begin{figure*}
	\centering
	{\includegraphics[scale = 0.6]{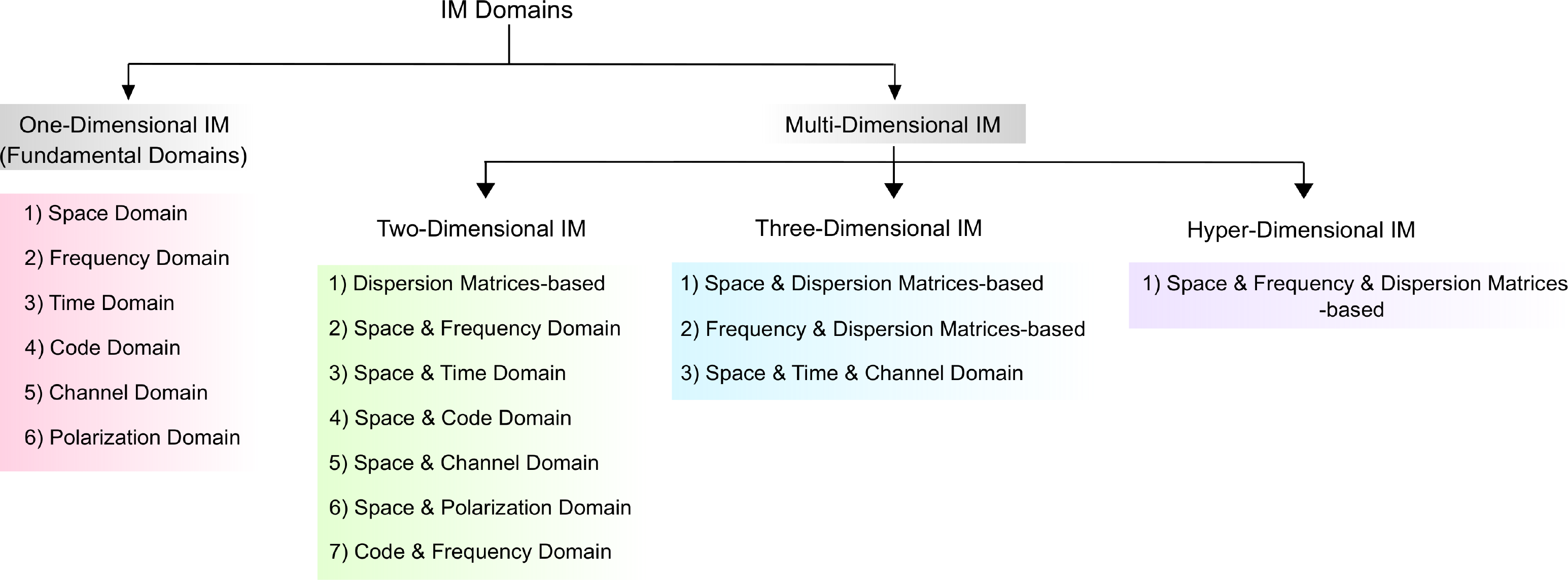}}
	\caption{\textcolor{black}{Dimensional-based categorization of the existing IM domains in the literature.}}
	\label{Fig:AppDomains}
\end{figure*}

\textcolor{black}{In spite of the fact that one-dimensional IM methods are well-known, a comprehensive overview of the multidimensional IM methods is lacking in the literature. In view of this, firstly, this section reveals the applied multidimensional IM domains in the literature and provides their dimensional-based categorization in detail, as illustrated in Fig.~\ref{Fig:AppDomains}. Later, the existing IM techniques are subsumed regarding the dimensional-based categorization. } \textcolor{black}{In Table~\ref{Table:super}, the right-angled triangle demonstrates the available IM options in the literature regarding their application domains, where the diagonal and off-diagonal cells correspond to one-dimensional and multidimensional IM schemes, respectively. Note that two-dimensional IM placed in diagonal cells is only DMs-based IM types, and their combinations with the other IM schemes are minimum three-dimensional IM. Also, the unfilled cells denote the unexplored multidimensional IM variants.}

\newcommand\notsotiny{\@setfontsize\notsotiny{6}{7}}
\newenvironment{tinyb}{\bgroup\tiny\bfseries\scshape}{\egroup}
\newcolumntype{C}[1]{>{\centering\let\newline\\\arraybackslash\hspace{0pt}}m{#1}}

\textcolor{black}{\subsection{One-Dimensional IM}}

\begin{table*} 
	\caption{\textcolor{black}{A comprehensive taxonomy of one-dimensional and multidimensional IM variants}}
	\centering	
	\fontsize{6.5pt}{7pt}
	\selectfont
	\begin{tabular}{|C{1.7cm}||C{1.8cm}C{2.3cm}C{1.7cm}C{1.8cm}C{1.6cm}C{1.9cm}C{1.7cm}} 
		\cline{1-2}
		\multirow{15}{*}{ \normalsize\textbf{Space}} & \multicolumn{1}{c|}{SSK$^{*}$\cite{956483}}& & & & & &  \\
    	& \multicolumn{1}{c|}{{GSSK}$^{*}$ \cite{4699782}} & & & & & &\\
	    & \multicolumn{1}{c|}{{SM}$^{*}$ \cite{4382913}}& & & & & & \\
        & \multicolumn{1}{c|}{{GSM}$^{*}$ \cite{5757786}} & & & & & &\\
        & \multicolumn{1}{c|}{{MA-SM}$^{*}$ \cite{6166339}}& & & & & & \\
        & \multicolumn{1}{c|}{{QSM}$^{*}$ \cite{6868290}}& & & & & &\\
        & \multicolumn{1}{c|}{{ESM}$^{*}$ \cite{7084604}} & & & & & &\\
		& \multicolumn{1}{c|}{{TCSM \cite{5508985}}}& & & & & & \\
		& \multicolumn{1}{c|}{{TC-QSM \cite{TCQSM1}}} & & & & & &\\
		& \multicolumn{1}{c|}{{STBC-SM$^{*}$ \cite{5672371}}}& & & & & & \\
		& \multicolumn{1}{c|}{{STBC-QSM \cite{8277677}} }& & & & & & \\
		& \multicolumn{1}{c|}{{PSM \cite{5956573}}}& & & & & & \\
		& \multicolumn{1}{c|}{{GPSM \cite{6644231}}}& & & & & & \\
		& \multicolumn{1}{c|}{ {GPQSM \cite{7467510}}}& & & & & & \\
		& \multicolumn{1}{c|}{{RIS-IM}$^{+}$ \cite{8981888}} & & & & & &\\
\cline{1-3}
\multirow{12}{*}{\normalsize\textbf{Frequency} } & \multicolumn{1}{c|}{}& \multicolumn{1}{c|}{{SIM-OFDM}\cite{5449882}} & & & & &  \\
& \multicolumn{1}{c|}{} & \multicolumn{1}{c|}{{ESIM-OFDM} \cite{6162549}} & & & & &\\
& \multicolumn{1}{c|}{}& \multicolumn{1}{c|}{{OFDM-IM} \cite{Basar1}}& & & & & \\
& \multicolumn{1}{c|}{} & \multicolumn{1}{c|}{{OFDM-ISIM} \cite{6841601}}  & & & & &\\
& \multicolumn{1}{c|}{{ISM-OFDM} \cite{7440707}}& \multicolumn{1}{c|}{{CI-OFDM-IM} \cite{7086323}} & & & & & \\
& \multicolumn{1}{c|}{{{GSFIM} \cite{7277106}}}& \multicolumn{1}{c|}{{OFDM-GIM} \cite{7112187}}& & & & &\\
& \multicolumn{1}{c|}{{GSTFIM} \cite{8487049}} & \multicolumn{1}{c|}{{DM-OFDM} \cite{7547943}} & & & & &\\
& \multicolumn{1}{c|}{{{GFDM-SFIM} \cite{8277679}}}& \multicolumn{1}{c|}{{MM-OFDM} \cite{7936676}} & & & & & \\
& \multicolumn{1}{c|}{} &\multicolumn{1}{c|}{ {GFDM-IM}  \cite{7848916}} & & & & &\\
& \multicolumn{1}{c|}{}& \multicolumn{1}{c|}{{{OFDM-I/Q-IM} \cite{7112187, 7230238}}}& & & & & \\
& \multicolumn{1}{c|}{}& \multicolumn{1}{c|}{{{ZTM-OFDM-IM}\cite{8254931}}}& & & & & \\
& \multicolumn{1}{c|}{}& \multicolumn{1}{c|}{{{L-OFDM-IM}\cite{8734769}}} & & & & & \\
\cline{1-4}
\multirow{8}{*}{ \normalsize\textbf{Time} } & \multicolumn{1}{c|}{}& \multicolumn{1}{c|}{} & \multicolumn{1}{c|}{}& & & &  \\  
& \multicolumn{1}{c|}{}& \multicolumn{1}{c|}{} & \multicolumn{1}{c|}{}& & & &  \\
& \multicolumn{1}{c|}{}& \multicolumn{1}{c|}{} & \multicolumn{1}{c|}{{SC-IM}  \cite{7738501}}& & & &  \\
 & \multicolumn{1}{c|}{{{TI-SM} \cite{7925922}}}& \multicolumn{1}{c|}{$\cdots$} & \multicolumn{1}{c|}{{FTN-IM} \cite{7973048}}& & & &  \\
 & \multicolumn{1}{c|}{{TI-SM-MBM} \cite{8110611}}& \multicolumn{1}{c|}{} & \multicolumn{1}{c|}{{DM-SCIM} \cite{8287922}}& & & &  \\
 & \multicolumn{1}{c|}{}& \multicolumn{1}{c|}{} & \multicolumn{1}{c|}{}& & & &  \\
 & \multicolumn{1}{c|}{}& \multicolumn{1}{c|}{} & \multicolumn{1}{c|}{}& & & &  \\
  \cline{1-5}
  \multirow{7}{*}{ \normalsize\textbf{Code} } & \multicolumn{1}{c|}{}& \multicolumn{1}{c|}{} & \multicolumn{1}{c|}{}& \multicolumn{1}{c|}{} & & &  \\ 
  & \multicolumn{1}{c|}{}& \multicolumn{1}{c|}{} & \multicolumn{1}{c|}{}& \multicolumn{1}{c|}{} & & &  \\
  & \multicolumn{1}{c|}{}& \multicolumn{1}{c|}{} & \multicolumn{1}{c|}{}& \multicolumn{1}{c|}{{{CIM-SS}} \cite{6994807}} & & &  \\
  & \multicolumn{1}{c|}{CIM-SM \cite{8404646TSP}}& \multicolumn{1}{c|}{ CFIM \cite{8792959}} & \multicolumn{1}{c|}{$\cdots$}& \multicolumn{1}{c|}{{{GCIM-SS}} \cite{7317808}} & & &  \\
  & \multicolumn{1}{c|}{}& \multicolumn{1}{c|}{} & \multicolumn{1}{c|}{}& \multicolumn{1}{c|}{{IM-OFDM-SS}\cite{8269169}}& & &  \\
  & \multicolumn{1}{c|}{}& \multicolumn{1}{c|}{} & \multicolumn{1}{c|}{}& \multicolumn{1}{c|}{} & & &  \\
  & \multicolumn{1}{c|}{}& \multicolumn{1}{c|}{} & \multicolumn{1}{c|}{}& \multicolumn{1}{c|}{}& & &  \\
  \cline{1-6}
  \multirow{8}{*}{ \normalsize\textbf{\textcolor{black}{Channel}} } & \multicolumn{1}{c|}{}& \multicolumn{1}{c|}{} & \multicolumn{1}{c|}{}& \multicolumn{1}{c|}{} & \multicolumn{1}{c|}{} & &  \\
  & \multicolumn{1}{c|}{}& \multicolumn{1}{c|}{} & \multicolumn{1}{c|}{}& \multicolumn{1}{c|}{} & \multicolumn{1}{c|}{} & &  \\ 
  & \multicolumn{1}{c|}{}& \multicolumn{1}{c|}{} & \multicolumn{1}{c|}{}& \multicolumn{1}{c|}{} & \multicolumn{1}{c|}{\textcolor{black}{{MBM} \cite{6620786}}}& &  \\
  & \multicolumn{1}{c|}{\textcolor{black}{{SM-MBM} \cite{8110611}}}& \multicolumn{1}{c|}{$\cdots$} & \multicolumn{1}{c|}{\textcolor{black}{{TI-MBM} \cite{8110611}} }& \multicolumn{1}{c|}{$\cdots$} & \multicolumn{1}{c|}{\textcolor{black}{{DMBM} \cite{7887715}}} & &  \\
  & \multicolumn{1}{c|}{\textcolor{black}{QCM \cite{8014408}}}& \multicolumn{1}{c|}{} & \multicolumn{1}{c|}{\textcolor{black}{{TI-SM-MBM} \cite{8110611}}}& \multicolumn{1}{c|}{}& \multicolumn{1}{c|}{\textcolor{black}{STCM \cite{7864471}} }& &  \\
  & \multicolumn{1}{c|}{}& \multicolumn{1}{c|}{} & \multicolumn{1}{c|}{}& \multicolumn{1}{c|}{} & \multicolumn{1}{c|}{	\textcolor{black}{ST-MBM \cite{8668573}} }& &  \\
  & \multicolumn{1}{c|}{}& \multicolumn{1}{c|}{} & \multicolumn{1}{c|}{}& \multicolumn{1}{c|}{}&\multicolumn{1}{c|}{}& &  \\
  \cline{1-7}
   \cellcolor{blue!8}& \multicolumn{1}{c|}{\cellcolor{blue!8}}& \multicolumn{1}{c|}{\cellcolor{blue!8}} & \multicolumn{1}{c|}{\cellcolor{blue!8}}& \multicolumn{1}{c|}{\cellcolor{blue!8}} & \multicolumn{1}{c|}{\cellcolor{blue!8}} &  \multicolumn{1}{c|}{\cellcolor{blue!8}\textcolor{black}{{DSM}$^{*}$ \cite{6879496}}}&  \\
  \cellcolor{blue!8}& \multicolumn{1}{c|}{\cellcolor{blue!8}}& \multicolumn{1}{c|}{\cellcolor{blue!8}} & \multicolumn{1}{c|}{\cellcolor{blue!8}}& \multicolumn{1}{c|}{\cellcolor{blue!8}} & \multicolumn{1}{c|}{\cellcolor{blue!8}} &  \multicolumn{1}{c|}{\cellcolor{blue!8}{STSK} \cite{5599264}}&  \\
  \cellcolor{blue!8}&  \multicolumn{1}{c|}{\cellcolor{blue!8}LMS-GSTSK\cite{8281511}}& \multicolumn{1}{c|}{\cellcolor{blue!8}} & \multicolumn{1}{c|}{\cellcolor{blue!8}}& \multicolumn{1}{c|}{\cellcolor{blue!8}} & \multicolumn{1}{c|}{\cellcolor{blue!8}}& \multicolumn{1}{c|}{\cellcolor{blue!8}{DSTSK} \cite{{5599264}}} &  \\
  \cellcolor{blue!8}& \multicolumn{1}{c|}{\cellcolor{blue!8}{MS-STSK}\cite{7494949}}& \multicolumn{1}{c|}{\cellcolor{blue!8}{OFDM-STSK-IM}\cite{8322306}} & \multicolumn{1}{c|}{\cellcolor{blue!8}}& \multicolumn{1}{c|}{\cellcolor{blue!8}} & \multicolumn{1}{c|}{\cellcolor{blue!8}} & \multicolumn{1}{c|}{\cellcolor{blue!8}{GSTSK} \cite{5703198}} &  \\
  \cellcolor{blue!8}{\normalsize\textbf{DMs}} & \multicolumn{1}{c|}{\cellcolor{blue!8} \textcolor{black}{JA-STSK \cite{8003424}}}& \multicolumn{1}{c|}{\cellcolor{blue!8}{MSF-STSK} \cite{7792569}} & \multicolumn{1}{c|}{\cellcolor{blue!8}$\cdots$}& \multicolumn{1}{c|}{\cellcolor{blue!8}$\cdots$}& \multicolumn{1}{c|}{\cellcolor{blue!8}$\cdots$}&  \multicolumn{1}{c|}{\cellcolor{blue!8}{STFSK}\cite{5688440}}&  \\
  \cellcolor{blue!8}& \multicolumn{1}{c|}{\cellcolor{blue!8}\textcolor{black}{JA-MS-STSK\cite{8003424}}}& \multicolumn{1}{c|}{\cellcolor{blue!8}} & \multicolumn{1}{c|}{\cellcolor{blue!8}}& \multicolumn{1}{c|}{\cellcolor{blue!8}} & \multicolumn{1}{c|}{\cellcolor{blue!8}}& \multicolumn{1}{c|}{\cellcolor{blue!8}{OFDM-STSK} \cite{6399190}} &  \\
 \cellcolor{blue!8} & \multicolumn{1}{c|}{\cellcolor{blue!8}}& \multicolumn{1}{c|}{\cellcolor{blue!8}} & \multicolumn{1}{c|}{\cellcolor{blue!8}}& \multicolumn{1}{c|}{\cellcolor{blue!8}}&\multicolumn{1}{c|}{\cellcolor{blue!8}}& \multicolumn{1}{c|}{\cellcolor{blue!8}{{{SFSK} \cite{8417812}}}} &  \\
 \cellcolor{blue!8} & \multicolumn{1}{c|}{\cellcolor{blue!8}}& \multicolumn{1}{c|}{\cellcolor{blue!8}} & \multicolumn{1}{c|}{\cellcolor{blue!8}}& \multicolumn{1}{c|}{\cellcolor{blue!8}}&\multicolumn{1}{c|}{\cellcolor{blue!8}}& \multicolumn{1}{c|}{\cellcolor{blue!8}\textcolor{black}{ LMG-SSTSK \cite{7448828} }} &  \\
  \cline{1-8}
   \multirow{6}{*}{ \normalsize\textbf{\textcolor{black}{Polarization}} } & \multicolumn{1}{c|}{}& \multicolumn{1}{c|}{} & \multicolumn{1}{c|}{}& \multicolumn{1}{c|}{} & \multicolumn{1}{c|}{} & \multicolumn{1}{c|}{}& \multicolumn{1}{c|}{} \\
  & \multicolumn{1}{c|}{}& \multicolumn{1}{c|}{} & \multicolumn{1}{c|}{}& \multicolumn{1}{c|}{} & \multicolumn{1}{c|}{}&\multicolumn{1}{c|}{} & \multicolumn{1}{c|}{} \\
  & \multicolumn{1}{c|}{\textcolor{black}{ SPSK \cite{SPSK}}}& \multicolumn{1}{c|}{$\cdots$} & \multicolumn{1}{c|}{$\cdots$}& \multicolumn{1}{c|}{$\cdots$} & \multicolumn{1}{c|}{$\cdots$} & \multicolumn{1}{c|}{$\cdots$}& \multicolumn{1}{c|}{\textcolor{black}{PolarSK\cite{8048033} }} \\
  & \multicolumn{1}{c|}{\textcolor{black}{DP-SM \cite{7248680,7795239}}}& \multicolumn{1}{c|}{} & \multicolumn{1}{c|}{}& \multicolumn{1}{c|}{}& \multicolumn{1}{c|}{}&\multicolumn{1}{c|}{} &\multicolumn{1}{c|}{\textcolor{black}{PM \cite{9028167}}}  \\
  & \multicolumn{1}{c|}{}& \multicolumn{1}{c|}{} & \multicolumn{1}{c|}{}& \multicolumn{1}{c|}{} & \multicolumn{1}{c|}{}&\multicolumn{1}{c|}{} & \multicolumn{1}{c|}{} \\
  & \multicolumn{1}{c|}{}& \multicolumn{1}{c|}{} & \multicolumn{1}{c|}{}& \multicolumn{1}{c|}{}&\multicolumn{1}{c|}{}&\multicolumn{1}{c|}{} & \multicolumn{1}{c|}{} \\
  \hline \hline
  \multirow{6}{*}{\rotatebox[origin=c]{40}{\parbox[c]{1.65cm}{\centering \normalsize \textbf{Modulation Domain}}}} & \multicolumn{1}{c|}{\multirow{6}{*}{{\normalsize\textbf{Space}}} }&  \multicolumn{1}{c|}{\multirow{6}{*}{\normalsize\textbf{Frequency}}} &  \multicolumn{1}{c|}{\multirow{6}{*}{\normalsize\textbf{Time}}} &  \multicolumn{1}{c|}{\multirow{6}{*}{\normalsize\textbf{Code}}} &  \multicolumn{1}{c|}{\multirow{6}{*}{\normalsize\textbf{\textcolor{black}{Channel}}}} &  \multicolumn{1}{c|}{\multirow{6}{*}{\normalsize\textbf{DMs}}}&  \multicolumn{1}{c|}{\multirow{6}{*}{\normalsize\textbf{\textcolor{black}{Polarization}}}}  \\
   &\multicolumn{1}{c|}{} & \multicolumn{1}{c|}{} & \multicolumn{1}{c|}{}& \multicolumn{1}{c|}{}& \multicolumn{1}{c|}{}& \multicolumn{1}{c|}{}& \multicolumn{1}{c|}{}\\
  & \multicolumn{1}{c|}{}& \multicolumn{1}{c|}{}& \multicolumn{1}{c|}{}& \multicolumn{1}{c|}{}&\multicolumn{1}{c|}{} &\multicolumn{1}{c|}{} & \multicolumn{1}{c|}{}\\
  & \multicolumn{1}{c|}{}& \multicolumn{1}{c|}{} & \multicolumn{1}{c|}{}&\multicolumn{1}{c|}{}&\multicolumn{1}{c|}{} & \multicolumn{1}{c|}{}&\multicolumn{1}{c|}{} \\
  & \multicolumn{1}{c|}{}& \multicolumn{1}{c|}{}&\multicolumn{1}{c|}{} &\multicolumn{1}{c|}{} & \multicolumn{1}{c|}{}&\multicolumn{1}{c|}{} & \multicolumn{1}{c|}{}\\
  & \multicolumn{1}{c|}{}& \multicolumn{1}{c|}{} & \multicolumn{1}{c|}{} & \multicolumn{1}{c|}{}& \multicolumn{1}{c|}{}& \multicolumn{1}{c|}{}& \multicolumn{1}{c|}{}\\
  \hline
	\end{tabular}
	\begin{tablenotes}\footnotesize
		\item[] Note - $(.)^{*}$: IM via antennas, $(.)^{+}$: IM via RIS \\
	\end{tablenotes}
	\label{Table:super}
\end{table*}

\textcolor{black}{One-dimensional IM corresponds to fundamental IM techniques that lay the foundations for multidimensional IM types. As illustrated in Fig.~\ref{Fig:AppDomains}, space, frequency, time, code, channel, and polarization domains are elaborated under this category.}

\subsubsection{\textcolor{black}{Space Domain IM}} \label{Sec:spaceIM}

\textcolor{black}{Two different physical entities consisting of antennas, and reconfigurable intelligent surfaces (RIS) are evaluated in the context of space domain IM.}

Spatial multiplexing (SMX) and spatial diversity (SD) are well-known techniques for boosting transmission rate through sending independent information bits over independent channels and increasing reliability through emitting the same information bits over independent channels for conventional $N_t \times N_r$ MIMO systems, respectively \cite{738086,752125}. {$N_t$} and {$N_r$} represent the number of transmitter (Tx) and receiver (Rx) antennas, respectively. However, 1) hardware complexity, 2) strict synchronization requirement between Tx antennas, and 3) decoding complexity should be alleviated to reap the advantages of MIMO systems. Firstly, activation of {$N_t$} Tx antennas at each transmission interval requires {$N_t$} RF chains, which might be impractical for mMTC devices. Secondly, all data symbols should be transmitted at the same time, thus inter-antenna synchronization (IAS) is needed. {Thirdly, the Rx is subject to a heavy decoding process due to the active {$N_t$} Tx antennas.}

\paragraph{\textcolor{black}{IM via Antennas}}

Space domain IM is introduced via SSK which utilizes a single antenna out of {$N_t$} Tx antennas \cite{956483,5165332}. The index of the active antenna conveys $m = \log_2(N_t)$ information bits, while the antenna itself does not carry $M$-ary symbol. There are $N_t$ different combinations of the information bits to decide the active antenna. For {the $i$-th combination}, transmission vector $\mathbf{x_i}$ presents the status of $N_t$ antennas, and it is expressed as
\begin{equation}
\mathbf{x_i} \triangleq \big[0 ~~ 0 ~~ 0 ~~ 1 ~~ 0 ~~\cdots~~ 0\big]^T, 
\label{Eq:Eq1}
\end{equation}
where the active antenna has unit transmission power while 0 refers to the inactive antennas. 
SSK attains a logarithmic increase on SE with {$N_t$} while SE of conventional SMX methods linearly increases with {$N_t$}. Thus, achieving higher data rates through SSK can be impractical due to the need for higher number of Tx antennas. Generalized SSK (GSSK) allows the {utilization of} multiple Tx antennas to carry the information bits \cite{4699782}. For $N_k$ active antennas, $\floor{\log_2\binom{N_t}{N_k}}$ data bits are conveyed by the indices of multiple active antennas. Hence, $\mathbf{x_i}$ corresponds to
\begin{equation}
\mathbf{x_i} \triangleq \big[0 ~~ \frac{1}{\sqrt{N_k}} ~~\cdots ~~ 0 ~~ \frac{1}{\sqrt{N_k}} ~~ 0 ~~\cdots~~ 0\big]^T. 
\label{Eq:Eq2}
\end{equation}
Since multiple Tx antennas are active, IAS is a necessity for GSSK. Otherwise, the system performance is affected by IAI. Moreover, channels between the activated Tx and Rx antennas should be as independent as possible to achieve a performance gain via spatial selectivity. Thus, the distance between Tx antennas in an array should be more than half of the wavelength $(\frac{\lambda}{2})$.                                                                                                                                                                                                    

The {invention} of the SM is an important breakthrough that not only paves the way for the introduction of the general IM concept to the wireless communication realm but also sheds light on its development \cite{4382913, 4149911, 4601434, 7416621}. Besides conveying information bits via the index of active Tx antenna, SM also performs conventional $M$-ary symbol transmission. In this case, the transmission vector $\mathbf{x_i}$ is expressed as
\begin{equation}
\mathbf{x_i}\triangleq \big[0 ~~ 0 ~~ 0 ~~ s_l ~~ 0 ~~\cdots~~ 0\big]^T, 
\label{Eq:Eq3}
\end{equation}
where $s_l \in \mathcal{S}$, where $\mathcal{S}$ is the set of $M$-ary symbols $\mathcal{S} = \{s_0 ~~ s_1 ~~ \cdots ~~ s_{M-1}\}$.
For each transmission interval, $\log_2(N_t)$ and $\log_2(M)$ bits are carried by the active antenna index and $M$-ary symbol, respectively. Thus, the number of transmitted bits per channel use (bpcu) for SM is

\begin{equation}
\eta = \log_2(N_t) + \log_2(M)~\text{[bpcu]}.
\label{Eq:SM}
\end{equation}

SM provides better SE than SSK while protecting the zero IAI feature. To improve both SE and achievable performance, GSM activates $N_k$ Tx antennas for the transmission of the same data symbol, given that $1 \leq N_k<N_t$ \cite{5757786}. Thus, $\mathbf{x_i} \triangleq \big[0 ~~ s_l ~~\cdots ~~ 0 ~~ s_l ~~ 0 ~~\cdots~~ 0\big]^T $, and the SE rises to 

\begin{equation}
\eta =\floor{\log_2\binom{N_t}{N_k}} + \log_2(M)~\text{[bpcu]}.
\label{Eq:MASM}
\end{equation}

The transmission of different data symbols through the activated antennas is performed by multiple active SM (MA-SM) \cite{6166339}. As a result of the efficient implementation of IM with the conventional QAM/PSK, the achievable rate increases to

\begin{equation}
\eta =\floor{\log_2\binom{N_t}{N_k}} + N_k\log_2(M)~\text{[bpcu]}.
\label{Eq:MASM}
\end{equation}

A new perspective to SM is introduced through QSM where in-phase and quadrature parts of complex data symbols are transmitted by two different Tx antennas \cite{6868290, 7574985}. The selection of two Tx antennas requires $2\log_2(N_t)$ data bits. 
Hence, the transmission rate for QSM equals
\begin{equation}
\eta = 2\log_2(N_t)+ \log_2(M) ~\text{[bpcu]}.
\end{equation}
Although not emphasized sufficiently in the literature, a particular strength of the QSM is that it exploits the spatial selectivity by conveying the real and imaginary parts of the data symbol separately. {In order to further boost the data rate of SM, ESM proposes the transmission of information bits by the use of two different QAM/PSK sets, i.e., $\mathcal{S}_1$ and $\mathcal{S}_2$, for the two active Tx antennas \cite{7084604}.} It should be ensured that the same number of data bits is transmitted at each transmission interval. Otherwise, error propagation occurs due to asynchronization between the data blocks. Therefore, higher order modulation  $\mathcal{S}_2$ is used when one of the two antennas is activated, while lower order modulation $\mathcal{S}_1$ is utilized in the presence of {two active antennas.} Moreover, the selected modulation types decide the BER performance of ESM. {If Euclidean distance between the symbols modulated with $\mathcal{S}_1$ and $\mathcal{S}_2$ is higher than that of SM, better BER performance is achieved than SM, and vice versa}. The aforementioned space domain IM types suffer from a lack of diversity gain. In \cite{5508985}, Trellis coded spatial modulation (TCSM) is presented with the implementation of Trellis coded modulation (TCM)  over the antenna combinations of SM. In this way, the spatial distance between antennas within the same subblock is maximized without increasing the power consumption. Also,  in \cite{TCQSM1},  TCM is incorporated with QSM (TC-QSM) to further improve the error performance of SM systems. On the other hand, in order to achieve transmit diversity gain for any number of Tx antennas, STBC with SM (STBC-SM) and STBC with QSM (ST-QSM)  are proposed in \cite{5672371} and \cite{8277677, 8411498}, respectively. ST-QSM divides the existing $N_t$ antennas into two subsets ($N_{t1}$ and $N_{t2}$) to carry two complex symbols. The real and imaginary parts of the symbols are transmitted over the subcarriers that are independently chosen from the first and second subsets. The total number of the active subcarriers in each set corresponds to ($N_{k1}$ and $N_{k2}$). 
\textcolor{black}{Moreover, IM concept is applied to Rx antennas via preprocessing/precoding of the transmission vector with the knowledge of CSI at Tx and named precoded SM (PSM) or receiver SM \cite{6240160,5956573}. Generalised precoding-aided SM (GPSM) and QSM (GPQSM) are introduced in \cite{6644231} and \cite{7467510}, respectively. GPSM corresponds to SM at Rx, while GPQSM is the QSM with multiple active antennas at Rx.}


\paragraph{\textcolor{black}{IM via RISs}} 

\textcolor{black}{
RIS concept has been extensively investigated in the past few years. Intelligent surfaces consist of small, low cost and a high number of passive elements which control the reflection features of the incoming signals. For a comprehensive overview of RIS concept, interested readers are referred to \cite{nadeem2019asymptotic, taha2019enabling, zhao2019survey}. RIS-assisted IM concept is introduced in \cite{8981888}. It is shown that IM can be applied on the passive elements as well as Tx and Rx antennas.}

\subsubsection{\textcolor{black}{Frequency Domain IM}}

Indexing of the subcarriers in the frequency domain is proposed to improve both SE and EE of the conventional OFDM systems. Subcarrier index modulation OFDM (SIM-OFDM) divides incoming data bits into two parts \cite{5449882}. On-off keying data bits decide the status of  {$N_{sc}$} subcarriers in an OFDM block, and the remaining bits are conveyed through $N_a$ subcarriers whose status is on. However, the inconsistent number of the total bits per OFDM block results in error propagation and degrades the BER performance of SIM-OFDM. Enhanced SIM-OFDM (ESIM-OFDM) splits the OFDM block to {$N_{sc}/2$} subblocks with two subcarriers, and it only activates a single subcarrier ($N_a = 1$) per subblock to avoid error propagation \cite{6162549}. Inspired by the SM, SIM-OFDM and ESIM-OFDM are the early attempts for frequency domain IM. However, their performances are not satisfactory, and their implementations are impractical. Hence, the general concept for frequency domain IM is firstly introduced by OFDM-IM \cite{Basar1}.

In OFDM-IM, available subcarriers are partitioned into {$N_G$} subblocks, and each subblock includes {$N_b = N_{sc}/N_G$} subcarriers. $N_a$ subcarriers out of $N_b$ subcarriers are activated according to $p_1 = \floor{\log_2\binom{N_b}{N_a}}$ bits. The remaining $p_2= N_a{\log_2(M)}$ bits are utilized to modulate the active subcarriers. The number of transmitted bits per OFDM-IM subblock is 

\begin{equation}
p = p_1+ p_2 = \floor{\log_2\binom{N_b}{N_a}} + N_a{\log_2(M)}.
\end{equation}

Then, OFDM-IM subblocks are concatenated to generate an OFDM block, and the remaining process is the same as conventional OFDM. Inverse fast Fourier transform (IFFT) is applied to the OFDM block, and cyclic prefix (CP) is added to avoid inter-symbol interference (ISI). Thus, the SE of OFDM-IM is
\begin{equation}
\begin{aligned}
\eta =\frac{N_G}{N_{sc}+N_{cp}-1}\bigg({\floor{\log_2\binom{N_b}{N_a}}} \\+  {N_a{\log_2(M)}}\bigg)~\text{[bits/s/Hz]},
\label{Eq:OFDMIM}
\end{aligned}
\end{equation}
where $N_{cp}$ is the CP size in the frequency domain. At Rx, {maximum likelihood} (ML) detector is used for joint estimation of the active subcarriers and the QAM/PSK symbols after CP removal and fast Fourier transform (FFT) process. However, ML detector is impractical for large $N_{sc}$ values. Hence, in \cite{Basar1} log-likelihood ratio (LLR) detector is proposed for OFDM-IM. In order to both reduce correlation and exploit frequency diversity, interleaving for an OFDM block is employed by OFDM with interleaved subcarrier index modulation (OFDM-ISIM) \cite{6841601}. Lower correlation between the active subcarriers improves the detection performance at Rx, and {consequently the BER.} {Coordinate interleaved OFDM-IM (CI-OFDM-IM) achieves an additional diversity gain through the transmission of real and imaginary parts of a complex data symbol over two active subcarriers via the CI orthogonal design.} Therefore, CI-OFDM-IM provides higher reliability than both OFDM-IM and OFDM-ISIM \cite{7086323}. \textcolor{black}{Additionally, OFDM with I/Q index modulation {(OFDM-I/Q-IM)} utilizes different information bits to generate the I/Q parts of data symbols \cite{7112187, 7230238}.}

In OFDM-IM, $N_a$ value is fixed for all OFDM subblocks. {On the other hand, OFDM with generalized index modulation (OFDM-GIM) allows varying $N_a$ values for the different subblocks to enhance the SE of OFDM-IM \cite{7112187, 7470953}.}  Further SE improvement is achieved with DM-OFDM that uses two different  QAM/PSK sets $\mathcal{S}_1$ and $\mathcal{S}_2$ for $N_a$ and $N_b-N_a$ subcarriers, respectively \cite{7547943}. In this way, all the subcarriers are modulated within a subblock. Hence, the achieved SE by DM-OFDM equals
\begin{equation}
\begin{aligned}
\eta =\frac{N_G}{N_{sc}+N_{cp}-1}\bigg({\floor{\log_2\binom{N_b}{N_a}}} +  {N_a{\log_2(M_1)}}\\ + { (N_b-N_a){\log_2(M_2)}}\bigg)~\text{[bits/s/Hz]}
\end{aligned}
\end{equation}
where $M_1$ and $M_2$ are the constellation size of $\mathcal{S}_1$ and $\mathcal{S}_2$, respectively. Inspired by DM-OFDM, two promising schemes including multiple-mode OFDM-IM (MM-OFDM) and zero-padded tri-mode index modulation aided OFDM (ZTM-OFDM-IM) are introduced in the literature \cite{7936676, 8254931}. MM-OFDM uses multiple QAM/PSK sets within a subblock to enhance the SE, while ZTM-OFDM-IM performs fractional subcarrier activation by two different QAM/PSK sets. \textcolor{black}{In order to further increase the SE of the OFDM-IM systems,  in \cite{8734769}, layered OFDM-IM (L-OFDM-IM) is proposed by division of $p$ incoming bits into $L$ layers, where $N_{a_L}$ out of $N_{b_L}$ subcarriers are activated, given that $ N_b = N_{b_L}+ N_{a_L}(L-1)$.}  
 
The aforementioned frequency domain IM types are based on OFDM technology. In \cite{7848916}, IM is applied to generalized frequency division multiplexing (GFDM), instead of OFDM. {GFDM performs block-based transmission over $T$ time slots, and each block consists of $K$ sub-symbols composed by $N_{sc}$ subcarriers.} Moreover, each block can include different number of sub-symbols. GFDM alleviates the strict synchronization requirement of OFDM since non-orthogonal pulse shaping is allowed. In this regard, GFDM with IM (GFDM-IM) combines the benefits of GFDM with IM flexibility. 


\subsubsection{\textcolor{black}{Time Domain IM}}
Inspired by the frequency domain IM, single carrier with IM (SC-IM) is proposed in the time domain \cite{7738501}. A SC block with $K_s$ symbols is divided into $K_G$ subblocks which consist of $K_b = K_{s}/K_G$ symbols. Data transmission is performed at the time intervals corresponding to active $K_a$ symbols, and the remaining $K_b-K_a$ symbols are set to zero. SC subblocks are concatenated to generate a SC block, and then CP is added before its transmission over a multi-path channel. The SE of SC-IM is
\begin{equation}
\begin{aligned}
\eta= \frac{K_{G}}{K+K_{cp}-1} \bigg(\floor{\log_2\binom{K_b}{K_a}} \\+ K_a{\log_2(M)}\bigg)~\text{[bits/s/Hz]},
\end{aligned}
\label{Eq:TI}
\end{equation}
where $K_{cp}$ refers to the CP size in the time domain. At Rx,  ML or LLR detector is utilized to find the non-zero symbols after CP removal and frequency domain equalization \cite{Basar1}. It is worth {mentioning} that interleaving at Tx is needed to tear the correlation between the active symbols if the channel is non-selective in time. Thus, de-interleaving is required at Rx. Faster-than-Nyquist signaling with IM (FTN-IM) has been proposed since the passive symbols in the SC block alleviate the effect of ISI \cite{7973048, 8093605}. Furthermore, dual-mode single carrier with index modulation (DM-SCIM) utilizes two different QAM/PSK sets for further increasing the SE of SC-IM, as in DM-OFDM \cite{8287922}. 

\subsubsection{\textcolor{black}{Code Domain IM}}

By taking the advantage of direct-sequence spread spectrum (DS-SS) technology, code index modulation SS (CIM-SS) has been proposed in \cite{6994807}. The information-bearing unit is the spreading code available in a predefined table of spreading codes. In \cite{6994807}, two orthogonal Walsh codes ($w_1$ and $w_2$) are stored in the look-up table. The incoming two bits are combined to generate a subblock, and one bit in each subblock chooses a code ($N_{ac}$) to spread the remaining bit {over a time duration}. {In-phase and quadrature parts} of a complex symbol are modulated by orthogonal Walsh codes. Generalized CIM-SS (GCIM-SS) uses the code table with $N_{ct}$ size, where $\floor{\log_2(N_{ct})}$ defines the number of bits required for choosing a code \cite{7317808, 7156412}. Hence, the SE of GCIM-SS is

\begin{equation}
\eta = \frac{1}{N_{ct}}\big(2\floor{\log_2(N_{ct})} + {\log_2(M)}\big)~\text{[bits/s/Hz]}.
\end{equation}

At Rx, distinct $N_{ct}$ correlators are used for {the in-phase and quadrature parts} of the complex symbol. The correlator that gives the maximum absolute value corresponds to the utilized code at Tx. {Later, de-spreading and conventional QAM/PSK demodulation are applied to obtain the transmitted information bits.} \textcolor{black}{CIM is also applied in the frequency domain with the aid of OFDM and named index modulated OFDM-SS (IM-OFDM-SS) \cite{8264822, 8269169}. In order to obtain diversity gain, IM-OFDM-SS spreads data symbol over several subcarriers via spreading codes. ML and maximum ratio combining (MRC)-based detectors are used at Rx. Also, a generalized framework for multi-user scenarios is introduced in \cite{8264822}.}


\subsubsection{\textcolor{black}{Channel Domain IM}}

{Media-based modulation (MBM) transmits information bits via different channel realizations {generated by} the on-off status of the available RF mirrors, which are located in the vicinity of the Tx antenna \cite{6620786, 7511273, 7416621, 7676245, BasarMBM}. In other words, each channel realization corresponds to a different point in the constellation diagram at the Rx. No additional energy is required to transmit the bits by MBM. Moreover, it is shown that $1 \times {N_r}$ single-input multiple-output (SIMO) systems with MBM can harvest the same energy as $N_t \times {N_r}$ MIMO systems, yielding $N_t = N_r$ \cite{6620786}. Unlike SSK, SE of MBM linearly increases with the number of RF mirrors {($N_{rf}$)}. Thus, the transmission rate of MBM with a single RF mirror activation ($N_{am}=1$) is 
\begin{equation}
\eta = N_{rf} + \log_2(M) ~\text{[bpcu]}.
\label{Eq:MBM}
\end{equation}
The main issue for MBM is the requirement of CSI at Rx. $2^{N_{rf}}$ channel realizations {need to be estimated} in the presence of ${{N_{rf}}}$ mirrors. Usually, the estimation of CSI is performed through training process. {However, it leads to severe signaling overhead for the system especially in the case of a higher number of RF mirrors.} To overcome this, differential MBM (DMBM) is proposed in the literature, where the estimation process is avoided by encoding consecutive data blocks at the cost of performance degradation \cite{7887715}. \textcolor{black}{In \cite{7864471} and \cite{8668573}, space-time channel modulation (STCM) and space-time MBM (ST-MBM) incorporate STBCs into channel domain IM for the purpose of achieving diversity gain. Specifically, STCM adopts Alamouti's STBC as the core, and ST-MBM amalgamates the Hurwitz-Radon family of matrices [32] with the MBM principles to allow a single RF chain-based transmission. }


\subsubsection{\textcolor{black}{Polarization Domain IM}}

\textcolor{black}{ In order to provide both higher multiplexing gain and higher SE for the single RF MIMO systems, polarization shift keying (PolarSK) is introduced in \cite{8048033}. PolarSK uses the available $P$ polarization states, i.e., linear polarization, circular polarization, and elliptic polarization, to transmit the incoming bits as in SSK. In a recent study, a novel IM scheme, i.e, polarization modulation (PM), utilizes polarization characteristics to carry extra information bits along with the complex data symbols. Specifically, not only vertical and horizontal polarizations but also the axial ratio and tilt angle of elliptic polarization are used for conveying the information bits through IM \cite{9028167}.}

\textcolor{black}{\subsection{Two-Dimensional IM}}

\textcolor{black}{Two-dimensional IM corresponds to the simultaneous activation of information-bearing units in two different dimensions, such as space \& frequency, and space \& time, as given in Fig.~\ref{Fig:AppDomains}.}

\subsubsection{\textcolor{black}{Dispersion Matrices-based IM}}

Space-time shift keying (STSK) introduces an innovative information-bearing unit, i.e., DMs, for conventional MIMO systems \cite{5684112,5599264}. STSK exploits the time domain along with the space domain through block-based transmission as $\mathbf{Y}= \mathbf{H}\mathbf{X}+ \mathbf{N}$, where $\mathbf{Y} \in \mathbb{C}^{N_r\times T}$, $\mathbf{H} \in \mathbb{C}^{N_t\times N_r}$, and $\mathbf{X} = \mathbf{D}s \in \mathbb{C}^{N_t\times T}$ denote the received block, the multi-path channel, and the transmitted block, respectively. $\mathbf{D} \in \mathbb{C}^{N_t\times T}$ refers to the DM to spread the $M$-ary symbol ($s$) over space and time dimensions, and $T$ is a block duration. A STSK block (${\mathbf{X} = \mathbf{D}s}$) is generated by the different combinations of $Q$ DMs with the $M$-ary symbols in a given modulation set. Also, SSK and SM can be assessed as the special cases of STSK, given that $T=1$. Thus, STSK provides diversity gain along with multiplexing gain by adjusting the number of DMs ($Q$), STSK block duration ($T$), and the number of Tx and Rx antennas ($N_t, N_r$). To exemplify, a single DM and the modulation set with $M = M_1$ complex symbols, or two DMs and the modulation set with the $M = M_1/2$ complex symbols can generate different STSK blocks to transmit $\frac{\log_2{(QM)}}{T}$ bits \cite{5599264}. It should be noted that the correlation between $Q$ DMs should be as low as possible to improve the detection performance at Rx. This is one of the ongoing research areas pertaining to the design of DMs \cite{6466367, 6378497}. Generalized STSK (GSTSK) is developed to choose $P$ DMs at each transmission interval \cite{5703198}. Hence, the achievable rate by GSTSK is
\begin{equation}
\eta =\frac{\floor{\log_2\binom{Q}{P}}+ P\log_2{(M)}}{T}~\text{[bpcu]}.
\label{Eq:STSK}
\end{equation}
The BER performance of STSK is affected by ISI under frequency-selective channel conditions. {Therefore, space-frequency shift keying (SFSK) proposes the utilization of the conventional $F$-FSK to spread the data symbol in space, time, and frequency, instead of $M$-QAM/PSK  \cite{5688440}. A SFSK block corresponds to the multiplication of the $F$-FSK symbol with the DM.} At Rx, square-law and ML detectors are used to detect the active frequencies and the DM, respectively. {Moreover, space-time-frequency shift keying (STFSK) amalgamates STSK and SFSK. The information bits are modulated by $M$-QAM/PSK, $F$-FSK, and the index of active DM.}

To exploit the robustness of OFDM against the frequency-selective channels, STSK has been combined with OFDM, and named OFDM-STSK \cite{6399190}. {Before the conventional OFDM transmission, $J = N_{sc}/T$ STSK blocks of size $N_t \times T$ are concatenated, where it is assumed that $N_{sc}$ is the multiple of $T$.} Thereafter, IFFT is applied, followed by CP addition. In other words, STSK block are modulated by OFDM. In this way, each column of the STSK block is transmitted by a subcarrier, corresponds to the frequency-flat channel. The transmission rate of STSK and OFDM-STSK is equal, given that $N_{sc}\gg N_{cp}$.
Different from the SFSK in \cite{5688440}, OFDM-based SFSK (OFDM-SFSK) approach is proposed in \cite{8417812}, where the data symbol is spread over space and frequency dimensions. Indeed, OFDM-SFSK and OFDM-STSK follow the same idea of achieving robustness against time-varying OFDM channels. Differently, DMs in OFDM-SFSK are generated by the circular shifting of sparse vectors that also provides robustness against ICI for OFDM systems. 
\textcolor{black}{In \cite{7448828}, the layered multi-group steered STSK (LMG-SSTSK) is proposed for multi-user MIMO downlink systems by combining OFDM, STSK, and Tx beamforming.} Moreover, differential SM (DSM) avoids heavy channel estimation by differentially encoding two successive data blocks at Tx \cite{6879496}. For this purpose, DSM exploits the time domain along with the space domain through block-based transmission as in STSK. In DSM, it is assumed that $T = N_t$. Each column of $\mathbf{X}$ corresponds to a transmission interval in which a single antenna is activated.}


\subsubsection{\textcolor{black}{Space \& Frequency Domain IM}}

Two transmit entities, i.e., antennas and subcarriers are used simultaneously to carry the information. Incoming bits are divided into three parts for antenna indexing, subcarrier indexing, and conventional $M$-ary modulation \cite{7440707,7414057,7277106}. SM-OFDM with subcarrier index modulation {(ISM-OFDM)} is proposed to alleviate the ICI impact for vehicle-to-everything (V2X) communication \cite{7440707}. Since a single antenna is activated at each transmission interval, the transmission rate of ISM-OFDM considering (\ref{Eq:SM}) and (\ref{Eq:OFDMIM}) equals 

\begin{equation}
\begin{aligned}
\eta = \frac{N_G}{N_{sc}+N_{cp}-1} \bigg(\floor{\log_2{N_t}} +\floor{\log_2\binom{N_b}{N_a}} \\ + {\log_2(M)}\bigg)~\text{[bpcu]}.
\label{Eq:ISMOFDM}
\end{aligned}
\end{equation}

{Instead of the conventional SM, generalized space-frequency IM (GSFIM) combines OFDM-IM with MA-SM in order to activate multiple Tx antennas and subcarriers at each transmission interval \cite{7277106}.} Regarding (\ref{Eq:MASM}) and (\ref{Eq:OFDMIM}), the achieved rate by GSFIM corresponds to the total number of transmitted bits by OFDM-IM and MA-SM.
%
Moreover, GSFIM has been evaluated in the context of GFDM, named GFDM with SFIM (GFDM-SFIM) that provides higher SE than GSFIM for a given BER performance \cite{8277679}.

\subsubsection{\textcolor{black}{Space \& Time Domain IM}}

Simultaneous indexing of the transmission entities in both space and time is evaluated in  \cite{7925922,8110611}. Considering the time slots, Tx antennas and RF mirrors as separate units, two different space\& time domain  IM schemes are presented: time-indexed SM (TI-SM) and time-indexed MBM (TI-MBM).

In TI-SM, Tx is equipped with $N_t$ antennas and one RF chain, while Rx contains $N_r$ antennas. As in (\ref{Eq:TI}), the active symbols for SC-IM are chosen by $\floor{\log_2\binom{K_b}{K_a}}$ bits. Then, $K_a\log_2(N_t)$ bits corresponding to (\ref{Eq:SM}) decide the active Tx antenna. 
{TI-MBM only requires a single Tx antenna supported by $N_{rf}$ RF mirrors. MBM is applied to transmit the additional bits, instead of SM. Hence, considering (\ref{Eq:MBM}), \textcolor{black}{The number of information bits conveyed by TI-MBM equals the total number of information bits carried by both SC-IM and MBM.}

\vspace{2mm}

\subsubsection{\textcolor{black}{Space \& Channel Domain IM}}

\textcolor{black}{SM with MBM (SM-MBM) and quadrature channel modulation (QCM) are employed through the combination of MBM with SM and QSM, respectively \cite{7925922, 8014408}. Basically, SM-MBM and QCM perform transmission through indexing both Tx antennas and RF mirrors. Therefore, $N_t$ antennas are equipped with $N_{rf}$ RF mirrors at Tx. The transmission rate of SM-MBM corresponds to $\eta =  \log_2(N_t) + N_{rf} + \log_2(M)~\text{[bpcu]}$. Considering the QSM principles, QCM transmists $\eta =  2\log_2(N_t) + N_{rf} + \log_2(M)~\text{[bpcu]}$.}

\vspace{2mm}

\subsubsection{\textcolor{black}{Space \& Code Domain IM}}

A novel MIMO transmission scheme is developed based on IM in space and code domains \cite{8404646TSP}. The transmission rate of CIM with SM (CIM-SM) is given by
\begin{equation}
\eta = \frac{1}{N_{ct}}\big[2\floor{\log_2N_{ct}}+\log_2{(N_t)}+ 2\log_2{(M)}\big]~\text{[bpcu]},
\end{equation}
which corresponds to the total number of bits conveyed by CIM and SM. Firstly, the Rx process of CIM is employed, followed by ML detector to decide the utilized antenna and the transmitted data symbols. 

\begin{figure*}
	\centering
	\includegraphics[scale = 0.40]{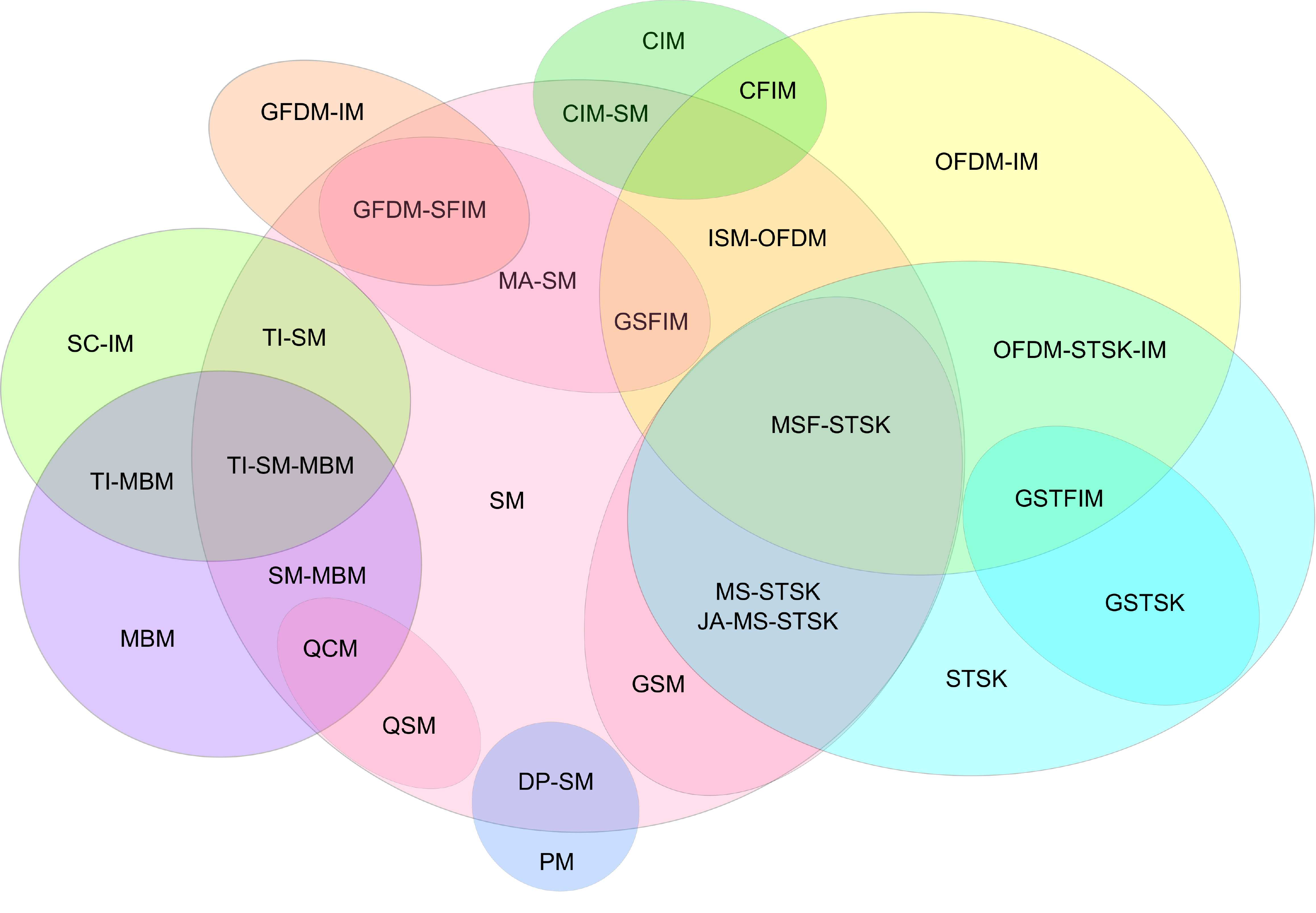}
	\caption{\textcolor{black}{Corresponding fundamental IM variants for the existing multidimensional IM schemes.}}
	\label{Fig:ımtypes}
\end{figure*}

\vspace{2mm}

\subsubsection{\textcolor{black}{Space \& Polarization Domain IM}}

\textcolor{black}{ In \cite{SPSK}, space-polarization shift keying (SPSK) is introduced via the utilization of dual-polarized antennas. Besides the active antenna, the utilized polarization type also carries information bits. Moreover, SM and PM are combined in SM with dual-polarized antennas (DP-SM) to avoid the spatial correlation in SM-MIMO systems \cite{7248680, 7795239}. As a result, the achievable SE is also increased since the space limitation in SM-MIMO systems due to the required distance between the adjacent antennas is alleviated.}

\vspace{2mm}

\subsubsection{\textcolor{black}{Code \& Frequency Domain IM}}

\textcolor{black}{Joint code-frequency index modulation (CFIM) is presented in \cite{8792959} by simultaneous indexing in frequency and code domains in order to support multi-user communication with low-power consumption.}

\textcolor{black}{\subsection{Three-Dimensional IM}}

\textcolor{black}{Three-dimensional IM types are the enhanced IM types that would serve diverse requirements of 5G and beyond networks. The existing three-dimensional IM types are the combinations of space \& DMs, frequency \& DMs, and time \& space \& frequency domains, as given in Fig.\ref{Fig:AppDomains}.}

\subsubsection{\textcolor{black}{Space \& Dispersion Matrices-based IM}}

The conventional STSK uses all of the available $N_t$ Tx antennas for a transmission. In order to enhance the system reliability, partial antenna activation for the transmission of STSK block is presented in multi-set STSK (MS-STSK) \cite{7494949}. Moreover, the columns of the STSK block corresponding to different time intervals are multiplied by  different phase shifts for reducing the correlation amongst the transmissions. \textcolor{black}{In \cite{8003424}, joint alphabet STSK (JA-STSK) and joint-alphabet MS-STSK (JA-MS-STSK) are performed by using a joint alphabet that corresponds to the utilization of different DMs and antenna combinations over multiple time slots for increasing the throughput gain of the STSK systems.} \textcolor{black}{In \cite{8281511}, a generalized framework that can accommodate all DM-based IM techniques is introduced and named layered multi-set GSTSK (LMS-GSTSK). Specifically, LDC, BLAST, SM, GSM, QSM, SSK, GSTSK, and MS-STSK can be implemented by the proper adjustment of LMS-GSTSK's parameters. Indeed, LMS-GSTSK provides adaptive dimensional IM due to its scalable structure.} 

\subsubsection{\textcolor{black}{Frequency \& Dispersion Matrices-based IM}}

CS-aided OFDM-STSK with frequency index modulation (OFDM-STSK-IM) is presented for further improving the SE and BER performance of OFDM-STSK \cite{7583706,8322306}. At first, incoming $m$ bits are divided into $N_G$ groups, and each group contains $\floor{\log_2\binom{N_{b}}{N_a}}$ and $\log_2{(QM)}$ bits to activate $N_a$ subcarriers and select a DM, respectively. Then, coordinate interleaved STSK blocks are mapped to active $N_a$ subcarriers. A virtual domain with $N_v$ subcarriers is introduced by CS for transmitting additional energy-free $\floor{\log_2\binom{N_{v}}{N_a}}$ bits per subblock, given that $N_v \gg N_b$. At Rx, the signal is first compressed from the virtual domain to the frequency domain and then ML detector is utilized to obtain the transmitted bits. \textcolor{black}{In \cite{8487049}, generalized space, time, and frequency index modulation (GSTFIM) combines GSTSK with OFDM-IM to achieve higher SE for STSK systems. }

\subsubsection{ \textcolor{black}{Space \& Time \& Channel Domain IM}}

\textcolor{black}{In \cite{8110611}, Time-indexed SM with MBM (TI-SM-MBM) allows joint utilization of SC-IM, SM, and MBM for the purpose of increasing SE. Accordingly,  the achieved SE by TI-SM-MBM equals the summation of each individual data rate.} 

\begin{table*}[h]
	
	\caption{\textcolor{black}{IM techniques with the aid of dimension(s) exploitation}}
	\centering
	\renewcommand{\arraystretch}{1.5}
	\scriptsize
	\begin{threeparttable}
		\begin{tabular}{|C{3.5cm}||c|c|c|c|c|c|c|c|c|C{2.5cm}|}
			\hline
			\multirow{2}{*}{\textbf{\textcolor{black}{IM Techniques}}}  & \multicolumn{4}{c|}{\textbf{\textcolor{black}{Exploited Domain(s)}} } & \multicolumn{5}{c|}{ \textbf{\textcolor{black}{IM Domain(s)}}} & \multirow{2}{*}{\textbf{\textcolor{black}{ IM Type}}}\\ 
			\cline{2-10}
			& \textbf{\textcolor{black}{Space}} & \textbf{\textcolor{black}{Time}} & \textbf{\textcolor{black}{Frequency}} & \textbf{\textcolor{black}{I/Q}} &\textbf{\textcolor{black}{Space}} & \textbf{\textcolor{black}{Time}} & \textbf{\textcolor{black}{Frequency}} & \textbf{\textcolor{black}{Code}} & \textbf{\textcolor{black}{Channel}} & \\
			\hline \hline
			\textcolor{black}{ QSM$^1$ \cite{5672371}} & & & & \textcolor{black}{ \ding{51}} & \textcolor{black}{ \ding{51}} &  & & & &\textcolor{black}{One-dimensional} \\
			\hline
			\textcolor{black}{ STBC-SM \cite{5672371}} &\textcolor{black}{ \ding{51}} & \textcolor{black}{\ding{51}} & & &\textcolor{black}{\ding{51}} &  & & & &\textcolor{black}{One-dimensional} \\
			\hline
			\textcolor{black}{ST-QSM\cite{8277677}}& \textcolor{black}{\ding{51}} & \textcolor{black}{\ding{51}} & & \textcolor{black}{ \ding{51}} & \textcolor{black}{\ding{51}}  & & & & &\textcolor{black}{One-dimensional}  \\
			\hline
			\textcolor{black}{ OFDM-I/Q-IM \cite{7112187}} & & & & \textcolor{black}{ \ding{51}} &  &  &\textcolor{black}{ \ding{51}} & & &\textcolor{black}{One-dimensional} \\
			\hline
			\textcolor{black}{IM-OFDM-SS\cite{8264822}}	& & & \textcolor{black}{\ding{51}} & & & & &\textcolor{black}{\ding{51}} & & \textcolor{black}{One-dimensional}  \\
			\hline
			\textcolor{black}{STCM \cite{7864471}} 	&\textcolor{black}{ \ding{51}} & \textcolor{black}{\ding{51}} & & & &  & & & \textcolor{black}{\ding{51}} & \textcolor{black}{One-dimensional} \\
			\hline
			\textcolor{black}{ST-MBM \cite{8668573}}	& \textcolor{black}{\ding{51}} & \textcolor{black}{\ding{51}} & & & & & & & \textcolor{black}{\ding{51}}& \textcolor{black}{One-dimensional}  \\
			\hline
			\textcolor{black}{SFSK \cite{5688440}}	& & & \textcolor{black}{\ding{51}}& & \textcolor{black}{\ding{51}} &\textcolor{black}{\ding{51}} & & & &\textcolor{black}{\textcolor{black}{Two-dimensional} } \\
			\hline
			\textcolor{black}{STFSK \cite{5688440}}	& & & \textcolor{black}{\ding{51}}& & \textcolor{black}{\ding{51}} &\textcolor{black}{\ding{51}} & & & &\textcolor{black}{\textcolor{black}{Two-dimensional}} \\
			\hline
			\textcolor{black}{OFDM-STSK \cite{6399190}}	& & & \textcolor{black}{\ding{51}}& & \textcolor{black}{\ding{51}} &\textcolor{black}{\ding{51}} & & & &\textcolor{black}{\textcolor{black}{Two-dimensional}} \\
			\hline
			\textcolor{black}{OFDM-SFSK \cite{8417812}}	& & & \textcolor{black}{\ding{51}} & &\textcolor{black}{\ding{51}} &\textcolor{black}{\ding{51}} & & & & \textcolor{black}{\textcolor{black}{Two-dimensional}} \\
			\hline
		\end{tabular}
		\begin{tablenotes}\footnotesize
			\item[] \textcolor{black}{${^1}$It is also valid for QSM-based IM techniques, such as QCM and GPQSM.}
		\end{tablenotes}
	\end{threeparttable}
	\label{Tab:FewType} 
\end{table*}

\textcolor{black}{\subsection{Hyper-Dimensional IM}}

\textcolor{black}{Hyper-dimensional IM types are relatively less investigated in the literature when compared to the lower-dimensional IM types due to their complex structure, as shown in Fig.~\ref{Fig:AppDomains}.} 

\subsubsection{\textcolor{black}{Space \& Frequency \& Dispersion Matrices-based IM}}
\textcolor{black}{Multi-space-frequency STSK (MSF-STSK) combines OFDM-IM, GSM, and STSK to attain interference immunity and diversity gain \cite{7792569}. Besides conventional $M$-ary symbols, incoming bits are carried by the indices of active DMs in space \& time domains, active subcarriers in the frequency domain, and active antennas in the space domain.} 

{Fig.~\ref{Fig:ımtypes} illustrates the multidimensional IM techniques and their constituent single domain ones. Since GSM, MA-SM and QSM are the advanced versions of SM, they are shown with the same color. According to Fig.~\ref{Fig:ımtypes}, one-dimensional IM schemes without intersection regions provide insights about possible novel multidimensional IM schemes.}

\textcolor{black}{
\begin{remark}
IM schemes, such as QSM, STBC-SM, IM-OFDM-SS, SFSK, OFDM-STSK, and OFDM-I/Q-IM, exploit different domain(s) alongside the IM domain(s) in order to serve diverse user demands. To exemplify, STBC-SM exploits space and time dimensions for the purpose of achieving transmit diversity gain for SM systems. In the same vein, OFDM-STSK utilizes the frequency domain aiming to overcome the ISI encountered in STSK transmission, while IM is applied in space and time domains. On the other hand, QSM and OFDM-I/Q-IM utilize I and Q dimensions for enhancing the SE of OFDM-IM systems. Therefore, these IM techniques are categorized considering the number of domains in which IM is employed. Table~\ref{Tab:FewType} illustrates the IM techniques that provide dimension(s) exploitation to alleviate the shortcomings of a particular IM type.
\end{remark}
}

\section{Appropriate IM techniques for Next-Generation Services}

\textcolor{black}{Although the variety of IM types promises appealing trade-offs amongst SE, EE, BER, and flexibility, the integration of diverse IM techniques into the current communication systems bring different challenges for the Tx and Rx sides of modern communication systems due to the requirement of numerous hardware design and signal processing techniques. In this section, promising IM types are subsumed considering the requirements of eMBB, mMTC, and URLLC. Thereafter, the advantages and disadvantages of a given IM domain are quantified for establishing a clear distinction between them, and its fidelity is evaluated in terms of practical implementation.} 

\begin{table*}
	\caption{\textcolor{black}{Data rate and computational complexity assessment of space domain IM techniques}}
	\centering
	\renewcommand{\arraystretch}{1.5}
	\scriptsize
	\begin{threeparttable}
		\begin{tabular}{|C{2.4cm}||C{2.6cm}|C{4.75cm}|p{6.25cm}|} 
			\hline
			\textcolor{black}{\textbf{{IM Techniques}}}  & \textcolor{black}{\textbf{\# RF Chain ($N_{rc}$})} & \textcolor{black}{\textbf{{Data Rate}} \text{[bpcu]}} & \textcolor{black}{\makecell{\textbf{{ Computational Complexity at Rx}}}} \\
			\hline \hline
			\multirow{2}{*}{\textcolor{black}{SM \cite{4382913}}} & \multirow{2}{*}{\textcolor{black}{1}} &   \multirow{2}{*}{\textcolor{black}{$ \log_2(N_t) + \log_2(M)$}} & \textcolor{black}{MRC $ \rightarrow 2N_rN_t-N_t$ }  \\
			&  &  & \textcolor{black}{Opt.  $ \rightarrow N_tM(3N_r+1)$}  \\
			\hline
			\textcolor{black}{GSM \cite{5757786}}&  \textcolor{black} {$1 \leq N_{rc}$ = $N_k<N_t$} &  \textcolor{black}{ $ \floor{\log_2\binom{N_t}{N_k}} + \log_2(M)$ }&  \textcolor{black}{ML $\rightarrow N_r 2^{\floor{\log_2\binom{N_t}{N_k}}}M(N_k+2)$ } \\
			\hline
			\multirow{2}{*}{\textcolor{black}{MA-SM \cite{6166339}}}&\multirow{2}{*}{ \textcolor{black} {$1 \leq N_{rc}$ = $N_k<N_t$} }& \multirow{2}{*}{ \textcolor{black}{$ \floor{\log_2\binom{N_t}{N_k}} + N_k\log_2(M)$}} &  \textcolor{black}{ ML $ \rightarrow N_r 2^{\floor{\log_2\binom{N_t}{N_k}}}M^{N_k}(N_k+2)$}\\
			&  &  &  \textcolor{black}{LC} $  \textcolor{black}{\rightarrow (2 N_k^3 +  3N_k^2-5N_k)/6+ N_rN_k(2N_k+1)} $\\
			\hline
			\multirow{2}{*}{ \textcolor{black}{QSM$^1$ \cite{5757786, 7574985}}}& 	\multirow{2}{*}{ \textcolor{black}{1}} & 	\multirow{2}{*}{\textcolor{black}{$ {\log_2(N_t^2)} + \log_2(M)$} }& \textcolor{black}{ML $\rightarrow 4N_rN_t^2M$ } \\
			& & & \textcolor{black}{LC $\rightarrow 4N_t^3 + 4N_{MPA}^2MN_r$}\\
			\hline
			\multirow{2}{*}{\textcolor{black}{ESM$^2$\cite{7084604}}} &  \multirow{2}{*}{\textcolor{black}{1, 2}} & \multirow{1}{*}{\textcolor{black}{$ \floor{\log_2\big(\binom{N_t}{N_{k_1}}+\binom{N_t}{N_{k_2}}+\binom{N_t}{N_{k_3}}\big)} + $}} & \multirow{2}{*}{\textcolor{black}{ML {$\rightarrow {{\binom{N_t}{N_{k_1}}}} M_1^{N_{k_1}} + { \binom{N_t}{N_{k_2}} {M_2}^{N_{k_2}}} + \binom{N_t}{N_{k_3}} {M_3}^{N_{k_3}} $ }}}\\ 
			& & \multirow{1}{*}{\textcolor{black}{$\log_2(M_1)$}}& \\
			\hline 
			{\textcolor{black}{TCSM \cite{5508985}}} &  {\textcolor{black}{1}} & {\textcolor{black}{$ \frac{1}{R}\big({\log_2(N_t)} + \log_2(M)\big)$, ($R$ = Code Rate)} }&  {\textcolor{black}{Opt. $ \rightarrow N_tM(3N_r + 1)$} }  \\
			\hline
			\textcolor{black}{TC-QSM \cite{TCQSM1}} & \textcolor{black}{1} &  \textcolor{black}{$ \frac{1}{R}\big({\log_2(N_t^2)} + \log_2(M)\big)$}& \textcolor{black}{LC $\rightarrow 4N_t^3 + 4N_{MPA}^2MN_r$}\\
			\hline
			\multirow{2}{*}{\textcolor{black}{STBC-SM \cite{5672371}}} & \multirow{2}{*}{ \textcolor{black}{2}} & \multirow{2}{*}{\textcolor{black}{$\frac{1}{2}\floor{\log_2{\binom{N_t}{N_k}}} + \log_2(M), (N_k=2) $}} &  \textcolor{black}{ ML $ \rightarrow 2^{\floor{\log_2\binom{N_t}{2}}}M^{2}$} \\
			& & & \textcolor{black}{ LC $ \rightarrow 2^{\floor{\log_2\binom{N_t}{2}}}M$} \\
			\hline
			\multirow{1}{*}{\textcolor{black}{ST-QSM \cite{8277677, 8411498}}}& \multirow{1}{*}{\textcolor{black}{$2 \leq N_{rc}<  N_t$} }  & \multirow{1}{*}{\textcolor{black}{${\log_2(N_{t1}) + \log_2(N_{t2})+\log_2(M)}$}}& \textcolor{black}{ML $ \rightarrow  M^{2} 2^{\floor{\log_2(N_{t1})}} +M^{2}2^{\floor{\log_2(N_{t2})}}$}\\
			\hline
			\textcolor{black}{{PSM} \cite{5956573}}& \textcolor{black}{1}  & {\textcolor{black}{$ \log_2(N_r) + \log_2(M)$}} & \textcolor{black}{ML $ \rightarrow  N_rM$}\\ 
			\hline
			\textcolor{black}{{GPSM} \cite{6644231}} & \textcolor{black}{ $1 \leq N_{rc}$ = $N_k<N_t$} & \textcolor{black}{ $ \floor{\log_2\binom{N_r}{N_k}} + N_k\log_2(M)$ } & \textcolor{black}{ML $ \rightarrow  2^{\floor{\log_2\binom{N_r}{N_k}}}M^{N_k}$}  \\ 
			& & & \textcolor{black}{ LC $ \rightarrow N_r+3N_kM+M$}\\
			\hline
			\textcolor{black}{{GPQSM} \cite{7467510}} & \textcolor{black}{ $1 \leq N_{rc}$ = $N_k<N_t$} & \textcolor{black}{ $ 2\floor{\log_2\binom{N_r}{N_k}} + N_k\log_2(M)$ } & \textcolor{black}{ML $ \rightarrow  2^{2\floor{\log_2\binom{N_r}{N_k}}}M^{N_k}$} \\
			& & & \textcolor{black}{ LC $ \rightarrow 10N_r+3N_rM+M$}\\
			\hline
		\end{tabular}
		\begin{tablenotes}\footnotesize
			\item [] \textcolor{black}{$^1$$N_{MPA}$ corresponds to the number of the most probable active antenna indices ($ 2 \leq N_{MPA}<  N_t$).}
			\item[] \textcolor{black}{$^2$The generalization of the computational complexity for ESM is not possible due to the variable number of  antenna combinations. Here, the complexity calculation is given in order to achieve the same data rate  ($ 8~\text{[bpcu]}$) with SM ($M=64QAM, N_t = 4, N_r =N_r$). Thus, $\mathcal{S}_1, \mathcal{S}_2$, and $\mathcal{S}_3$ are used with the size of  $M_1 = 16 $, and $M_2 = M_3 = 4$, respectively, while $N_{k_1}=1$, and  $ N_{k_2}= N_{k_3}=2N_{k_1} = 2$.}
		\end{tablenotes}
	\end{threeparttable}
	\label{Tab:CompRateComplexityS} 
\end{table*}

\begin{table*}
	\caption{\textcolor{black}{Data rate and computational complexity assessment of frequency domain IM techniques}}
	\centering
	\renewcommand{\arraystretch}{1.4}
	\scriptsize
	\begin{threeparttable}
		\begin{tabular}{|C{2.65cm}|| C{3.3cm}|C{4.65cm}|p{5.35cm}|} 
			\hline
	{\textcolor{black}{\textbf{{IM Techniques}}}}  & \textcolor{black}{\textbf{\# Active Subcarriers ($N_{a}$})} & \textcolor{black}{\textbf{{Data Rate}} \text{[bpcu]}} & \textcolor{black}{\makecell{\textbf{{ Computational Complexity at Rx}}}} \\
     \hline \hline
	\multirow{2}{*}{\textcolor{black}{{OFDM-IM} \cite{Basar1}}}&  \multirow{2}{*}{\textcolor{black} {$1 \leq N_a <N_b$}} &  \multirow{2}{*}{\textcolor{black}{${N_G}\big({\floor{\log_2\binom{N_b}{N_a}}}+{N_a{\log_2(M)}}\big)$}}  & \textcolor{black}{ML $ \rightarrow  {N_G}2^{\floor{\log_2\binom{N_b}{N_a}}}M^{N_a}$} \\
		& & & \textcolor{black}{LLR $ \rightarrow  {N_G}{N_b}M$}\\
		    \hline 
       \multirow{2}{*}{\textcolor{black}{{OFDM-ISIM} \cite{6841601}}}&  \multirow{2}{*}{\textcolor{black} {$1 \leq N_a <N_b$}} &  \multirow{2}{*}{\textcolor{black}{${N_G}\big({\floor{\log_2\binom{N_b}{N_a}}}+{N_a{\log_2(M)}}\big)$}}  & \textcolor{black}{ML $ \rightarrow  {N_G}2^{\floor{\log_2\binom{N_b}{N_a}}}M^{N_a}$} \\
		    & & & \textcolor{black}{LLR $ \rightarrow  {N_G}{N_b}M$}\\
		    \hline
		   \multirow{2}{*}{\textcolor{black}{{CI-OFDM-IM} \cite{7086323} }}&  \multirow{2}{*}{\textcolor{black} {$1 \leq N_a <N_b$}} &  \multirow{2}{*}{\textcolor{black}{${N_G}\big({\floor{\log_2\binom{N_b}{N_a}}}+{N_a{\log_2(M)}}\big)$}}  & \textcolor{black}{ML $ \rightarrow  {N_G}2^{\floor{\log_2\binom{N_b}{N_a}}}M^{N_a}$} \\
		  & & & \textcolor{black}{LLR $ \rightarrow  {N_G}{N_b}M$}\\
		  \hline
		 \multirow{2}{*}{\textcolor{black}{{OFDM-GIM$^1$} \cite{7112187}}} & \multirow{2}{*}{\textcolor{black} {$1 \leq N_{a_\psi} <N_b$}} & \multirow{1}{*}{\textcolor{black}{$\sum_{\psi = 1}^{\Psi} {N_{G_\psi}} \big({\floor{\log_2\binom{N_b}{ N_{a_\psi}}}}+$}}& \textcolor{black}{ML $ \rightarrow  \sum_{\psi = 1}^{\Psi}{N_{G_\psi}}2^{\floor{\log_2\binom{N_b}{N_{a_\psi}}}}M^{N_{a_\psi}}$}  \\
		 & & \textcolor{black}{${ N_{a_\psi}{\log_2(M)}}\big)$}& \multirow{1}{*}{\textcolor{black}{LLR $ \rightarrow \sum_{\psi = 1}^{\Psi}{N_G}{N_b}M$}}  \\
		    \hline 
		\multirow{2}{*}{\textcolor{black}{{OFDM-I/Q-IM} \cite{7112187, 7230238}} }  &  \multirow{2}{*}{\textcolor{black}{$1 \leq N_a < N_b$}} & \multirow{2}{*}{\textcolor{black}{$2{N_G}\big({\floor{\log_2\binom{N_b}{N_a}}}+{N_a{\log_2(M)}}\big)$}} & \multirow{1}{*}{\textcolor{black}{ML $ \rightarrow  4{N_G}2^{\floor{\log_2\binom{N_b}{N_a}}}M^{N_a}$}} \\
		& & & \multirow{1}{*}{\textcolor{black}{LLR $ \rightarrow 4{N_G}{N_b}M $}} \\
		    \hline 
		\multirow{2}{*}{\textcolor{black}{{DM-OFDM} \cite{7547943}}} & \multirow{1}{*}{\textcolor{black}{$ N_{a_1} = N_{a_1}, $}} & {\textcolor{black}{${N_G}\big({\floor{\log_2\binom{N_b}{N_{a_1}}}} +  {N_{a_1}{\log_2(M_1)}} + $}} &  \multirow{1}{*}{\textcolor{black}{ML $ \rightarrow {N_G}2^{\floor{\log_2\binom{N_b}{N_{a_1}}}}{M_1}^{N_{a_1}}{M_2}^{(N_b-N_{a_1})} $}} \\
		& \textcolor{black}{$N_{a_2} = N_b-N_{a_1} $} & \textcolor{black}{${(N_b-N_{a_1}){\log_2(M_2)}}\big)$}& \multirow{1}{*}{\textcolor{black}{LLR $ \rightarrow {N_G}{N_b}{( M_1 + M_2)} $}}  \\
		    \hline 
		\multirow{2}{*}{\textcolor{black}{{MM-OFDM} \cite{7936676}}}& \multirow{2}{*}{\textcolor{black}{$N_a = N_b$}}& \multirow{2}{*}{\textcolor{black}{${N_G}\big({\floor{\log_2(N_b!)}}+{N_b{\log_2(M)}}\big)$}}& \textcolor{black}{ ML $ \rightarrow  {N_G}2^{\floor{\log_2(N_b!)}}$}\\
		& & & \textcolor{black}{LLR $ \rightarrow  \frac{{N_G}{N_b}M}{2}(N_b+1)$} \\
		    \hline 
	     	\multirow{2}{*}{\textcolor{black}{{GFDM-IM}  \cite{7848916}} } & 	\multirow{2}{*}{\textcolor{black}{$1 \leq N_a <N_b$}}& 	\multirow{2}{*}{\textcolor{black}{$T{N_G}\big(\floor{\log_2\binom{N_b}{N_a}}+{N_a{\log_2(M)}}\big)$}} & 	\multirow{2}{*}{\textcolor{black}{ML $ \rightarrow  T{N_G}2^{\floor{\log_2\binom{N_b}{N_a}}}M^{N_a}$}} \\
	     & & & \\
			\hline
	    	\multirow{2}{*}{\textcolor{black}{{ZTM-OFDM-IM}  \cite{7848916}}} & \multirow{1}{*}{\textcolor{black}{$ N_{a_1} = N_{a_1}, N_{a_2} = N_{a_2}$,}} & {\textcolor{black}{${N_G}\bigg( {N_{a_1}{\log_2(M_1)}} +  {N_{a_2}{\log_2(M_2)}} $}} &  \multirow{2}{*}{\textcolor{black}{ML $ \rightarrow {N_G}2^{\floor{\log_2\big(\binom{N_b}{N_{a}} \binom{N_a}{N_{a_1}}\big)}}{M_1}^{N_{a_1}}{M_2}^{N_{a_2}} $}} \\
	    	& \textcolor{black}{$N_{a_1} + N_{a_2} = N_a \leq N_b$}& \textcolor{black}{$\floor{\log_2\big(\binom{N_b}{N_{a}} \binom{N_a}{N_{a_1}}\big)}\bigg)$} & \\
	    \hline
	     	\multirow{3}{*}{\textcolor{black}{{L-OFDM-IM}\cite{8734769}}}  & \multirow{2}{*}{\textcolor{black}{$1 \leq N_{a_L} <N_{b_L} <N_{b}$,}}&   \multirow{3}{*}{\textcolor{black}{${N_G}L\big({\floor{\log_2\binom{N_{b_L}}{N_{a_L}}}}+{N_{a_L}{\log_2(M)}}\big)$}}  &  \multirow{1}{*}{\textcolor{black}{ML $ \rightarrow  {N_G}\big({2^{\floor{\log_2\binom{N_{b_L}}{N_{a_L}}}}M^{N_{a_L}}}\big)^L$} }\\
	     	& & & \\
	    & \multirow{1}{*}{\textcolor{black}{$ N_b = N_{b_L}+ N_{a_L}(L-1)$}}& & \multirow{1}{*}{\textcolor{black}{LC $ \rightarrow  {N_G}\sum_{\psi=1}^{L}M(N_{b_L}+ N_{a_L}(\psi-1))$} }   \\
	    
	    \hline
		\end{tabular}
		\begin{tablenotes}\footnotesize
			\item [] \textcolor{black}{$^1$$\Psi$ is the size of the allowed number of different subcarriers within an OFDM-IM block.}
		\end{tablenotes}
	\end{threeparttable}
\label{Tab:CompRateComplexityF} 
\end{table*}

\begin{table*}
	\caption{\textcolor{black}{Data rate and computational complexity assessment of time, code, and channel domain IM techniques}}
	\centering
	\renewcommand{\arraystretch}{1.4}
	\scriptsize
	\begin{threeparttable}
		\begin{tabular}{|C{2.65cm}|| C{3.3cm}|C{4.65cm}|p{5.35cm}|} 
			\hline
			{\textcolor{black}{\textbf{{IM Techniques}}}}  & \textcolor{black}{\textbf{\# Active Symbols ($ K_a $})} & \textcolor{black}{\textbf{{Data Rate}} \text{[bpcu]}} & \textcolor{black}{\makecell{\textbf{{ Computational Complexity at Rx}}}} \\
			\hline \hline   
			\multirow{2}{*}{\textcolor{black}{{SC-IM}  \cite{7738501}}} &  \multirow{2}{*}{\textcolor{black} {$1 \leq K_a <K_b$}} &  \multirow{2}{*}{\textcolor{black}{${K_G}\big({\floor{\log_2\binom{K_b}{K_a}}}+{K_a{\log_2(M)}}\big)$}}  & \textcolor{black}{ML $ \rightarrow  {K_G}2^{\floor{\log_2\binom{K_b}{K_a}}}M^{K_a}$} \\
			& & & \textcolor{black}{LLR $ \rightarrow  {K_G}{K_b}M$}\\
			\hline 
				\multirow{2}{*}{\textcolor{black}{{FTN-IM} \cite{7973048}}} &  \multirow{2}{*}{\textcolor{black} {$1 \leq K_a <K_b$}} &  \multirow{2}{*}{\textcolor{black}{${K_G}\big({\floor{\log_2\binom{K_b}{K_a}}}+{K_a{\log_2(M)}}\big)$}}  & \textcolor{black}{ML $ \rightarrow  {K_G}2^{\floor{\log_2\binom{K_b}{K_a}}}M^{K_a}$} \\
			& & & \textcolor{black}{LLR $ \rightarrow  {K_G}{K_b}M$}\\
			\hline 
			\multirow{2}{*}{\textcolor{black}{{DM-SCIM} \cite{8287922}}} & \multirow{1}{*}{\textcolor{black}{$K_{a_1} = K_{a_1}, $}} & {\textcolor{black}{${K_G}\big({\floor{\log_2\binom{K_b}{K_{a_1}}}} +  {K_{a_1}{\log_2(M_1)}} + $}} &  \multirow{1}{*}{\textcolor{black}{ML $ \rightarrow {K_G}2^{\floor{\log_2\binom{K_b}{K_{a_1}}}}{M_1}^{K_{a_1}}{M_2}^{(K_b-K_{a_1})} $}} \\
			& \textcolor{black}{$K_{a_2} = K_b-K_{a_1} $} & \textcolor{black}{${(K_b-K_{a_1}){\log_2(M_2)}}\big)$}& \multirow{1}{*}{\textcolor{black}{LLR $ \rightarrow {K_G}{K_b}{( M_1 + M_2)} $}}  \\
			\hline \hline
			{\textcolor{black}{\textbf{{IM Techniques}}}}  & \textcolor{black}{\textbf{\# Active Codes ($N_{ac}$})} & \textcolor{black}{\textbf{{Data Rate}} \text{[bpcu]}} & \textcolor{black}{\makecell{\textbf{{ Computational Complexity at Rx}}}} \\
			\hline  
			\multirow{2}{*}{\textcolor{black}{{{CIM-SS}} \cite{6994807}}} & \multirow{2}{*}{\textcolor{black} {$ N_{ac} =1, N_{ct} = 2 $}} & \multirow{1}{*}{\textcolor{black}{$2\floor{\log_2(N_{ct})} + {\log_2(M)} = $}}& \textcolor{black}{ML $ \rightarrow  2N_{ct}M$}  \\
			& & \textcolor{black}{$2 + \log_2(M)$} & \textcolor{black}{LC $ \rightarrow  2N_{ct} + M$} \\
			\hline 
			\multirow{2}{*}{\textcolor{black}{{{GCIM-SS}} \cite{7317808}} }  &  \multirow{2}{*}{\textcolor{black}{$ N_{ac} =1, 2\leq N_{ct} $}} & \multirow{2}{*}{\textcolor{black}{$2\floor{\log_2(N_{ct})} + {\log_2(M)}$}} & \multirow{1}{*}{\textcolor{black}{ML $ \rightarrow  2N_{ct}M$}} \\
			& &  & \multirow{1}{*}{\textcolor{black}{LC $ \rightarrow  2N_{ct} + M$}}\\
			\hline 
			\multirow{2}{*}{\textcolor{black}{{IM-OFDM-SS}\cite{8269169}} }  &  \multirow{2}{*}{\textcolor{black}{$ N_{ac} =1, 2\leq N_{ct} $}} & \multirow{2}{*}{\textcolor{black}{$2\floor{\log_2(N_{ct})} + {\log_2(M)}$}} & \multirow{1}{*}{\textcolor{black}{ML $ \rightarrow  2N_{ct}M$}} \\
			& & & \multirow{1}{*}{\textcolor{black}{LC $ \rightarrow  2N_{ct} + M$}} \\
			\hline 
			\hline
			{\textcolor{black}{\textbf{{IM Techniques}}}}  & \textcolor{black}{\textbf{\# Active RF Mirrors ($N_{am}$})} & \textcolor{black}{\textbf{{Data Rate}} \text{[bpcu]}} & \textcolor{black}{\makecell{\textbf{{ Computational Complexity at Rx}}}} \\
			\hline
			\multirow{1}{*}{\textcolor{black}{{MBM} \cite{6620786}}}& \multirow{1}{*}{\textcolor{black}{$N_{am} = 1$}}& \multirow{1}{*}{\textcolor{black}{${N_{rf}}+{{\log_2(M)}}$}}& \textcolor{black}{ ML $ \rightarrow N_r2^{N_{rf}}M,  N_t = 1 $}\\
			\hline 
			\multirow{2}{*}{\textcolor{black}{STCM \cite{7864471}}} & \multirow{2}{*}{\textcolor{black}{$N_{am} = 1$}} & \multirow{2}{*}{\textcolor{black}{$2{N_{rf}}+{2{\log_2(M)}}$}} &  \multirow{1}{*}{\textcolor{black}{ML $ \rightarrow 4N_r2^{2N_{rf}}M^2, N_t = 2 $}} \\
			& & & \multirow{1}{*}{\textcolor{black}{LC $ \rightarrow 4N_r2^{2N_{rf}}M $}} \\
			\hline
		\end{tabular}
	\end{threeparttable}
	\label{Tab:CompRateComplexityTC}
\end{table*}

\subsection{Enabling IM Techniques for Next-Generation Services}
The presented IM techniques in Section \ref{Sec:Types} are categorized considering the requirements of three main services.  
\subsubsection{eMBB} \label{Section:eMBBDis}
The crucial requirement for eMBB is the efficient spectrum utilization, as explained in Section \ref{sec:eMMB}. Therefore, the IM schemes are assessed according to their SE performance. \textcolor{black}{Table \ref{Tab:CompRateComplexityS}, \ref{Tab:CompRateComplexityF}, and \ref{Tab:CompRateComplexityTC} present the data-rate and the computational complexity assessment of space, frequency, time, code and channel domains. The computational complexity at Rx is provided for both available low-complex (LC) and ML detectors, and is calculated in terms of complex multiplications. It is readily seen that one-dimensional main IM types in space, frequency, time, code, and polarization domains lead to a decrease in SE due to both the partial transmission and the logarithmic increase on SE with the number of active information-bearing entities. Additionally, in comparison to conventional schemes, the reduction in SE becomes suddenly high in the case of high-order modulation usage. To exemplify, OFDM-IM with ($N_b$ = 8, $N_a$ = 4) corresponds to $2^{\floor{\log_2\binom{8}{4}}}$ = 64 legitimate subcarrier combinations that enable the transmission of maximum number of bits through the subcarriers' indices, i.e., IM bits, for $N_b$ = 8. However, it results in \%12 and \%32 SE loss for $M=4$ and $M=16$, respectively with respect to OFDM. Hence, these types of IM require an additional mechanism that allows the transmission with a higher number of $M$-ary symbols in support of eMBB application and use-cases.}  

\textcolor{black}{ GSM increases the number of IM bits, from $\log_2(N_t)$ to $\floor{\log_2\binom{N_t}{N_a}}$, but the number of $M$-ary symbols remains the same and equals one \cite{Ref5}. Therefore, it offers a moderate improvement in SE with the assistance of multiple antenna activation. As given in Table \ref{Tab:CompRateComplexityS}, MA-SM achieves higher SE than GSM via transmitting different data symbols through these activated antennas at the expense of lower BER, which is not a primary concern for eMBB communication. It can be seen in Table \ref{Tab:CompRateComplexityS} that QSM and its advanced versions provide increment only for IM bits. ESM enables the transmission of the data bits by both the active antennas' indices and the constellation type. On the other hand, achieving higher data rates with SM-based IM types is challenging in microwave frequency bands, since incorporating a higher number of Tx antennas becomes infeasible for both BS and UEs because of the required distance ($\frac{\lambda}{2}$) between the consecutive antennas. In the light of these considerations, the fundamental types of space domain IM are far from satisfying the requirements of eMBB use-cases.}


\textcolor{black}{DM-SCIM provides a higher data rate than the classical SC-IM by modulating the inactive data symbols with different modulation types. Although DM-SCIM improves the SE of SC-IM types as given in Table \ref{Tab:CompRateComplexityTC}, its counterpart in the frequency domain, i.e., DM-OFDM, is superior to DM-SCIM owing to flexible resource allocation. ZTM-OFDM-IM combines OFDM-IM and DM-OFDM in order to boost the SE of OFDM-IM systems. As seen in Table \ref{Tab:CompRateComplexityF}, it provides a significant improvement in the number of information bits carried by the indices, but it poses a limit for competing with conventional schemes under high-order modulation conditions. MM-OFDM enables not only the modulation of all subcarriers by using multiple QAM/PSK sets but also the utilization of permutations of subcarriers' combinations. In other words, DM-OFDM and MM-OFDM activate all the subcarriers as in conventional OFDM and transmit IM bits as well. The number of legitimate subcarrier combinations is increased  from  $\floor{\log_2{(N_b)}}$ to $\floor{\log_2{(N_b!)}}$ by MM-OFDM-IM. OFDM-GIM provides a degree of freedom to control the number of IM bits and $M$-ary symbols adaptively. Thus, it supports both OFDM-IM and OFDM transmissions. In other words, L-OFDM-IM facilitates the improvement in both IM bits and $M$-ary symbols at the cost of the exponential increase in the processing complexity with $L$, as given in Table \ref{Tab:CompRateComplexityF}.}

The SE of MBM linearly increases with the number of RF mirrors. \textcolor{black}{Due to the linear increase in SE with $N_{rf}$, SM-MBM and TI-MBM provide higher data rates compared to TI-SM. TI-SM and TI-MBM need to utilize higher modulation orders than SM-MBM for the sake of achieving the same SE. However, the use of higher modulation orders leads to degradation of the BER performance of TI-SM and TI-MBM.} However, a high number of RF mirrors leads to high training overhead to estimate $2^{N_{rf}}$ channel realizations. Therefore, DMBM can be a candidate solution to satisfy the demand of high SE, since it removes the channel estimation at Rx, which is the UE in downlink transmission. \textcolor{black}{On the other hand, receiver SM provides opportunities in downlink transmission for both reducing cost and increasing the EE at the UE side. GPSM can support the same throughput as conventional MIMO systems with the same processing complexity at Rx side \cite{6644231, 6877673, 8765330}.}  

\textcolor{black}{Multidimensional IM types are more appealing for the purpose of fulfilling the requirements of eMBB service. GSFIM performs transmission through antenna indices, subcarrier indices, and $M$-ary symbols. Also, conventional MIMO-OFDM corresponds to the special case of GSFIM. It is shown that \%20 rate gain can be achieved by GSFIM under the conditions of ($N_t$ = 32, $N_r$ = 32 and $M$ = 4). In this regard, its advanced version GSTFIM can also be considered for eMBB use-cases.} Additionally, current OFDM technology is one of the promising solutions for eMBB applications and use-cases. However, OFDM suffers from ICI in case of high mobility scenarios. To provide eMBB communications for the high mobility classes defined in Table \ref{tab:LatTable2}, instead of the conventional OFDM, ISM-OFDM can alleviate the effect of ICI without compromising the SE.

\textcolor{black}{
	\begin{remark}
Amongst one-dimensional IM types, frequency domain IM types can compete with the conventional OFDM in terms of SE due to its flexible structure.  The advanced versions of OFDM-IM, such as DM-OFDM, MM-OFDM, L-OFDM, are conducive to support eMBB service, even if high-order modulation types are considered. Space domain IM types are easily defeated by conventional SMX schemes due to not only the logarithmic increase with $N_T$ but also the transmission of a limited number of $M$-ary symbols. In order to overcome this limitation, MBM is deemed to be promising, but it leads to the monumental complexity at Rx side, which can not be handled by UE in downlink transmission.
	\end{remark}
}

\subsubsection{mMTC}
Researchers in both academia and industry have been seeking technologies to provide large coverage area, low power consumption and low cost for mMTC services where latency, data rate, and reliability are not primary concerns, as explained in Section \ref{sec:mMTC}. In essence, IM provides high EE owing to the energy-free carried information bits by the indices of the transmit entities. For example, OFDM-IM with ($N_b$ = 4 and  $N_a$ = 2) transmits $p_1$ = 2 bits by the subcarriers' indices and $p_2$ = 2 bits by the modulated subcarriers with binary phase-shift keying (BPSK). However, the classical OFDM requires 4 active subcarriers to transmit $p_1+p_2$ = 4 bits. Hence, OFDM-IM with ($N_b$ = 4 and  $N_a$ = 2) harvests 50\% of the Tx power to transmit the same number of data bits. Utilization of the same Tx power for the OFDM-IM and conventional OFDM significantly extends the coverage area for OFDM-IM. \textcolor{black}{Besides the high EE, hardware and computational complexity originated by IM should be considered for mMTC applications and use-cases. Please note that the SE and complexity of a given IM scheme are dependent on each other. Thus, Table {\ref{Tab:CompRateComplexityS}, \ref{Tab:CompRateComplexityF}, and \ref{Tab:CompRateComplexityTC}} provide the computational complexity of IM types considering the given SE. For instance, L-OFDM-IM and OFDM-IM offer the same complexity and SE, while $L$ = 1.} 

\textcolor{black}{One-dimensional space domain IM types including SSK and SM significantly reduce the hardware complexity due to the use of a single RF chain at Tx, as given in Table \ref{Tab:CompRateComplexityS}. In recent studies, it has been also demonstrated that SSK can be implemented even with a simple RF signal generator \cite{BasarMBM}. In this way, further reduction is achieved at both Tx and Rx sides. Therefore, SSK and SM provide a high EE, low hardware complexity at Tx, and low computational complexity at Rx for MIMO systems.} {Due to the increased antenna combinations, GSM and MA-SM require multiple RF chain activation and IAS at Tx and leads to the more complex Rx than that of SM.} 

The EE and computational complexity of frequency domain IM variants is dependent on the block size, the subblock size and the number of active subcarriers. Mainly, the ML detector is used for the detection of information bits carried by the indices of active subcarriers and $M$-ary symbols. SIM-OFDM suffers from a complex Rx due to the block-based ML detection {\cite{5449882}. \textcolor{black}{For example, in order to activate a quarter of the subcarriers in SIM-OFDM systems, ML detector should search over $2^{\floor{\log_2\binom{128}{32}}}$ = $2^{100}$ subcarrier combinations to find the correct active subcarriers for ($N_{sc}$= 128, $N_a$ = 32)}. Thus, OFDM-IM divides the OFDM block into the subblocks to reduce the number of possible combinations for the ML detector. However, a larger subblock size still causes a higher complexity at Rx. Hence, OFDM-IM with the LLR detector is proposed in the literature. \textcolor{black}{For OFDM-IM with ($N_{sc}$ = 128, $N_b$ = 8, $N_a$ = 2, $M$ = 2), LLR detector reduces the computational complexity four times than that of ML detector, as given in Table \ref{Tab:CompRateComplexityF}. Amongst one-dimensional IM types, spectral-efficient schemes such as DM-OFDM, MM-OFDM, L-OFDM-IM, and ZTM-OFDM-IM lead to the increased processing complexity at Rx. \textcolor{black}{Due to the adaptive number of subcarrier activation, OFDM-GIM loses the inherent advantages of IM including EE and reliability.} In fact, the achieved EE is limited due to the activation of the majority of existing subcarriers, i.e., $N_b/ N_a$ $\simeq$ 1. In OFDM-IM, the obtained $N_b/ N_a$ power can be utilized for achieving a higher BER performance by increasing the power per subcarrier or can be harvested for achieving a higher EE by keeping the power per subcarrier as in OFDM.} In \cite{ShpirtveBen}, the power level is utilized to provide robustness against IUI in asynchronous mMTC networks, where the sporadic nature of mMTC originates time offset between the UEs and destroys the orthogonality among them. To satisfy the requirements of NB-IoT given in Table \ref{tab:LatTable2}, the use of direct conversion Rx is proposed for the NB-IoT devices due to its simple structure \cite{7130774}. However, the direct conversion causes in-phase and quadrature imbalance (IQI), and severely degrades the performance of the conventional OFDM. In \cite{Shpirt}, it is shown that the presence of inactive subcarriers in OFDM-IM allows easy estimation and compensation of the IQI.}

\begin{table*}
	
	\caption{The comparison of space and channel domain IM techniques}
	\centering
	\renewcommand{\arraystretch}{1.4}
	\scriptsize
	\begin{threeparttable}
		\begin{tabular}{|C{2.5cm}||m{4.5cm}|m{4.5cm}|c|c|c|c|c|c|c|}
			\hline
			\multirow{2}{*}{\textbf{{IM Techniques}}}  & \multirow{2}{*}{\textbf{{Pros}}} & \multirow{2}{*}{\textbf{{Cons}}} & \multicolumn{7}{c|}{Performance Metrics} \\ \cline{4-10} 
			& &  & 1 & 2 & 3 & 4 & 5 &6 & 7  \\
			\hline \hline
			SM \cite{4382913}& Single antenna activation, use of one RF chain at Tx, IAI free transmission & Logarithmic increase on SE with $N_t$, limited data rate &${\downarrow}$&${\uparrow}$&${\downarrow}$&${\downarrow}$&${\downarrow}$&${\downarrow}$& ${\uparrow}$ \\
			\hline
			GSM \cite{5757786}& Multiple antenna activation to transmit the same data symbol, IAI free transmission & 
			Logarithmic increase on SE with $N_t$, limited data rate, requirement of multiple RF chains and IAS at Tx &${\downarrow}$&${\uparrow}$&${\uparrow}$&${\downarrow}$&${\downarrow}$&${\downarrow}$&${\uparrow}$ \\
			\hline
			MA-SM \cite{6166339} & Multiple antenna activation to transmit the different data symbols, higher SE than SM/GSM &Logarithmic increase on SE with $N_t$, requirement of multiple RF chains and IAS at Tx, complex Rx process&${\uparrow}$ &${\uparrow}$&${\uparrow}$&${\uparrow}$&${\downarrow}$&${\uparrow}$&${\downarrow}$\\
			\hline
			QSM \cite{6868290} & Activation of two Tx antennas to transmit in-phase and quadrature parts of a complex symbol, exploiting spatial selectivity & Logarithmic increase on SE with $N_t$ &${\uparrow}$& ${\uparrow}$ & ${\downarrow}$ & ${\downarrow}$ &${\downarrow}$ & ${\downarrow}$ & ${\uparrow}$\\
			\hline
			ESM \cite{7084604} & Activation of a single and two Tx antennas for high order and low order modulations, increased Euclidean distance &  Requirement of multiple RF chains and IAS at Tx, complex Rx process &${\uparrow}$ & ${\uparrow}$ & ${\uparrow}$ & ${\uparrow}$ & ${\downarrow}$ & ${\uparrow}$ & ${\downarrow}$\\
			\hline
			STBC-SM \cite{7084604} & Use of two RF chains at Tx, achieving diversity and coding gains &  Limited data rate & ${\downarrow}$ & ${\uparrow}$ & ${\downarrow}$ & ${\downarrow}$ & ${\uparrow}$ & ${\uparrow}$ & ${\uparrow}$\\
			\hline
			MBM \cite{6620786} & Linear increase on SE with $N_{rf}$, robustness against channel fading & Estimation of $2^{N_{rf}}$ channel realizations, high training overhead, complex Rx process & ${\uparrow}$ & ${\uparrow}$ & ${\downarrow}$ & ${\uparrow}$ & ${\downarrow}$ &${\downarrow}$& ${\uparrow}$ \\ 
			\hline
			
			DMBM \cite{6620786} &  Avoiding channel estimation at Rx & Increased processing time due to feedback mechanism, an error performance loss & ${\uparrow}$ & ${\uparrow}$ & ${\downarrow}$ & ${\uparrow}$ & ${\downarrow}$ &${\downarrow}$& ${\uparrow}$ \\ 
			\hline
		\end{tabular}
		\begin{tablenotes}\footnotesize
			\item[] Performance Metrics - 1: SE, 2: EE, \textcolor{black}{3: Hardware Complexity at Tx, 4: Computational Complexity at Rx}, 5: Tx Diversity Gain, 6: Multiplexing Gain, and 7: Interference Immunity 
			\item[] Sign - ${\uparrow}$: High ${\downarrow}$: Low 
		\end{tablenotes}
	\end{threeparttable}
	\label{Tab:SpaceCompIMTypes} 
\end{table*}

\begin{table*}
	
	\caption{The comparison of frequency, time and code domain IM techniques}
	\centering
	\renewcommand{\arraystretch}{1.4}
	\scriptsize
	\begin{threeparttable}
		\begin{tabular}{|C{2.5cm}||m{5.1cm}|m{5.1cm}|c|c|c|c|c|c|c|}
			\hline
			\multirow{2}{*}{\textbf{{IM Techniques}}}  & \multirow{2}{*}{\textbf{{Pros}}} & \multirow{2}{*}{\textbf{{Cons}}} & \multicolumn{5}{c|}{Performance Metrics} \\ \cline{4-10} 
			& &  & 1 & 2 & 3 & 4 & 5 \\
			\hline \hline
			OFDM-IM \cite{Basar1} & Robustness against ICI, hardware impairments including CFO and IQI & Lower SE than conventional OFDM for high-order modulations & ${\downarrow}$ & ${\uparrow}$ & ${\downarrow}$  & ${\downarrow}$ & ${\uparrow}$ \\
			\hline
			OFDM-ISIM \cite{6841601} & Exploiting frequency diversity with the aid of interleaving, reliable transmission & Lower SE than conventional OFDM for high-order modulations & ${\downarrow}$ & ${\uparrow}$ & ${\downarrow}$  & ${\downarrow}$ & ${\uparrow}$ \\
			\hline
			CI-OFDM-IM  \cite{7086323} & Transmission of in-phase and quadrature parts of a symbol by different subcarriers, exploiting the real frequency diversity, reliable transmission & Lower SE than conventional OFDM for high-order modulations & ${\downarrow}$ & ${\uparrow}$ & ${\downarrow}$ & ${\downarrow}$ & ${\uparrow}$ \\
			\hline
			OFDM-GIM \cite{7112187} & Allowing different number of active subcarriers for different subblocks, higher SE than OFDM-IM & Two-stage Rx process, error propagation, sensitivity to hardware impairments  & ${\uparrow}$ & ${\downarrow}$ & ${\uparrow}$  &  ${\uparrow}$ & ${\downarrow}$ \\
			\hline
			DM-OFDM \cite{7547943} & Use of two different modulation schemes within a subblock, higher SE than OFDM-IM & Sensitivity to hardware impairments  & ${\uparrow}$ & ${\downarrow}$ & ${\downarrow}$ & ${\downarrow}$ & ${\downarrow}$ \\
			\hline
			GFDM-IM \cite{7848916} & Relaxed time and frequency synchronization & Complex Rx process & ${\uparrow}$ & ${\uparrow}$ & ${\downarrow}$ & ${\uparrow}$ & ${\uparrow}$ \\
			\hline
			SC-IM \cite{7738501} & Robustness against ISI & Higher PAPR than conventional SC due to inherent sparsity & ${\downarrow}$ & ${\uparrow}$ & ${\downarrow}$ & ${\downarrow}$ & ${\uparrow}$ \\
			\hline
			CIM-SS \cite{6994807} & Higher SE compared to DS-SS with lower energy consumption & Logarithmic increase on SE with $N_c$, complex Rx process &${\downarrow}$  & ${\uparrow}$ & ${\downarrow}$ & ${\uparrow}$ & ${\uparrow}$ \\
			\hline
		\end{tabular}
		\begin{tablenotes}\footnotesize
			\item[] Performance Metrics - 1: SE, 2: EE, \textcolor{black}{3: Computational Complexity at Tx, 4:  Computational Complexity at Rx}, and 5: Interference Immunity 
			\item[] Sign - ${\uparrow}$: High ${\downarrow}$: Low
		\end{tablenotes}
	\end{threeparttable}
	\label{Tab:FTCCompIMTypes} 
\end{table*}

\begin{table*}

	\caption{The comparison of multidimensional IM techniques}
	\centering
	\renewcommand{\arraystretch}{1.4}
	\scriptsize
	\begin{threeparttable}
		\begin{tabular}{|C{2.5cm}|m{4.5cm}|m{4.5cm}|c|c|c|c|c|c|c|}
			\hline
			\multirow{2}{*}{\textbf{{IM Techniques}}}  & \multirow{2}{*}{\textbf{{Pros}}} & \multirow{2}{*}{\textbf{{Cons}}} & \multicolumn{7}{c|}{Performance Metrics} \\ \cline{4-10} 
			& &  & 1 & 2 & 3 & 4 & 5 &6 & 7  \\
			\hline \hline
			
			ISM-OFDM \cite{7440707} & Robustness against IAI and ICI & Logarithmic increase on SE with $N_t$ and $N_{sc}$& ${\downarrow}$ & ${\uparrow}$ & ${\uparrow}$ & ${\uparrow}$ &${\downarrow}$ & ${\downarrow}$ & ${\uparrow}$ \\
			\hline
			GSFIM \cite{7277106}&  Robustness against ICI, multiple antenna activation to transmit the different data symbols &  Logarithmic increase on SE with $N_t$ and $N_{sc}$, requirement of multiple RF chains and IAS at Tx & ${\uparrow}$ & ${\uparrow}$ & ${\uparrow}$ & ${\uparrow}$ &${\downarrow}$ & ${\uparrow}$ & ${\downarrow}$ \\ 
			\hline
			GFDM-SFIM \cite{8277679} & Relaxed time and frequency synchronization & Complex transceiver process  & ${\uparrow}$ & ${\uparrow}$ & ${\uparrow}$& ${\uparrow}$ & ${\downarrow}$ & ${\uparrow}$ &${\downarrow}$ \\
			\hline
			TI-SM \cite{7925922} & Robustness against IAI and ISI & Limited data rate & ${\downarrow}$ & ${\uparrow}$  & ${\downarrow}$ & ${\downarrow}$  &${\downarrow}$  & ${\downarrow}$ & ${\uparrow}$  \\
			\hline
			SM-MBM \cite{7925922} & Robustness against IAI, linear increase on SE with $N_{rf}$ & Estimation of $2^{N_{rf}}$ channel realizations, high training overhead, complex Rx process & ${\uparrow}$ & ${\uparrow}$ & ${\downarrow}$ & ${\uparrow}$  & ${\downarrow}$ & ${\downarrow}$ & ${\uparrow}$\\
			\hline
			TI-MBM \cite{7925922} & Robustness against ISI, linear increase on SE with $N_{rf}$ & Estimation of $2^{N_{rf}}$ channel realizations, high training overhead, complex Rx process & ${\uparrow}$ & ${\uparrow}$ & ${\downarrow}$  & ${\uparrow}$ &${\downarrow}$ & ${\downarrow}$  & ${\uparrow}$\\
			\hline
			TI-SM-MBM \cite{7925922} & Robustness against ISI and IAI, linear increase on SE with $N_{rf}$ & Estimation of $2^{N_{rf}}$ channel realizations, high training overhead, complex Rx process & ${\uparrow}$ & ${\uparrow}$ & ${\uparrow}$& ${\uparrow}$ &${\downarrow}$  & ${\downarrow}$  & ${\uparrow}$\\
			\hline
			CIM-SM \cite{8404646TSP} & Higher SE than SM &  Complex Rx process & ${\uparrow}$ & ${\uparrow}$& ${\uparrow}$& ${\uparrow}$ & ${\downarrow}$& ${\downarrow}$ & ${\uparrow}$ \\
			\hline
			DSM \cite{6879496} & Robustness against IAI, avoiding channel estimation at Rx & Increased processing time due to feedback mechanism &${\uparrow}$ & ${\uparrow}$ & ${\downarrow}$ & ${\downarrow}$ & ${\uparrow}$ & ${\uparrow}$ & ${\uparrow}$ \\
			\hline
			STSK \cite{5599264}  & Providing diversity gain as well as multiplexing gain & Requirement of $N_{rf} = N_t$ RF chains at Tx,  complex Rx design, sensitivity to ISI & ${\downarrow}$ &  ${\downarrow}$ & ${\uparrow}$ & ${\uparrow}$ &${\uparrow}$ & ${\uparrow}$ & ${\downarrow}$ \\
			\hline
			OFDM-STSK \cite{7494949} & Robustness against ISI  & Requirement of $N_{rf} = N_t$ RF chains at Tx, complex Rx design & ${\downarrow}$ & ${\downarrow}$ & ${\uparrow}$& ${\uparrow}$ &${\uparrow}$ & ${\uparrow}$ & ${\uparrow}$ \\
			\hline
			OFDM-SFSK \cite{8417812} & Robustness against ISI and ICI with the aid of sparse DMs  & Requirement of $N_{rf} = N_t$ RF chains at Tx, complex Rx process & ${\downarrow}$ &  ${\downarrow}$ & ${\uparrow}$ &${\uparrow}$ &${\uparrow}$ & ${\uparrow}$ & ${\uparrow}$ \\
			\hline
			OFDM-STSK-IM \cite{8322306}& Robustness against ISI and ICI, high SE with the aid of CS,  & Requirement of $N_{rf} = N_t$ RF chains at Tx, complex Rx process & ${\uparrow}$  & ${\downarrow}$ & ${\uparrow}$  & ${\uparrow}$ & ${\uparrow}$ & ${\uparrow}$ & ${\uparrow}$ \\
			\hline
			MS-STSK \cite{7494949} 
			& Increased SE by partial antenna activation, use of $N_{rf} \leq N_t$ RF chains at Tx &  Complex Rx process & ${\uparrow}$ &  ${\uparrow}$ & ${\downarrow}$& ${\uparrow}$ &${\uparrow}$ & ${\uparrow}$ & ${\downarrow}$ \\
			\hline
			MSF-STSK \cite{7792569} & Robustess against ISI, ICI and IAI, increased SE by partial antenna activation, use of $N_{rf} \leq N_t$ RF chains at Tx & Complex Rx process  & ${\uparrow}$ & ${\uparrow}$ & ${\downarrow}$ & ${\uparrow}$  & ${\uparrow}$& ${\uparrow}$ & ${\uparrow}$\\
			\hline
		\end{tabular}
		\begin{tablenotes}\footnotesize
			\item[] Performance Metrics - 1: SE, 2: EE, \textcolor{black}{3: Hardware and Computational Complexity at Tx, 4: Hardware and Computational Complexity at Rx}, 5: Tx Diversity Gain, 6: Multiplexing Gain, and 7: Interference Immunity 
			\item[] Sign - ${\uparrow}$: High ${\downarrow}$: Low 
		\end{tablenotes}
	\end{threeparttable}
	\label{Tab:CompIMTypes} 
\end{table*}

For UL transmission, SC-IM provides higher EE than the conventional SC due to the transmission of additional informationbits through IM \cite{8698792}. However, the inherent sparsity in the time domain leads to higher PAPR than that of classical SC. Multidimensional IM types have the complex transceiver structure for the detection of active entities in multiple domains. Thus, among the multidimensional IM techniques, TI-SM, which has a moderate number of active entities in time and space, can be considered for mMTC.

The complexity of Rx in MBM is dependent on the number of RF mirrors, which determines both the SE and the system reliability via CSI estimation accuracy. Hence, it provides a trade-off among SE and complexity while ensuring high EE. The SE of MBM with a low number of RF mirrors is limited, but it significantly reduces the Rx complexity since the number of estimated channel realizations exponentially decreases with $N_{rf}$. \textcolor{black}{{In recent studies, a CS-based detection mechanism with low complexity is proposed at Rx to exploit the inherent sparsity of RF mirrors \cite{8363012}. Besides, in \cite{9063516}, sparse user activity in mMTC, i.e., the intermittent and sporadic characteristics of mMTC, is exploited to improve the detection performance at Rx.}}  

\textcolor{black}{
\begin{remark}
One-dimensional space domain IM types can be considered as potential candidates for mMTC service due to low hardware complexity at TX and consequently low power consumption. MBM with a reasonable number of RF mirrors is also appropriate to obtain high EE at the UE side since no CSI is required at Tx. Partial activation in OFDM-IM and SC-IM provides robustness against IUI caused by asynchronous mMTC networks along with the high EE. Additionally, the channel and polarization domain IM types are more appealing when the UE is insufficient to accommodate multiple Tx antennas without spatial correlation between them.
\end{remark}
}

\subsubsection{URLLC}
As explained in Section \ref{sec:URLLC1}, URLLC is the most challenging service due to the simultaneous yet conflicting demands of ultra-reliability and low-latency. In order to achieve the BLER values given in Table \ref{tab:LatTable2}, IM schemes that provide diversity gain, interference immunity, and robustness against hardware impairments, such as carrier frequency offset (CFO) and IQI, are the promising solutions for URLLC. Achieving high reliability via IM requires sufficient selectivity between the active entities in a given IM domain. Thus, space domain IM techniques require $\frac{\lambda}{2}$ separation distance between Tx antennas to improve the detection performance at Rx. Proper detection of the active antennas provides a high probability for the correct estimation of both the index bits and the $M$-ary symbols. Otherwise, the system performance severely reduces due to the high correlation between the utilized antennas. GSM enhances the BER performance of the SM by the transmission of the same data over the multiple active antennas. MA-SM requires an advanced Rx design to avoid IAI, which reduces reliability. However, its complex transceiver structure causes a long processing time. QSM exploits the spatial selectivity via the transmission of in-phase and quadrature parts of the complex data symbol separately. Hence, QSM ensures the achievement of a better BER performance than the conventional SMX and SM \cite{6868290}. Also, STBC-SM, ST-QSM, TCSM, and TCQSM improve the performance of SM by additional diversity and coding gains. In MBM, the correct estimation of the transmitted bits relies on the exact CSI at Rx. Hence, the estimation error in CSI can lead to catastrophic BER performance. Even though DMBM removes the necessity of CSI at Rx, it reduces the system reliability and leads to latency due to the feedback mechanism for the encoding of two consecutive data blocks.

For the sake of supporting the desired reliability, frequency domain IM schemes require the exploitation of frequency diversity via proper mapping of the information bits to the subcarriers. In the conventional OFDM-IM, incoming data bits are directly mapped to the subcarriers within a subblock. Thus, the high correlation between the active subcarriers degrades the detection performance at Rx. Mapping of the data bits to the subcarriers located in different subblocks reduces the correlation between the activated subcarriers and enhances the BER performance. Therefore, OFDM-ISIM and CI-OFDM-IM attain higher reliability than the classical OFDM-IM by means of interleaving. Furthermore, CI-OFDM-IM provides an additional diversity gain with the aid of coordinate interleaving. For a given SE, DM-OFDM can provide better BER performance than OFDM-IM by proper selection of the used modulation sets \cite{7547943}. However, DM-OFDM, MM-OFDM and OFDM-GIM are sensitive to the hardware and channel impairments since there is no idle subcarrier \cite{8904248}. Due to the non-orthogonal transmission in GFDM-IM, attaining the ultra-reliability becomes a challenge. Moreover, resolving the ICI and ISI at Rx requires a complex Rx design that leads to extra latency. Recently, an interference immune OFDM-IM-based NOMA scheme is proposed in \cite{8703780} to alleviate the effect of the collisions due to GF access of the URLLC UE. 

\begin{figure*}
	\centering
	{\includegraphics[scale = 0.48]{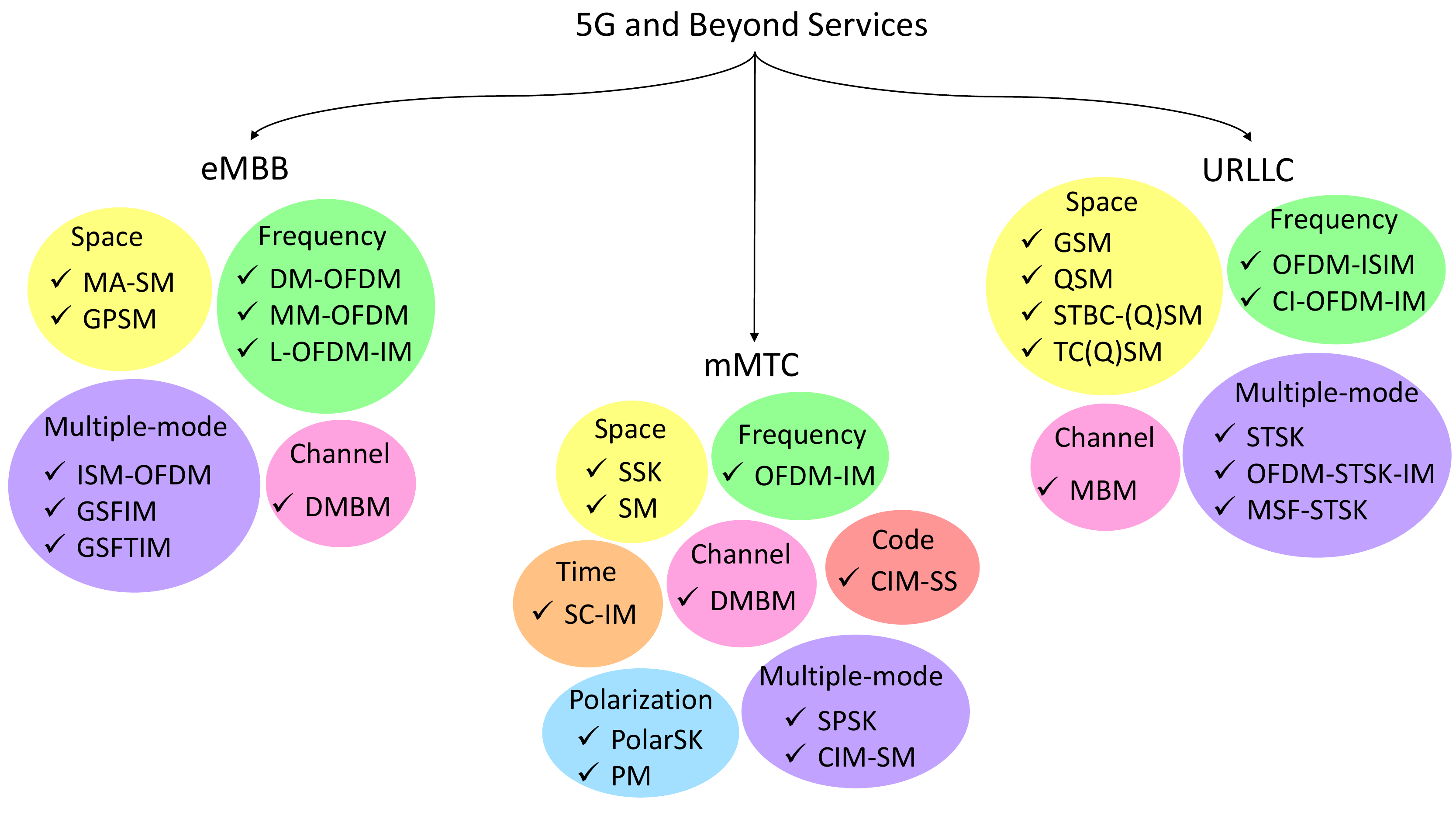}}
	\caption{Promising IM variants for 5G and beyond services.}
	\label{Fig:last}
\end{figure*} 

If wireless channel is time-invariant, the interleaving of data symbols is a necessity for SC-IM to provide ultra-reliability. However, interleaving in the time domain requires storage and thus causes a long latency. Although space domain IM techniques provide multiplexing gain, they suffer from a lack of diversity gain which is of great importance for achieving ultra-reliability. Likewise, conventional SMX and STBC techniques maximize the multiplexing gain and diversity gain, respectively. STSK, which encompasses SMX, STBC, and SM, provides an attractive trade-off among complexity, multiplexing gain, and diversity gain through the spreading a given symbol in space and time dimensions. In this way, it is shown that STSK provides a significant improvement in BER performance \cite{5599264}. To have robustness against frequency-selective channels, STSK is combined with conventional OFDM through OFDM-STSK. It should be noted that OFDM-STSK performs IM on DMs and exploits the frequency domain to improve system performance. MS-STSK increases the reliability of STSK by combining it with GSM. However, it suffers from ISI under dispersive channels. OFDM-IM and MS-STSK are coupled in MSF-STSK to provide robustness against ISI and ICI. So, DM-based IM types can provide the ultra-reliability by diversity gain, but the duration of the STSK block should be adjusted considering the latency constraint. \textcolor{black}{In \cite{7925922}, it is shown that {multidimensional IM schemes} provide better BER performance than one-dimensional schemes for the same SE. This is due to a lower number of active entities in a single domain, which corresponds to a reduced correlation between them and consequently better detection performance at Rx.} Lastly, in \cite{8981888}, it is shown that RIS-aided IM at Rx can provide up to 15 dB gain in the BER performance compared to the conventional MIMO schemes. Therefore, the RIS-aided IM concept is a promising approach for URLLC, but its usage for URLLC is still an open research area in the literature. 

\textcolor{black}{
\begin{remark}
The direct comparison of IM techniques in terms of BER performance is a difficult task since they should be assessed under the same SE. However, different QAM/PSK sets should be utilized for the achievement of the same SE and thus the performance of given IM type is directly affected. Therefore, IM techniques that can provide diversity gain are more prominent for URLLC. Additionally, frequency domain IM techniques provide the exploitation of frequency selectivity with the aid of interleaving, while conventional OFDM systems require a coding scheme at Tx, that leads to significant complexity at Rx, to further improve the BER performance.
\end{remark}
}

\textcolor{black}{In view of the above considerations, space, frequency, time, channel, code and multidimensional IM variants are compared in Table \ref{Tab:SpaceCompIMTypes}, \ref{Tab:FTCCompIMTypes} and  \ref{Tab:CompIMTypes} in terms of SE, EE, Tx complexity, Rx complexity, transmit diversity gain, multiplexing gain, and interference immunity. The number of active entities given in Table \ref{Tab:CompRateComplexityS}, \ref{Tab:CompRateComplexityF} and \ref{Tab:CompRateComplexityTC} determines the interference immunity of a given IM scheme, and also defines the hardware complexity at Tx for space domain IM techniques. Lastly, Fig. \ref{Fig:last} illustrates the categorization of promising IM variants for eMBB, mMTC, and URLLC.}

\subsection{Key Advantages and Disadvantages of Different IM Domains}

\begin{figure*}
	\centering
	{\includegraphics[scale = 0.40]{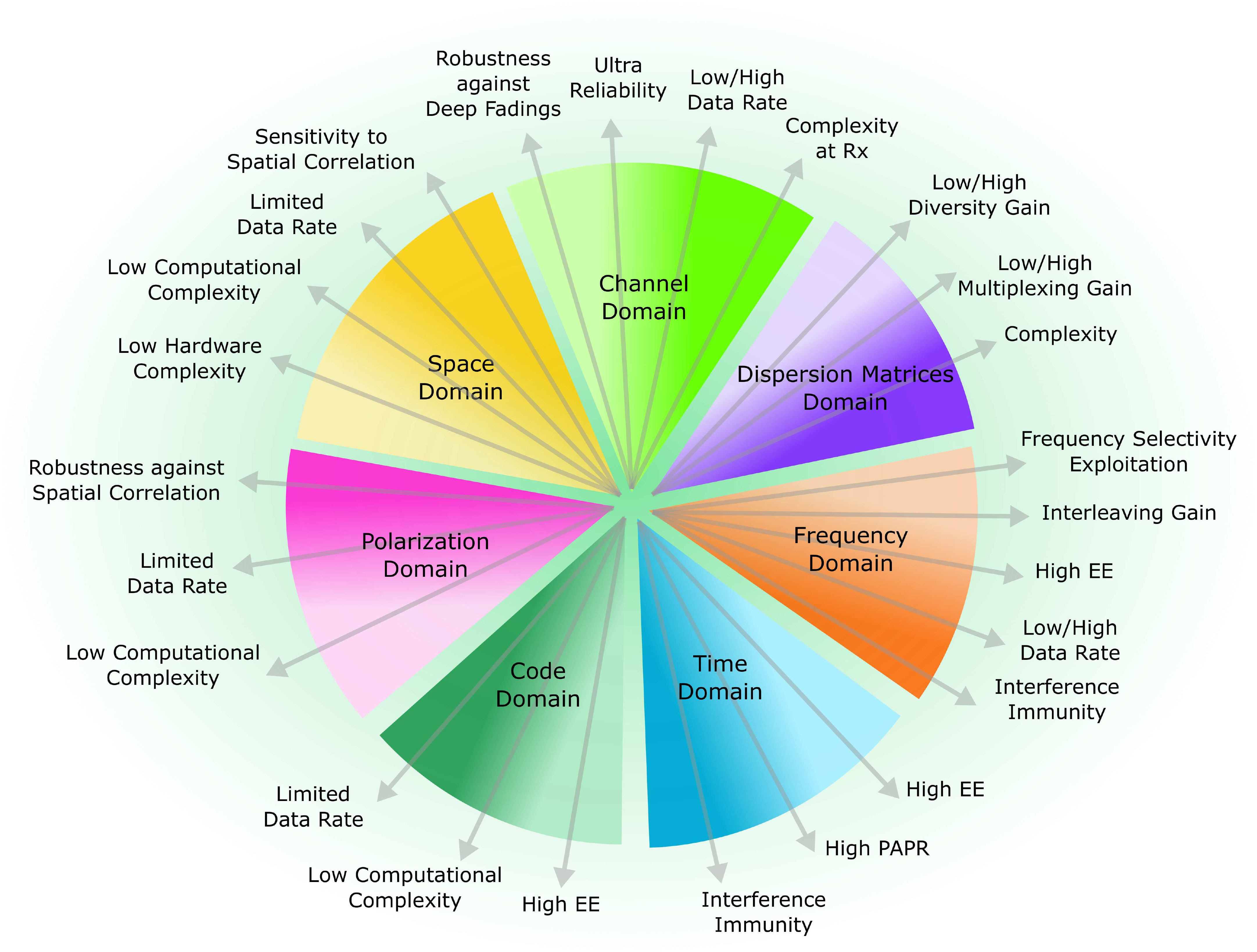}}
	\caption{\textcolor{black}{The key advantages and disadvantages of the existing IM domains.}}
	\label{Fig:IMDomainAdv}
\end{figure*}

{In modern communication systems, the hardware complexity of the Tx is dependent on the number of used RF chains and antennas. Also, the required channel estimation and detection mechanisms in the Rx lead to additional computational complexity, especially in the presence of a higher number of antennas.} The conventional MIMO systems, such as V-BLAST, have a complex structure due to the utilization of all Tx antennas and thus require an advanced Rx design. This structure makes its modification difficult for the diverse demands of 5G and beyond networks. 

{In general}, space domain IM schemes have similar transmission structure, i.e., the transmitted signals can be expressed in the vector form as in  (\ref{Eq:Eq1}), (\ref{Eq:Eq2}), and (\ref{Eq:Eq3}), where ML is commonly utilized at Rx for the detection of active antenna indices and $M$-ary symbols, as given in Table \ref{Tab:CompRateComplexityS}. It is worth noting that the complexity of ML detector exponentially increases with $N_t$, {while the achieved SE enhancement is logarithmic.} Moreover, a higher number of Tx antennas is required to achieve higher data rates, but it leads to a larger transceiver size and higher power consumption. \textcolor{black}{Therefore, space domain IM types are suitable for low data rates, high EE, and latency-critical transmission where the conventional MIMO technologies suffer. The spatial correlation between Tx antennas hurdles the performance of space domain techniques. In order to overcome this, its combination with polarization and channel domain IM types is promising.} Considering the main benefits and shortcomings of space domain IM types given in Table \ref{Tab:SpaceCompIMTypes}, GSM, MA-SM, QSM, and ESM can be easily integrated into a given MIMO system for the purpose of meeting a wide range requirements of 5G and beyond.

Frequency, time and code domain IM types do not require additional hardware for practical implementation, as presented in Table \ref{Tab:FTCCompIMTypes}. The complexity of conventional OFDM-IM using the LLR detector is in the same order as that of OFDM in terms of complex multiplications, as given in Table \ref{Tab:CompRateComplexityF} \cite{Basar1}. {In OFDM-ISIM and CI-OFDM-IM, block-interleaving, corresponding to the multiplication of OFDM block with an interleaving matrix, is easily applied in the frequency domain in order to break the correlation between the active subcarriers.} \textcolor{black}{Although DM-OFDM, MM-OFDM, and L-OFDM-IM result in a higher complexity than OFDM-IM, LC detectors are introduced and they can provide higher BER than OFDM due to the exploitation of frequency diversity}. \textcolor{black}{In this regard, frequency domain IM variants can easily serve different applications and use-cases with the proper type and parameter selection. }

\textcolor{black}{In contrast to the frequency domain, implementation of the interleaving in the time domain requires extra storage and increases the complexity at Tx. On the other hand, high reliability is difficult to achieve without interleaving gain. Thus, time domain IM types provide a low data rate while maintaining high EE, which is the primary concern for mMTC. In the similar vein, code domain and polar domain IM offers limited data rate, but their combination with space domain IM types is one of the candidate solutions for avoiding spatial correlation between the active antennas, while improving the SE.} 

{Channel domain IM provides a trade-off between ultra-reliability and complexity. The implementation of channel domain IM with the moderate number of RF mirrors is suitable for SIMO systems since the channel estimation at Rx becomes impractical in case of a higher number of RF mirrors.}

The major concern for multidimensional IM types is their multi-stage transceiver architecture.
Table \ref{Tab:CompIMTypes} compares the reviewed multidimensional IM types in terms of SE, EE, complexity, transmit diversity gain, multiplexing gain, and interference immunity. {Specifically, the complexity of ML detection tremendously increases due to the numerous possibilities of the active entities in different domains.} A moderate Rx design can be achieved for ISM-OFDM with a lower number of Rx antennas. \textcolor{black}{Fig. \ref{Fig:IMDomainAdv} summarizes the key advantages and disadvantages of available IM domains. Although multidimensional IM types are proposed for increasing the SE of IM systems, it is important to emphasize that proper combination of one-dimensional IM types can serve different requirements simultenously, as illustrated in Fig. \ref{Fig:IMDomainAdv}.}

\textcolor{black}{\section{Potential Challenges and Future Directions}}

\textcolor{black}{In this section, potential challenges and future perspectives on the integration of IM into beyond 5G technologies are anticipated. Although the research world of academia and industry has shown a substantial and explicit emphasis on the capacity and reliability of IM technologies, its utilization in a flexible manner needs to be fully explored in emerging wireless networks.}

\textcolor{black}{\subsection{Cognitive Radio Network through IM}}

\textcolor{black}{ CR has been heavily studied and considered as an enabling technology for dynamic spectrum access aiming to relax the problem of spectrum scarcity via a shared wireless channel between licensed and unlicensed users, i.e., primary user (PU) and secondary user (SU), respectively, and not to limit radio resources only to the license holders \cite{4796930, akyildiz2008survey, 5723803}. In the literature, three fundamental spectrum access methods named interweave, underlay, and overlay are adopted for CR technology \cite{goldsmith2009breaking}. Basically, SU is asked to sense and determine the possible availability of spectrum in which it can transmit without causing any interruption or endangering the legitimate user. The primary concern in such a system is handling the mutual interference originated by the overlapping of PU and SU. Here IM promises to control the interference through the fractional utilization of the available resources with different activation probabilities and with adaptive transmission power. On this basis, the application of SM in CR is relatively more understood in the literature \cite{7417481,7565136}, while the superiority of different IM techniques against conventional schemes used in CR networks remains unknown in the literature. The authors in  \cite{alizadeh2016performance} exploit the space domain to allow CR communications with the improvement of the performance of PU. SU performs a retransmission of the data symbols of PU over conventional $M$-ary symbols, while its own data are conveyed by the activated antenna indices. In line with this, a novel dual-hop SM technique is proposed in \cite{9055380}, where relay conveys its own information. In recent studies, frequency domain IM is exploited for opportunistic spectrum sharing. In \cite{8891788}, SU senses the inactive subcarriers of PU with OFDM-IM and performs its transmission over these subcarriers. Moreover, the transmission of PU is supported by means of an amplify-and-forward relay in \cite{8854299}. Flexible utilization of IM in CR reveals that the advantages of IM dominate its shortcomings when it is fully exploited. To exemplify, SE loss for SU becomes negligible as the reliability of PU is preserved or enhanced compared with conventional OFDM based technologies. Lastly, the utilization of different signal dimensions and types will be more prominent for the CR scenarios, where PU and SU have different priorities. Spectrum sharing between LTE license-assisted access (LTE-LAA) and IEEE 802.11 Wi-Fi systems is an active research area in the literature that can be considered as an example scenario \cite{8241371, 8468986}. It can be inferred from the above-mentioned scenarios that the study of IM in CR is at an incipient stage considering its multidimensional application, and there exist a plethora of trade-offs between PU and SU, which need to be exhibited by academia and industry.}

\textcolor{black}{\subsection{Cooperative Networks through IM}}

\textcolor{black}{Cooperative communication is an alternative enabling approach of exploring spatial diversity, which is achievable via the transmission of data from uncorrelated channels \cite{nosratinia2004cooperative}. In other words, the transmitted signal is exposed to various channel fades and consequently reducing the possibility of facing deep fading. In particular, a cooperative user not only transmits its own data but also conveys the data of its corresponding partner, in order to improve diversity gain, coverage area, and data rate. Thus, investigation of mutual interest between cooperative networks and IM  has brought a new interest in the literature. A significant reliability gain is achieved from the combination of SM and cooperative networks compared to conventional $M$-ary modulation \cite{altin2016performance}. Power allocation strategy is developed for frequency domain IM in cooperative networks to maximize the network capacity \cite{8269167}. SM-aided cooperative NOMA scheme is investigated in order to provide effective multiple access while ensuring the low complexity and low power consumption \cite{8636968}. In the same vein, information-bearing units of OFDM-IM, i.e., the utilized subcarriers and their indices, are used to carry the data of both paring users, respectively, in \cite{9078069}. In this way, IUI is avoided in cooperative-based NOMA, alleviating the receiver complexity at the cost of a reduction in SE. The sparse representation of OFDM-IM is proposed to mitigate and control self-interference at relaying user while performing full-duplex (FD) communication in cooperative systems \cite{8476574}. Furthermore, in \cite{8579221}, it is shown that the error performance of FD relay with media-based modulation outperforms classical $M$-ary constellations due to extra channel diversity under the same SE. As observed from the aforementioned literature, a limited number of studies are available in the literature for the utilization of different IM domains, and a gap exists in the works assessing the performance of multidimensional IM in cooperative networks. Therefore,  it is beneficial to give more attention to IM-aided cooperative networks for improving coverage area,  cell-edge performance, and data rate under moderate system complexity and power consumption.}
	
\textcolor{black}{\subsection{Non-orthogonal Transmission through IM}}

\textcolor{black}{Non-orthogonal communications, where the data message of one or multiple users is exposed to interference, is a classical problem of communication systems. Previously, interference was undesired and always avoided. In today's communication systems, it is difficult to avoid due to the increased number of users, and its compensation leads to high complexity at Rx. In fact, the non-orthogonal transmission is intentionally generated in emerging technologies including NOMA, GF access, and FD communication. However, having effective multiple access under non-orthogonal conditions is still a conundrum in the literature, which encourages the researchers to seek new strategies that allow the utilization of available resources by multiple users and not causing heavy computational complexity at Rx side. One of the well-known approaches is power domain NOMA, where multiplexing is employed in the power domain at Tx and successive interference cancellation (SIC) is utilized for demultiplexing at Rx. Despite all the efforts, power domain NOMA leads to catastrophic reliability when the overlapped users have similar channel gains \cite{7676258}. The partial utilization of available resources in IM concept enables the power control in active subcarriers and relaxes the dependency of NOMA on channel gain difference  \cite{8352624, 9027834}. Different than classical NOMA with wide-band interference, IM-based NOMA results in sparse narrow-band interference, which is why it offers better radio resource utilization. Hence, there is a growing interest in IM-based multiple access (IMMA) \cite{8809839, 8681607}. Unfortunately, the existing studies do not fully utilize the flexibility of  IM and are very limited considering the richness of IM techniques. For example, in GF access over the resource pool, where multiple users perform transmission, adjustment of different OFDM-IM parameters, such as subcarrier activation ratio and the number of active subcarriers, will allow the control of interference caused by the collision between multiple users. In this sense, it is worthy to put further research endeavors for having a plethora of gains such as high reliability, low latency, and low complexity through IM in non-orthogonal transmission. }

\textcolor{black}{\subsection{Security in/with IM}}

\textcolor{black}{A doubtless trust is expected in beyond 5G wireless networks such as vehicular communications, health servicing, or other critical data information carriers. However, the broadcast nature of wireless systems makes the privacy and secrecy of these designs suspicious. For this reason, physical layer security (PLS) has emerged as a new powerful alternative that can complement or even replace cryptography-based approaches \cite{hamamreh2018classifications}. Basically, in the literature, the exploitation of channel features and the design of specific transceiver architectures are utilized in order to both provide reliable communication for the desired user and prevent the data detection by unwanted users, i.e., eavesdroppers. Although the possibility of ensuring reliability and security simultaneously has long excited the researchers, the performance of existing IM techniques is not elaborated from the perspective of security as it is ignored during 5G standardizations. Moreover, the mapping of data bits into the information-bearing entities in IM allows a chance to exploit the channel characteristics and consequently achieve the secrecy gap between the desired user and eavesdropper. For instance, in OFDM-IM, channel-based randomization is explored for the mapping of information bits to the entities in order to confuse the eavesdropper \cite{lee2017secure}.  The secrecy gap is increased through the joint decision of modulation type and activation ratio regarding the SNR level at Rx side \cite{furqan2018adaptive}. In the space domain, secrecy enhancement is offered for the data carried by antenna indices and data symbols via the rotation of the antenna indices and constellation symbols at the legitimate transmitter \cite{jiang2017secrecy}. In the case of time-division duplexing, the channel information of the legitimate user can not be learned by eavesdropper since the channel is reciprocal and not feed to Tx. It is expected that the exploitation of the reciprocal channel feature with time domain IM can provide further improvement in security. Additionally, PLS approaches based on artificial noise and transceiver impairments need to be investigated for different IM domains. Thus, considering the aforementioned possible applications of IM, enormous research potential challenges exist in this newly emerging research field of IM.}


\textcolor{black}{\subsection{Intelligent Wireless Communications through IM}}

\textcolor{black}{\enquote{Intelligent} and \enquote{smart} are the keywords to define the expectations from beyond 5G networks and consequently are used frequently in recent studies. A smart network is expected to adapt itself to user requirements, environment, and channel conditions. In such a system, the scalable structure of IM becomes more attractive. Here there are two standpoints that IM can be the pioneer for the smartness or assist the existing intelligent network. One of the main features of intelligent networks is the channel control capability. Channel domain IM, i.e., MBM, enables to create different channel realizations to transmit the data bits \cite{BasarMBM}. These channel states allow us to have plenty of advantages including low complexity, high data rate, and ultra-reliability. However, the channel control capability of MBM is only dependent on the incoming bits. Therefore, channel domain modulation should be exploited not only for data transmission but also for the control of channel characteristics, such as delay and Doppler spread in order to overcome the problems of wireless communication systems. Although this requires profound thinking, channel domain modulation can be considered as a candidate solution to have intelligent communications over future systems.}
\textcolor{black}{ Furthermore, the recently reputed RIS concept controls the propagation environment to increase the quality of service at Rx \cite{lis1, 8796365, 8466374, 8801961}. In recent studies, IM is coupled with RIS-empowered communication to exploit the advantages of both technologies. In this regard, three scenarios including IM over Tx antennas, IM over RIS, and IM over Rx antennas are introduced in \cite{8981888}. RIS-assisted beam-index modulation \cite{gopi2020intelligent} and SSK \cite{canbilen2020reconfigurable} are proposed for avoiding line-of-sight blockage in mmWave frequency bands and achieving high EE with high reliability, respectively. Since the utilization of RIS is an early stage, its amalgamation with IM needs an intensive research effort to speculate the potential benefits and drawbacks. Therefore, it is beneficial to give substantial attention to design smart communication systems via the combination of RIS and IM concepts.}

\textcolor{black}{\subsection{Investigation of Novel IM Techniques}}

\textcolor{black}{As presented throughout the survey, a plethora of studies on the IM concept is available in the literature. The majority of the studies provide their comparison with the well-adopted systems in terms of SE and BER. It should be noted that users had more or less the same priorities in previous wireless communication systems so that service providers had guaranteed their satisfaction. The emerge of different necessities hurdles the simultaneous happiness of the existing users in a network. IM concept reveals a novel perspective to wireless systems, where the physical entities can convey information. Moreover, it is adaptive and controlled by the incoming data bits. In other words, the rationale behind IM arises from the flexibility. Thus, it is essential and beneficial to overview the general picture and investigate novel IM technologies to serve multiple requirements simultaneously.
	For 6G and beyond networks, it is envisioned that there will be applications and use cases that can not be categorized under the eMBB,  mMTC, and URLLC.  Thus, the multidimensional application of IM  opens the door for the system design to promote multiple demands. As can be seen in Fig. \ref{Fig:ımtypes}, IM application on both RF mirrors and subcarriers, i.e., corresponds to MBM with frequency domain IM, is not explored yet, and can be coupled to jointly support eMBB and URLLC services. Also, the potential of available IM techniques is not well-understood for serving the different combinations of these services. In addition to the foregoing approaches, sparse design in each dimension provides a degree of freedom in order to reap the advantages of multidimensional IM types and to facilitate the implementation of IM in current communication systems \cite{8284057}. For instance, in \cite{8656571}, CS is integrated into Tx and Rx side of two dimensional IM scheme for the sake of increasing the data rate and system flexibility, while reducing the complexity of ML detector at Rx. Moreover, CS-aided IM are investigated in the literature for SM \cite{8107720}, QSM \cite{8680042}, STBC-QSM \cite{8411498}, STSK \cite{8493475, 8322306}. In recent study \cite{wang2020deep}, neural network-based IM detector is introduced to relax ML complexity. As noticed from the above-reviewed literature, many CS-based research studies have been focusing on space domain IM types, but its applications for the other IM domains are missing in the literature. Moreover, the utilization of neural network-based strategies, machine learning, and deep learning for multidimensional IM types alongside one-dimensional IM types are noteworthy and virgin research area for the researchers worldwide. Hence, the researchers need to make further efforts on IM transceiver design to facilitate its implementation and the detection of active entities and $M$-ary symbols at Rx.}

\vspace{2mm}

\section{Conclusion}

This study has presented the recent research progress on one-dimensional and multidimensional IM techniques. Moreover, considerations regarding the utilization of these IM schemes for next-generation wireless communication are provided. Specifically, IM schemes presented in the literature have been firstly grouped regarding their application dimensions. Later, we have categorized these IM techniques considering the requirements of eMBB, mMTC, and URLLC including  SE, EE, system complexity, reliability, and latency.

In addition to the information bits carried by QAM/PSK, IM carries extra bits by utilizing the indices of information-bearing transmit entities. As a result, an increase in the number of utilized IM entities such as time slots, subcarriers, Tx antennas, and RF mirrors offers a high SE for IM-based schemes. Hence, multidimensional IM options have been considered as promising solutions for eMBB applications and use-cases. At this point, we conclude that further research on the joint utilization of other new possible entities will be beneficial to enhance SE of IM systems.

Additional bits conveyed over the indices of active entities do not require extra power for transmission and thus provide a high EE for IM-based systems.  Transceiver complexity is increased by the distribution of information bits over multiple domains, which specifically increases the Rx complexity. Thus, mainly one-dimensional IM variants appear as candidate solutions for mMTC applications and use-cases. Additionally, IM techniques with simple Rx structure, such as DMBM and TI-SM, can be also considered.

High diversity gain, interference immunity, and fast processing time are the main priorities for URLLC applications and use-cases. Thus, IM schemes, such as GSM, QSM, STBC-SM, and STSK are suitable candidates.

To sum up, it has been demonstrated that the presented categorization of IM techniques considering broad range demands of next-generation wireless systems can be considered as a reference point for new solutions in 5G and beyond technologies.

\section*{Acknowledgment}

This work was supported in part by the Scientific and Technological Research Council of Turkey (TUBITAK) under Grant 218E035.

\bibliographystyle{IEEEtran}
\bibliography{seda}



\begin{IEEEbiography}[{\includegraphics[width=1in,height=1.25in,clip,keepaspectratio]{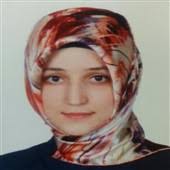}}]{\textbf{Seda~Do\u{g}an}} received the B.Sc. degree in electronics and telecommunication engineering from Kocaeli University, Kocaeli, Turkey, in 2015. She is currently pursuing the Ph.D. degree as a member of the Communications, Signal Processing, and Networking Center (CoSiNC) at Istanbul Medipol University, Istanbul, Turkey. Her research interests include index modulation, millimeter-wave frequency bands, non-orthogonal multiple accessing (NOMA), and random access techniques for next-generation wireless networks. 
\end{IEEEbiography}

\begin{IEEEbiography}[{\includegraphics[width=1in,height=1.25in,clip,keepaspectratio]{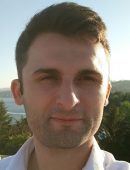}}]{\textbf{Armed Tusha}} 
	received the B.Sc. degree in electrical and electronics engineering from Istanbul Sehir University, Istanbul, Turkey, in 2016, and the M.Sc. degree from Istanbul Medipol University, Istanbul, Turkey, in 2018. He is currently pursuing the Ph.D. degree as a member of the Communications, Signal Processing, and Networking Center (CoSiNC) at Istanbul Medipol University. His research interests include index modulation, digital communications, signal processing techniques, multicarrier
	schemes, orthogonal/non-orthogonal multiple accessing (OMA/NOMA) in wireless networks, and channel coding.
\end{IEEEbiography}

\begin{IEEEbiography}[{\includegraphics[width=1in,height=1.25in,clip,keepaspectratio]{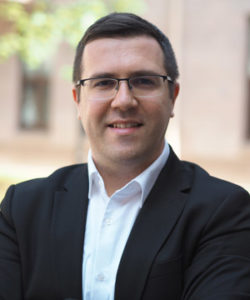}}]{\textbf{Ertugrul Basar}} (S'09-M'13-SM'16) received the B.S. degree (Hons.) from Istanbul University, Turkey, in 2007, and the M.S. and Ph.D. degrees from Istanbul Technical University, Turkey, in 2009 and 2013, respectively. He is currently an Associate Professor with the Department of Electrical and Electronics Engineering, Ko\c{c} University, Istanbul, Turkey and the director of Communications Research and Innovation Laboratory (CoreLab). His primary research interests include MIMO systems, index modulation, intelligent surfaces, waveform design, visible light communications, and signal processing for communications.
	Recent recognition of his research includes the Science Academy (Turkey) Young Scientists (BAGEP) Award in 2018, Mustafa Parlar Foundation Research Encouragement Award in 2018, Turkish Academy of Sciences Outstanding Young Scientist (TUBA-GEBIP) Award in 2017, and the first-ever IEEE Turkey Research Encouragement Award in 2017.
	Dr. Basar currently serves as a Senior Editor of the \textsc{IEEE Communications Letters} and the Editor of the \textsc{IEEE Transactions on Communications} and \textit{Physical Communication}.
\end{IEEEbiography}

\begin{IEEEbiography}[{\includegraphics[width=1in,height=1.25in,clip,keepaspectratio]{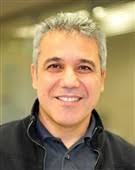}}]{\textbf{H\"{u}sey\.{i}n Arslan}}
	(S'95-M'98-SM'04-F'15) received the B.S. degree from Middle East Technical University, Ankara, Turkey, in 1992, and the M.S. and Ph.D. degrees from Southern Methodist University, Dallas, TX, USA, in 1994 and 1998, respectively. From 1998 to 2002, he was with the Research Group, Ericsson Inc., NC, USA, where he was involved with several projects related to 2G and 3G wireless communication systems. Since 2002, he has been with the Electrical Engineering Department, University of South Florida, Tampa, FL, USA. He has also been the Dean of the College of Engineering and Natural Sciences, Istanbul Medipol University, since 2014. He was a part-time Consultant for various companies and institutions, including Anritsu Company, Morgan Hill, CA, USA, and The Scientific and Technological Research Council of Turkey (T\"{U}B\.{I}TAK). His research interests are in physical layer security, mmWave communications, index modulation, small cells, multicarrier wireless technologies, co-existence issues on heterogeneous networks, aeronautical (high-altitude platform) communications, in vivo channel modeling, and system design. 
\end{IEEEbiography}

\end{document}